\documentclass[twocolumn,twocolappendix]{aastex631}

\usepackage{natbib}
\usepackage{graphicx}	
\usepackage{amsmath}	
\usepackage{amssymb}	
\usepackage{bm}	
\usepackage{subfigure}
\usepackage{longtable}
\usepackage{array}
\usepackage{booktabs}
\usepackage{multirow}
\usepackage{xcolor}
\usepackage{lipsum}

\newcommand{\msun}{\,{M_{\odot}}}

\newcommand{\sgrb}{\textit{s}GRB }
\newcommand{\sgrbs}{\textit{s}GRBs }
\newcommand{\sgrbss}{\textit{s}GRB's }

\shorttitle{Late engine's cocoon, and the blue KN}
\shortauthors{Hamidani, Kimura, Tanaka, Ioka}

\begin{document}
\title{Late Engine Activity in Neutron Star Mergers and Its Cocoon: An Alternative Scenario for the Blue Kilonova}

\correspondingauthor{Hamid Hamidani}
\email{hhamidani@astr.tohoku.ac.jp}

\author[0000-0003-2866-4522]{Hamid Hamidani}
\affiliation{Astronomical Institute, Graduate School of Science, Tohoku University, Sendai 980-8578, Japan}

\author[0000-0003-2579-7266]{Shigeo S. Kimura}
\affiliation{Frontier Research Institute for Interdisciplinary Sciences, Tohoku University, Sendai 980-8578, Japan}
\affiliation{Astronomical Institute, Graduate School of Science, Tohoku University, Sendai 980-8578, Japan}

\author[0000-0001-8253-6850]{Masaomi Tanaka}
\affiliation{Astronomical Institute, Graduate School of Science, Tohoku University, Sendai 980-8578, Japan}

\author[0000-0002-3517-1956]{Kunihito Ioka}
\affiliation{Yukawa Institute for Theoretical Physics, Kyoto University, Kyoto 606-8502, Japan}

\begin{abstract}
Follow-up observations of short gamma-ray bursts (\textit{s}GRBs) have continuously unveiled late extended/plateau emissions, attributed to jet launch due to late engine activity, the nature of which remains enigmatic. 
Observations of GW170817 confirmed that \textit{s}GRBs are linked to neutron star (NS) mergers, and discovered a kilonova (KN) transient. 
Nevertheless, the origin of the early ``blue" KN in GW170817 remains unclear. 
Here, we investigate the propagation of late jets in the merger ejecta. 
By analytically modeling jet dynamics, we determine the properties of the jet heated cocoon, and estimate its cooling emission. 
Our results reveal that late jets generate significantly brighter cocoons compared to prompt jets, primarily due to reduced energy loss by adiabatic cooling. 
Notably, 
with typical late jets,
emission from the cocoon trapped inside the ejecta can reproduce the blue KN emission. 
We estimate that the forthcoming Einstein Probe mission will detect the early cocoon emission with a rate of $\sim 2.1_{-1.6}^{+3.2}$ yr$^{-1}$,
and that optical/UV follow-ups in the LIGO-VIRGO-KAGRA O5 run will be able to detect $\sim 1.0_{-0.7}^{+1.5}$ cocoon emission events.
As an electromagnetic counterpart, this emission provides an independent tool to probe NS mergers in the Universe, complementing insights from \textit{s}GRBs and gravitational waves.
\end{abstract}

\keywords{Gamma-ray bursts (629), Neutron stars (1108), Hydrodynamics (1963), Relativistic fluid dynamics (1389), Relativistic jets (1390),  Gravitational waves (678), X-ray transient sources (1852)}

\section{Introduction}
\label{sec:1}
\textit{Short} gamma-ray bursts (\textit{s}GRBs) have traditionally been classified based on the short duration ($<$ 2 s) of their hard gamma-ray emission (\citealt{1993ApJ...413L.101K}), also referred to as ``prompt" emission.
\textit{s}GRBs have theoretically been associated with binary neutron star (BNS; also black hole neutron star: BH-NS) mergers (\citealt{1986ApJ...308L..43P}; \citealt{1986ApJ...308L..47G}; \citealt{1989Natur.340..126E}).
In this scenario, the merger produces a compact object surrounded by an accretion disk.
This system is short lived, and is considered as the central engine that powers relativistic jets and explain the short nature of \textit{s}GRBs' prompt emission.
Hence, \textit{s}GRBs have been regarded as a tool to study merging BNS systems in the Universe.

Additionally, BNS mergers have been proposed as a site for the rapid neutron-capture process (r-process) nucleosynthesis thanks to the neutron rich content of NSs \citep{1989Natur.340..126E}.
Simulations showed that a substantial amount of mass ($\sim 10^{-3} - 10^{-2} M_\odot$) is ejected during the merger and ideal conditions for r-process are achieved (\citealt{1999PhRvD..60j4052S}; \citealt{2000PhRvD..61f4001S}; \citealt{2013PhRvD..87b4001H}; \citealt{2013ApJ...773...78B}; \citealt{2016MNRAS.460.3255R}; etc.).
Later on, it has been suggested that radioactive decay of freshly synthesised heavy elements can power a unique ultraviolet/optical/infrared transient called macronova/kilonova (KN hereafter) [\citealt{1998ApJ...507L..59L}; \citealt{2005astro.ph.10256K}; \citealt{2010MNRAS.406.2650M}].
The KN was expected to take a red color, due to the high opacity of heavy elements synthesized (i.e., lanthanides; \citealt{2013ApJ...775...18B,2013ApJ...775..113T,2013ApJ...774...25K}).

In 2017, a BNS merger event (GW170817) was observed for the first time with two LIGO GW detectors assisted by the VIRGO detector (\citealt{2017PhRvL.119p1101A}).
About $1.7$ s after the merger, a \sgrb (GRB 170817A) was detected (\citealt{2017ApJ...848L..13A}). 
Follow-up observations in X-ray (\citealt{2017ApJ...848L..20M}), optical (\citealt{2018NatAs...2..751L}), and radio (\citealt{2018Natur.561..355M,2019MNRAS.489.1919T}) confirmed the presence of a relativistic jet.
This further confirmed that \sgrbs are produced by BNS mergers (\citealt{1986ApJ...308L..43P}; \citealt{1986ApJ...308L..47G}; \citealt{1989Natur.340..126E}).

About 10 h after the merger, 
follow-up observations through the entire electromagnetic (EM) spectrum discovered a transient, referred to as AT2017gfo, similar to the theoretically predicted r-process powered KN  (\citealt{2017ApJ...848L..33A}; \citealt{Chornock:2017sdf}; \citealt{Coulter:2017wya}; \citealt{2017ApJ...848L..29D}; \citealt{2017Sci...358.1570D}; \citealt{Kilpatrick:2017mhz}; \citealt{2017Sci...358.1559K}; \citealt{2017ApJ...848L..18N}; \citealt{Pian:2017gtc}; \citealt{2017Natur.551...75S}; \citealt{Shappee:2017zly}; \citealt{2017ApJ...848L..16S}; \citealt{2017PASJ...69..102T}; \citealt{2017PASJ...69..101U}; \citealt{2017ApJ...848L..24V}; etc.). 

However, simple fitting of the early part revealed the need for a blue KN component (requiring $\sim 0.2\:c$, $\sim 0.02\:\msun$, and $\kappa\sim0.1–1$ cm${^2}$ g$^{-1}$; see \citealt{2017Sci...358.1559K,2017ApJ...851L..21V}) with an unexpectedly fast velocity and low opacity compared to theoretical predictions (e.g., see \citealt{2014MNRAS.441.3444M,2017PhRvD..96l3012S,2018ApJ...860...64F}).
Different scenarios have been suggested to explain this component, e.g.,
wind from a rapidly rotating and highly magnetized hyper-massive NS remnant (i.e., magnetar; \citealt{2018ApJ...856..101M}), 
post-merger (viscous/MHD) disk wind (\citealt{2017PhRvD..96l3012S,2020ApJ...901..122F}),
jet wind interaction (\citealt{2021MNRAS.500.1772N,2023arXiv230312284C}; also see \citealt{2023MNRAS.523.2990S}),
geometrical distribution with realistic opacities 
(\citealt{2018ApJ...865L..21K}), etc. 
Still, the origin of this blue KN component is under debate.

\begin{figure*}
    \centering
    \includegraphics[width=0.69\linewidth]{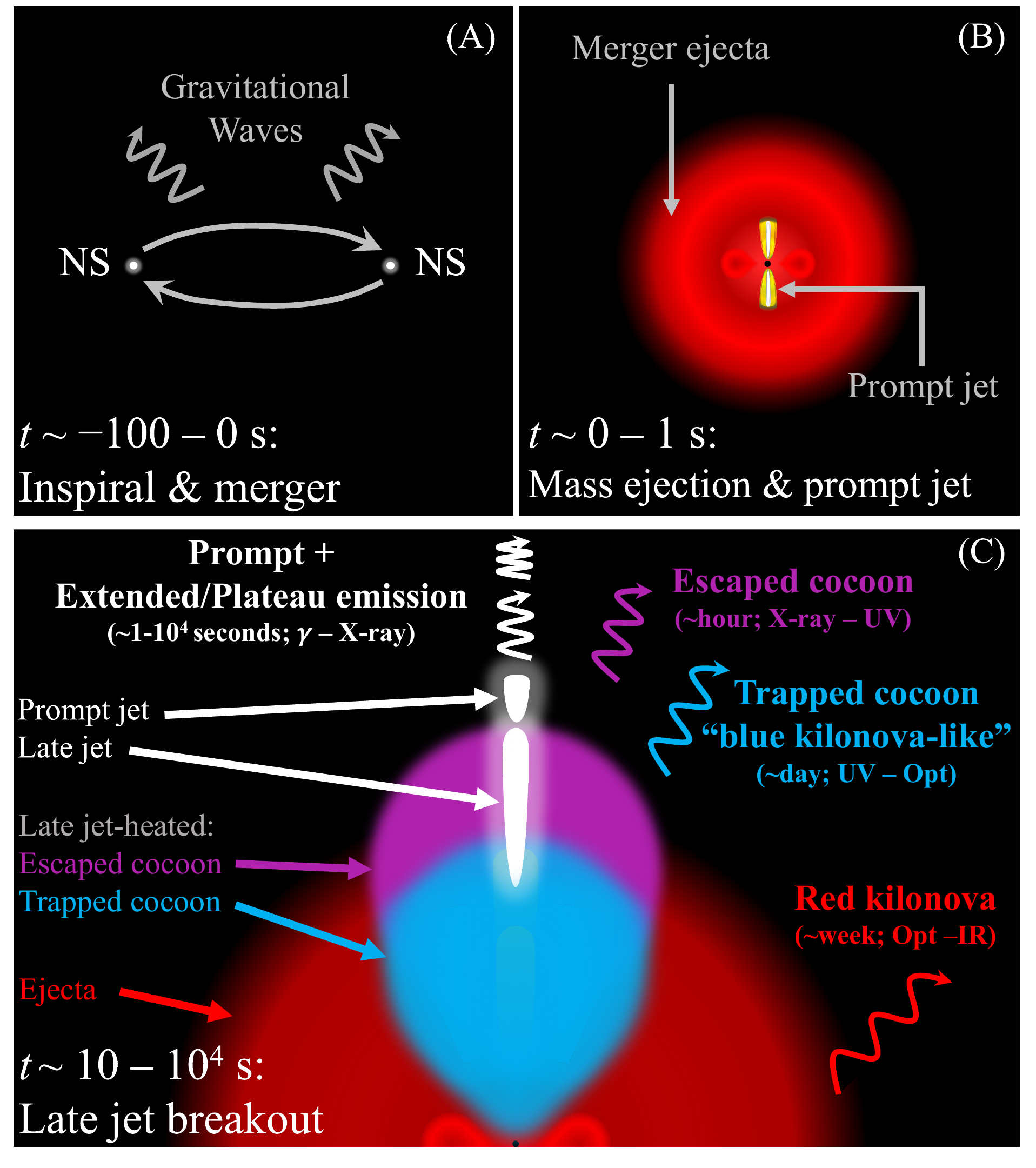} 
  \caption{Schematic illustration of a late jet heated cocoon in BNS mergers, and its cooling emission. 
  After the merger (A), mass is ejected and a \textit{s}GRB's prompt jet is launched within $\sim 1$ s (B). 
  At much later times, $\sim 10-10^4$ s, jets associated with extended/plateau emission, commonly observed in \textit{s}GRBs (see Figure \ref{fig:intro}), are launched (C).
  These jets (white) shock the ejecta (red), heating it up producing a hot cocoon.
  There are two cocoon components: i) a component that is fast enough to escape the ejecta, named ``escaped cocoon" (purple); and ii) a component that is slower and is trapped inside the ejecta, named ``trapped cocoon" (light blue).
  Cooling emission from the escaped cocoon is expected to produce soft X-ray to UV photons.
  For the trapped cocoon heated by late jets, cooling emission could be similar to a blue KN component.
  The r-process heated ejecta and its red KN emission (red) is also shown.
  }
  \label{fig:f0} 
\end{figure*}

In addition to the merger ejecta produced by the merger, there is the \textit{s}GRB-jet launched by the merger remnant.
The \sgrbss prompt emission is produced by emission from this jet in the line-of-sight (\citealt{1975NYASA.262..164R}).
In NS mergers, mass is ejected dynamically before the merger (tidal interaction) and immediately after the merger ($\lesssim$ 10 ms; \citealt{2013PhRvD..87b4001H,2019ARNPS..69...41S}). 
Hence, as shown in Figure \ref{fig:f0} (panel B), the jet is initially surrounded by the dense ejecta, and it had to propagate through this dense ejecta to produce the observed \sgrb prompt emission (panel C).
This is known to produce a jet shock that heats up a part of the ejecta forming a ``cocoon" component (in AGNs: \citealt{1989ApJ...345L..21B}; in collapsars \citealt{1997ApJ...479..151M}; \citealt{1999ApJ...524..262M}; \citealt{2003MNRAS.345..575M}; and in \sgrbs \citealt{2014ApJ...784L..28N,2014ApJ...788L...8M,2015ApJ...813...64D,2020MNRAS.495.3780N,2021MNRAS.502.1843N,2021MNRAS.500..627H,2022MNRAS.517.1640G}).
Hence, the jet, with its large energy budget, is another element to take into account in NS mergers (\citealt{2017ApJ...834...28N,2017Sci...358.1559K,2018MNRAS.473..576G}).

This prompt jet heated cocoon has been proposed to explain the blue KN in GW170817/AT2017gfo \citep{2018ApJ...855..103P,2018PTEP.2018d3E02I}.
\cite{2018ApJ...855..103P}, claimed that AT2017gfo shows evidence for the prompt jet cocoon.
However, this contradicts simulation results, showing that for prompt \textit{s}GRB-jets most of the cocoon is trapped inside the ejecta (see Figure \ref{fig:f0}) and only a small fraction of its mass ($<10\%$) manage to escape ($\sim 10^{-4}-10^{-6}\msun$; see Table 2 in \citealt{2023MNRAS.520.1111H}).
This small mass implies that the timescale of emission from the escaped cocoon is too short ($\sim 1$ h) to explain the blue KN ($\sim 1$ d), and by the time the escaped cocoon is transparent, the trapped cocoon has adiabatically cooled down to the point of being too dim to be relevant (\citealt{2023MNRAS.520.1111H,2023MNRAS.524.4841H}).
Therefore, parameters of the prompt jet's cocoon assumed in \cite{2018ApJ...855..103P} are questionable (in particular mass as $M\sim 0.01 M_{\odot}$).

\begin{figure*}
    \centering
    \begin{subfigure}
    \centering
    \includegraphics[width=0.49\linewidth]{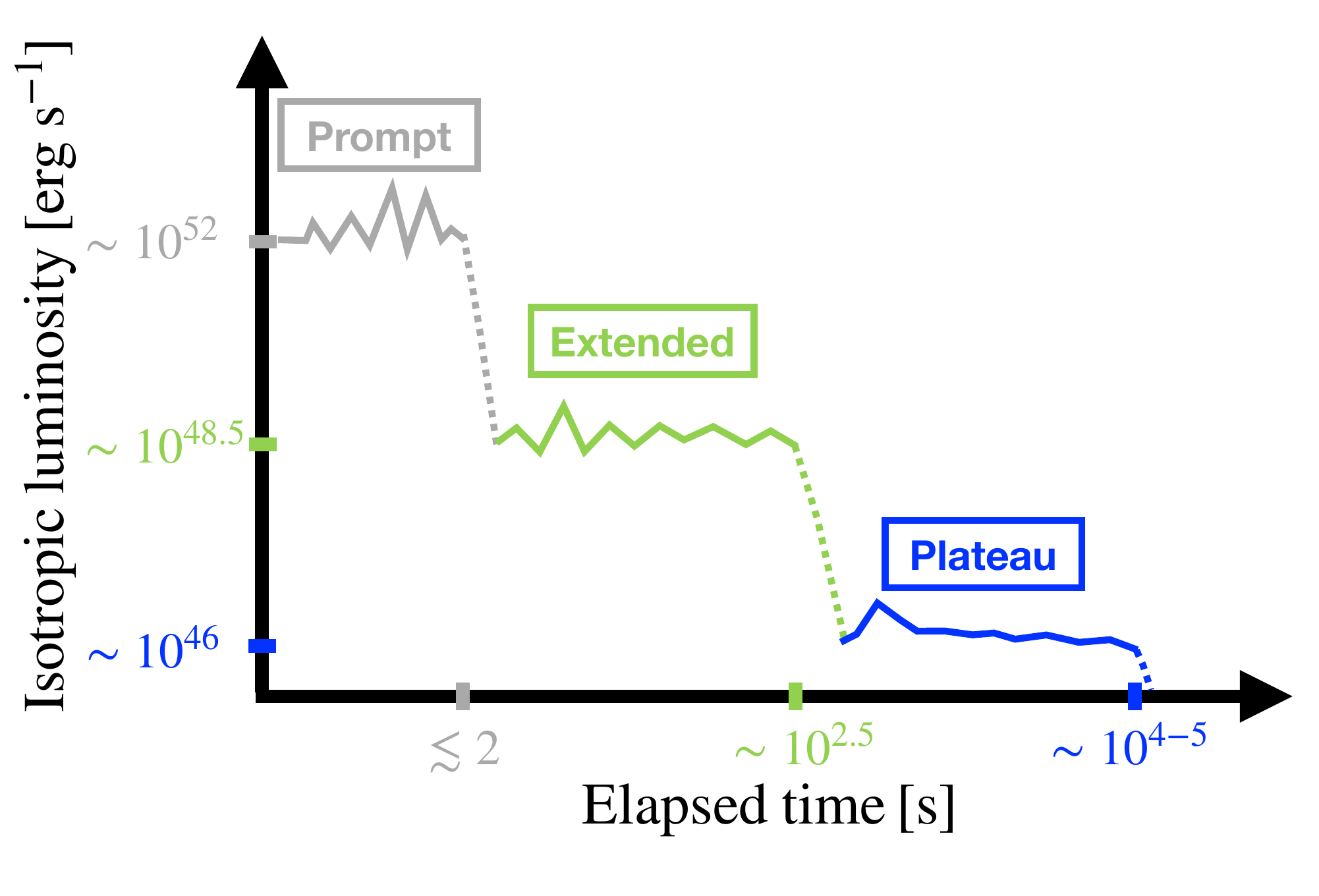}
    \includegraphics[width=0.49\linewidth]{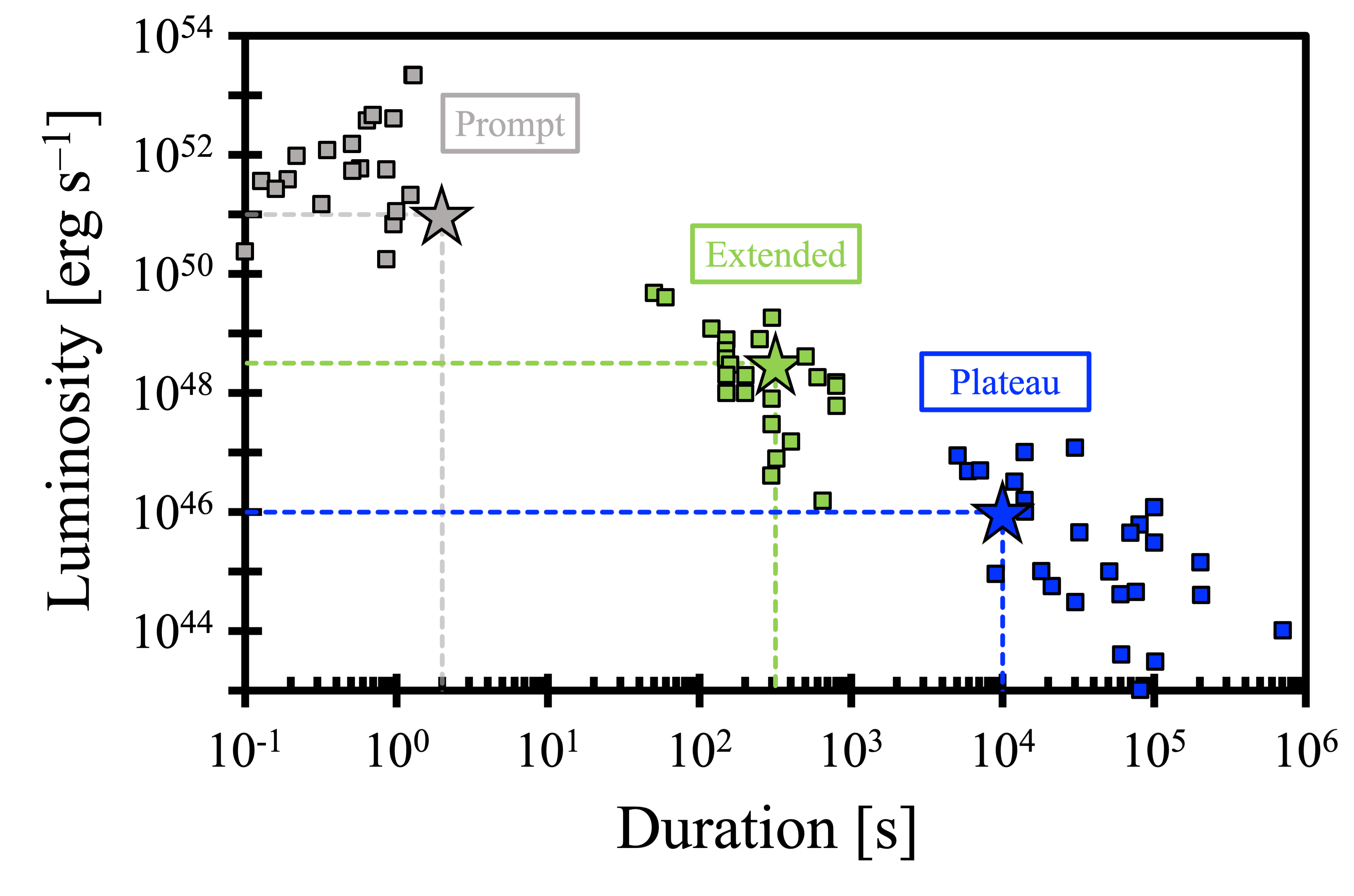}
    \centering
  \end{subfigure}
  \caption{On left: a schematic light curve of an observed \sgrb with late engine activity. 
  Prompt (light grey), extended (green), and plateau (blue) phases are shown, in terms of their observed isotropic equivalent luminosities, and time (since the prompt burst trigger time). 
  Dashed lines illustrate 
  the transition between the different phases, which is often interpreted as the end of energy injection (\citealt{2005ApJ...631..429I}).
  On the right: A sample of observed (squares) \sgrbss prompt (grey), extended (green), and plateau (blue) events as a function of their observed luminosity and duration (with redshift measurement; taken from \citealt{2017ApJ...846..142K} and \citealt{2018ApJ...852L...1Z}).
  Typical luminosities (and duration), indicated with dashed lines and star symbols, are adopted throughout this paper (assuming $\eta_\gamma \sim 0.1$; see §\ref{sec:3} and Table \ref{tab:models}).
  }
  \label{fig:intro}
\end{figure*}

After the prompt phase, interaction of the prompt-jet with the circumstellar medium produces a softer emission, from X-ray to radio, called ``afterglow" (\citealt{
1998ApJ...497L..17S}).
This emission is expected to decay over time following a simple power-law function (e.g., see Figure 1 in \citealt{2006ApJ...642..354Z}). 
However, follow up observations in X-ray show an excess luminosity after the prompt phase in most \textit{s}GRBs (\citealt{2005ApJ...635L.133B,2006ApJ...643..266N}; etc.).
As shown in Figure \ref{fig:intro}, these phases (referred to as ``extended" and ``plateau") are unusually flat over time, in terms of observed luminosity, before decaying (also see Figure 1 in \citealt{2013MNRAS.431.1745G}; Figure 1 in \citealt{2017ApJ...846..142K}; Figures 1 and 2 in \citealt{2019ApJ...877..147K}; although time evolution can be different for some events: \citealt{2015MNRAS.452..824K,
2016ApJ...829....7L,2022ApJ...936L..10Z}).
As a result, explaining these phases with afterglow modeling is challenging, and their origin has generally been attributed to late engine activity  (\citealt{2005ApJ...631..429I}).
\cite{2015ApJ...804L..16K} showed that late engine activity can be explained with late time mass accretion (see their Figure 2). 
In a recent study, \cite{2023ApJ...958L..33G} showed using self-consistent end-to-end simulations that late engine activity (extended phase) naturally arises as the disk transitions into a magnetically arrested disk (MAD) state after the prompt phase (see their Figure 1).

These late phases have been identified in many \sgrbs ($\sim 70\%$ of Swift/BAT \sgrbs according to \citealt{2017ApJ...846..142K}).
Also, recent observations of GRB 211211A and GRB 230307A (long GRBs but with potential kilonova candidate) show evidence of late emission in the form of a soft tail (for GRB 211211A: \citealt{2022Natur.612..223R}; 
\citealt{2022Natur.612..228T,2022Natur.612..232Y}; and for GRB 230307A:
\citealt{2023arXiv230702098L};
\citealt{2023arXiv230705689S};
\citealt{2023arXiv230800633G}; 
\citealt{2023arXiv230800638Y}).
Hence, in BNS mergers, after the prompt phase, late engine activity is expected to continue in most cases.

The observed luminosity and duration of both the prompt phase and these subsequent late phases are shown in Figure \ref{fig:intro} (right panel) for a sample of \sgrbs (taken from: \citealt{2017ApJ...846..142K} and \citealt{2018ApJ...852L...1Z}).
The observed luminosities and duration show some scatter, however, 
in general, compared to the prompt emission that shines at $\sim 10^{51}-10^{52}$ erg s$^{-1}$ for $\lesssim 2$ s,
the observed extended emission is on the order of $\sim 10^{48}-10^{49}$ erg s$^{-1}$ and lasts for $\sim 10^{2.5}$ s; while the plateau emission is about $\sim 10^{46}$ erg s$^{-1}$ for $\sim 10^{4}-10^6$ s.

Details about this late engine activity are not clear.
The flat (or shallow) temporal behaviour of the observed luminosity during these two late phases (extended and plateau) requires extra energy.
Energy injection at the decelerating forward shock has been suggested to explain this phase (\citealt{2006ApJ...642..354Z,2006ApJ...642..389N,2006MNRAS.366.1357P}).
However, this cannot explain cases where this phase is followed by a steep decay (\citealt{2007ApJ...665..599T,2007ApJ...670..565L}).
Also, this would require extreme radiation efficiencies in the prompt phase (e.g., see \citealt{2006MNRAS.370.1946G}; \citealt{2006A&A...458....7I}).
Additionally, these phases (the extended emission in particular) show highly variable and flaring features (e.g., see Figure 3 in \citealt{2005ApJ...635L.133B}; Figure 1 in \citealt{2017ApJ...846..142K}).
In \cite{2020MNRAS.493..783M}, using the compactness argument, it has been found that the late emission (extended phase in particular) must have a Lorentz factor $\Gamma \gtrsim 10$ (see their Table 1).
For the plateau emission, the situation is more enigmatic. 
The external forward shock model has been invoked to explain the plateau emission (\citealt{2006ApJ...642..354Z}; interaction between the prompt and the extended outflow: \citealt{2020MNRAS.493..783M}; also see  \citealt{2019ApJ...883...48L} for GRB 160821B; and other scenarios: \citealt{2020ApJ...893...88O}; \citealt{2020MNRAS.492.2847B}), 
but in many cases it is interpreted as late central engine (BH or magnetar in some cases) activity (e.g., \citealt{2013MNRAS.431.1745G}; \citealt{2015ApJ...804L..16K}; \citealt{2019ApJ...872..118B}; 
\citealt{2023ApJ...958L..33G}; etc.). 
Therefore, based on these arguments, late emission phases are typically interpreted as indicators of 
late engine activity launching jet outflow
(\citealt{2005ApJ...631..429I}; \citealt{2015ApJ...804L..16K}; \citealt{2023ApJ...958L..33G}).

Despite observations hinting toward late engine activity in many \textit{s}GRBs, with a few exceptions, the impact of such late engine activity on the KN has not been subject to thorough investigation (\citealt{2015ApJ...802..119K,2016ApJ...818..104K,2018PTEP.2018d3E02I,2018ApJ...861...55M}). 
In particular, late jet launch is expected to shock and heat-up the ejecta, producing a cocoon, at a much larger radius (than in the prompt jet case).
Adiabatic cooling is much less efficient in the case of later activity, and the cocoon is expected to be much brighter at later times.
Such emission powered by late engine activity could give hints about the central engine of \textit{s}GRBs, which is still poorly understood.

In this paper, we investigate late jet launch and propagation through BNS merger ejecta at later times (see Figure \ref{fig:f0}).
We analytically estimate the properties of late jet-cocoons (using \citealt{2021MNRAS.500..627H,2023MNRAS.520.1111H}), and calculate their cooling emission analytically (using \citealt{2023MNRAS.524.4841H}).
We found that late jets' cocoon can produce a bright blue KN component, especially for wide jets, and for jets that fail to penetrate the ejecta (i.e., choked).
We also confirm that such a cocoon is a bright source of soft X-rays immediately after the breakout \citep{2023MNRAS.524.4841H} [see Figure \ref{fig:f0}].
Our estimates suggest that the cocoon emission will be detected on a yearly basis, and in the near future, emission from the cocoon can be a new set of eyes to probe BNS mergers.

This paper is organized as follows. 
Analytic modeling for jet propagation in the merger ejecta is presented in §\ref{sec:2}.
Analytic modeling for the cocoon cooling emission is presented in §\ref{sec:3}.
Results are presented in §\ref{sec:4}, and discussed in §\ref{sec:5}.
Finally, a conclusion is presented in Section \ref{sec:6}.
Details related to our analytic model can be found in appendices §\ref{ap:jet}, \ref{ap:es}, \ref{ap:cooling}, \ref{ap:trapped}, and \ref{ap:KN}.

\section{Dynamics}
\label{sec:2}

\subsection{Overview}
\label{sec:Overview}
As shown in Figure \ref{fig:f0}, the picture of a BNS after the merger time ($t_m$) is as follows.
A compact merger remnant with an accretion disk is formed; we refer to it as the central engine.
This central engine is surrounded by the expanding merger ejecta.
The prompt \textit{s}GRB-jet is launched soon after the merger, delayed for a certain time ($\sim 1$ s for GRB 170817A; \citealt{2017ApJ...848L..13A}).
This jet is active for a certain duration ($\sim 2$ s for \textit{s}GRBs), and is characterized by its opening angle and power.
This jet propagates through the merger ejecta.
This propagation produces a cocoon.
Once the jet reaches the outer edge of the ejecta, both the jet and the cocoon break out.

At later times ($\sim 1-10$ s), more mass is ejected (i.e., post-merger ejecta; \citealt{2018ApJ...860...64F}).
Also, after the prompt phase, late jets are launched in two phases: 
the extended phase and the plateau phase (see Figure \ref{fig:intro}).
Jet parameters are different in each phase.
These late jets are expected to propagate through the ejecta in a manner similar to the prompt jets; 
forming a cocoon and breaking out to produce the extended/plateau emission (full arguments behind this reasoning are presented in §\ref{sec:late jet approx}).
A small fraction of the cocoon is fast enough to break out of the ejecta, and the rest is trapped inside the ejecta \citep{2023MNRAS.524.4841H} [see panel C in Figure \ref{fig:f0}].

In GW170817/GRB 170817A, late emission has not been observed.
It should be noted that this could be due to the large off-axis viewing angle ($\sim 20^\circ -30^\circ \gg \Gamma_{jet}^{-1}$).
Therefore, a late jet in GW170817/GRB 170817A can not be excluded from the lack of extended/plateau emission.
Also, the presence of a late jet, with its smaller energy budget relative to the prompt jet, would not necessarily contradict afterglow (and VLBI; \citealt{2018Natur.561..355M}) observations (see \citealt{2018ApJ...866L..16M} for a similar argument).
Additionally, the poor early localization in GW170817/GRB 170817A is another issue that could explain the lack of EM counterparts related to e.g., late jet breakout.
While,  the two recent nearby GRBs (GRB 211211A and GRB 230307A) associated with a KN, show evidence of late emission in the form of a soft tail (\citealt{2022Natur.612..223R,2023arXiv230702098L}; also see \citealt{2023arXiv230705689S}). 
The high energy emission in GRB 211211A, delayed by $\sim 10^3$ s, further suggest late engine activity and late jet launch (\citealt{2022Natur.612..236M}).

Parameters of the ejecta and the jet are set in the following (§\ref{sec:para ejecta} and §\ref{sec:para}; respectively).
Given these parameters, the dynamics of the jet-cocoon are determined analytically (§\ref{sec:pre-breakout} and §\ref{sec:post-breakout}) following the analytic model (and approximations) in \cite{2021MNRAS.500..627H,2023MNRAS.520.1111H,2023MNRAS.524.4841H}.
In particular, we derive the breakout time [see equation (\ref{eq:tb})], and use it to find the cocoon's total mass and energy (in §\ref{sec:pre-breakout}), and the mass and energy of the cocoon that escapes the ejecta (in §\ref{sec:post-breakout}).

\subsection{Initial conditions: the ejecta}
\label{sec:para ejecta}
The time since the merger is taken as $t-t_m=t$ ($t_m=0$ for simplicity).
The ejecta is assumed to be expanding homologously:
\begin{equation}
    r(t) = c\beta t.
    \label{eq:r=vt}
\end{equation}
The ejecta is assumed to have a total mass $M_e$.
In terms of mass-density, we assume a simple spherical wind-like function $\rho(\beta,t) \propto r^{-n}$ with $n=2$, and  $M_{e}=\int_{\beta_0}^{\beta_m}4\pi(ct)^3 \rho(\beta,t) \beta^2 d\beta$;
where the maximum velocity is $\beta_m=\sqrt{3}/5$ (so that $\langle{\beta}\rangle\sim 0.2$; see \citealt{2013PhRvD..87b4001H}), and the minimum velocity is taken as $\beta_0=\beta_m/100\ll\beta_m$.
For the prompt jet we take the ejecta mass as $\sim 0.01 \msun$.
This is comparable to the expected dynamical ejecta immediately after the merger (\citealt{2013PhRvD..87b4001H}).
At later times, post-merger ejecta is expected to dominate ($\sim 1-10$ s after the merger; \citealt{2018ApJ...860...64F}).
Considering measurements of the ejecta mass from the late time KN emission in AT2017gfo, indicating a mass of $\sim 0.05\msun$ (also similar to the KNe associated with GRB 211211A and GRB 130603B; \citealt{2013ApJ...774L..23B,2013Natur.500..547T,2022Natur.612..223R}), we adopt the same ejecta mass at the time of late jet launch (see Table \ref{tab:models})\footnote{There is room for the ejecta mass, relevant for jet propagation to be even higher. This is because the jet is expected to propagate through bound material as well (before its eventual fall back).}.

\subsection{Initial conditions: the jet}
\label{sec:para}

The \textit{s}GRB-engine-powered polar jet can be defined by its opening angle $\theta_0$; and its power $L_j$ ($2L_j$ for two polar jets), or more practically its isotropic equivalent luminosity $L_{iso,0}={2L_j}/({1-\cos{\theta_0}})$.
We consider three independent engine phases: prompt, extended, and plateau; 
each producing a distinct relativistic jet.
The jet launch time is $t_0$; 
and the jet launch lasts until $t_e$ for a duration $t_e-t_0$.
As a convention, the subscript ``$_0$" is used to indicate parameters that are related to the central engine.

For simplicity, we assume an initial top-hat structure for the jet as launched by the central engine. 
This is reasonable considering that any initial complex jet structure (e.g., Gaussian) is expected to converge to a top-hat structure as the jet will strongly be collimated during its propagation across the dense medium (e.g., see \citealt{2023MNRAS.518.5145U}).
It should be noted that this is not inconsistent with observations of GW170817/GRB 170817A, as the more complex ``structured jet" is expected to emerge after the jet breakout (also see §\ref{sec:opening angle}).
Also, during the engine activity ($t_0<t<t_e$) at a given phase (prompt, extended, or plateau), we assume that the jet is continuously launched with a constant jet power and constant jet opening angle.

In the following (§\ref{sec:jet time}; §\ref{sec:opening angle}; §\ref{sec:jet initial luminosity}), these parameters are inferred from typical values observed for these three phases.

\subsubsection{Jet launch time and duration}
\label{sec:jet time}
Considering observations of GW170817 (\citealt{2017PhRvL.119p1101A}) and GRB 170817A (\citealt{2017ApJ...848L..13A}), the delay between the BNS merger and the prompt emission should be $<1.7$ s (more accurately $\lesssim 1.3$ s according to \citealt{2020MNRAS.491.3192H}).
Considering that jet propagation through the ejecta contributes to this delay, it is reasonable to take the delay between the merger and the jet launch as $t_0 -t_m = 1$ s.
Also, based on the bimodal distribution of GRBs, the duration of \sgrbss prompt emission ($t_e-t_b\lesssim t_e-t_0$) is $\lesssim 2$ s, and with the time it takes the jet to breakout ($t_b-t_0$) typically $\sim  0.5 - 1$ s, we roughly assume that the prompt jet duration as $t_e-t_0=2$ s.

For the later extended and plateau phases, the jet launch and quenching times are not well understood.
However, one can roughly use the time of the start of these phases from observations to infer the jet breakout.
Observations show that the extended emission starts $\sim 10$ s after the prompt phase, and the plateau emission starts $\sim 10^3$ s after the extended emission phase (not observed after GW170817, but this is most likely due to the off-axis line-of-sight; see §\ref{sec:Overview}).
Hence, based on this data and considering the time it takes the jet to breakout, we take the starting time of the extended emission's jet as a few seconds after the merger $t_0-t_m= 10^{0.5}$ s, and for the plateau phase we take $t_0-t_m= 10^{2.5}$ s.
In other words, it has been assumed that the engine activity is mostly continuous throughout these three hypothetical jet phases (although the luminosity drops by orders of magnitudes; see Table \ref{tab:models}).

Finally, the jet duration can simply be inferred from the observed duration of these phases as it can be seen in right panel of Figure \ref{fig:intro}. We estimated the average observed duration values and adopted them as ``typical" values (as shown with the dashed lines and the star symbols).
One exception is that for the prompt phase we adopt a typical duration of $2$ s, which is regarded as the typical duration of \textit{s}GRB jets (\citealt{1993ApJ...413L.101K}). This is because the breakout time can be comparable to this timescale, making the observed duration of \sgrbs much shorter than the actual timescale of the engine activity (\citealt{2013ApJ...764..179B}).

\subsubsection{Jet opening angle}
\label{sec:opening angle}
First, we assume that the post-breakout jet opening angle $\theta_j$ eventually converges to the jet's initial opening angle $\theta_0$.
In reality, after the breakout, there is a complex jet structure, composed of a jet-core and the cocoon in the wings (\citealt{2013ApJ...777..162M,2017ApJ...848L...6L,2017MNRAS.472.4953L,2018MNRAS.478..733L,2018PhRvL.120x1103L,2021MNRAS.500.3511G}). 
However, it has been shown that the jet-core can be approximated to have a top-hat structure, and its opening angle is comparable to $\theta_0$ at sufficiently late times after the breakout (\citealt{2021MNRAS.500.3511G,2023MNRAS.518.5145U} for prompt jets). 
Hence, here, with the jet opening angle $\theta_j\sim \theta_0$ we refer to the jet-core structure.

With measurements of ``jet breaks" in the afterglow light curve, the initial jet opening angle can be inferred. 
In \cite{2023ApJ...959...13R}, afterglow analysis for a sample of \sgrbs shows that while the typical value of the jet opening angle is $\theta_j\sim 6^\circ$, several \sgrbs are associated with narrow jets $\theta_j \sim 3^\circ$, and the jet opening angle can be as large as $\theta_j \gtrsim 15^\circ$  (e.g., GRB 211106A; see \citealt{2022ApJ...935L..11L,2023ApJ...959...13R}).
With the intention of representing these observed features, we adopt three jet models with different opening angles: narrow ($\theta_0=3^\circ$), typical ($\theta_0=6^\circ$) and wide ($\theta_0=15^\circ$) [see Table \ref{tab:models}].
It should be noted that, as prompt jets dominate in terms of jet energy, these opening angles (as inferred from observations) are expected to reflect prompt jets. 
Therefore, the actual opening angles of late jets are not clear. Here, for simplicity, similarity to the prompt jet have been assumed.

\subsubsection{Jet luminosity}
\label{sec:jet initial luminosity}
Not all the jet energy is released in the form of emission.
Hence, in order to infer the intrinsic jet power from observations, we introduce radiation efficiency $\eta_{\gamma}$ defined as:
\begin{equation}\label{eq:eta}
        \eta_\gamma = \frac{E_{\gamma,iso}}{E_{\gamma,iso}+E_{K,iso}}, 
\end{equation}
where $E_{\gamma,iso}$ is the isotropic equivalent radiated energy, and $E_{K,iso}$ is the (isotropic equivalent) kinetic energy of the jet after photon radiation.

For simplicity let us assume that this parameter ($\eta_\gamma$) is constant throughout the duration of a given emission phase.
For the prompt phase of \textit{s}GRBs, this radiation efficiency has been estimated in several studies (e.g., see \citealt{2015ApJ...815..102F}; \citealt{2023ApJ...959...13R}). 
According to \cite{2015ApJ...815..102F} typically $\eta_\gamma\sim 0.1-0.5$.
In more recent analysis \cite{2023ApJ...959...13R} suggests that this efficiency is typically $\eta_\gamma\sim 0.2$, which is comparable with $\eta_\gamma\sim 0.15$ found by \cite{2016MNRAS.461...51B} [for $\epsilon_B\sim 10^{-4}$].

For the extended and plateau phases, the radiation efficiency is not well studied.
However, \cite{2020MNRAS.493..783M} presented an evaluation of the radiation efficiency of late engine jets as $\sim 0.03$ on average, although the standard deviation is large.

Here, for simplicity, we assume the radiation efficiency for prompt, extended, and plateau jets to be the same, and 
we set it as:
\begin{equation}\label{eq:eta2}
\eta_\gamma\sim 0.1 .    
\end{equation}

Hence, from the above radiation efficiency [equation (\ref{eq:eta})] the initial luminosity of the jet can be found as a function of the observed luminosity $L_{obs}\approx E_{\gamma,iso}/(t_e-t_0)$ as (using the assumption $\theta_j\sim \theta_0$; see §\ref{sec:opening angle}):
\begin{equation}
    L_{iso,0} = L_{obs}/\eta_\gamma .  
    \label{eq:Liso0}
\end{equation}
Using equation (\ref{eq:eta2}), and adopting typical values for the observed luminosity and duration for each of the three phases (see right panel in Figure \ref{fig:intro}), the typical initial parameters of the jet can be found for the three phases as presented in Table \ref{tab:models}.

Note that the afterglow modeling requires several free parameters that are difficult to constrain (i.e., $n$, $\epsilon_e$ and $\epsilon_B$; jet geometry as well \citealt{2022Univ....8..612L}) and the results are often sensitive to the assumed values.
In particular, the estimated values of $\eta_\gamma$ are highly sensitive to the assumption that all the electrons are non-thermal, i.e., $f_e\sim 1$ (as $E_{K,iso}\propto f_e^{-1}$).
Recently, \cite{2019ApJ...884L..58U} indicated a low value of $f_e\sim 0.1$, which may suggest a lower value of $\eta_\gamma$. 
Hence, our conservative choice of $\eta_\gamma\sim 0.1$ is reasonable.

\subsection{Analytic model: pre-breakout}
\label{sec:pre-breakout}
\subsubsection{Breakout time and radius}
\label{sec:breakout}
We follow the analytic model in \cite{2020MNRAS.491.3192H} and \cite{2021MNRAS.500..627H} to solve the jet propagation, in particular the jet collimation by the ejecta and the cocoon self-consistently.
This model has been demonstrated to give analytic results consistent with numerical simulations (see Figures 3, 4, and 5 in \citealt{2021MNRAS.500..627H}; also see Figure 4 in \citealt{2020MNRAS.491.3192H}).
Eventually, the jet breakout time $t_b$ and cocoon properties at $t_b$ are determined.

The breakout time since the merger time ($t_m= 0$) in an expanding medium can be written as [for more details see equation (44) in \cite{2021MNRAS.500..627H}]:
\begin{equation}
t_b \approx \left[ \frac{(c\beta_m)^\frac{3}{4}}{(5-n)A_1}+ t_0^\frac{1}{4}\right]^{4} ,
\label{eq:tb}
\end{equation}
where $c$ is the speed of light, $\beta_m$ is the velocity of the outer ejecta, $n$ is the power-law index of the ejecta's density profile, $t_0$ is the delay between the merger and the jet launch.
$A_1$ is a constant that can be written as a function of the parameters of the jet and the ejecta, as
$A_1\approx N_s \left[\frac{\langle{\eta'}\rangle}{\langle{\chi}\rangle^2}\right]^\frac{1}{4}\left[\frac{2^{n+5}}{3-n}\frac{\:L_j\:c\beta_{m}}{\theta_0^4 M_{e}}        \right]^{\frac{1}{4}}$,
where $N_s$ is a calibration coefficient determined through comparison with numerical simulations ($N_s=0.46$ here)\footnote{We adopted the value of $N_s$ determined through numerical simulations of prompt jets. However, this could introduce uncertainty for late jets, as the true value of $N_s$ may differ (refer to §C.2 in \citealt{2021MNRAS.500..627H}).}; 
$\langle{\eta'}\rangle=(1-\langle{\beta_h}\rangle)\eta$ with $\eta=0.5$
here (equipartition of energy), and $\langle{\beta_h}\rangle=\frac{r_b-r_0}{c(t_b-t_b)}$ is the average jet head velocity\footnote{This expression of $\eta'$ has slightly been improved from the one in \cite{2021MNRAS.500..627H}, with its dependency on $\langle{\beta_h}\rangle$ taken into account.};
$\langle{\chi}\rangle$ is another free-parameter to account for the effect of the lateral expansion of ejecta\footnote{\label{foot:chi}$\langle{\chi}\rangle$ and $\langle{\eta'}\rangle$ are assumed constant and are determined iteratively (see \citealt{2021MNRAS.500..627H}).};
$L_j$ is the jet power (one side) and $\theta_0$ is its initial opening angle;
and $M_e$ is the total mass of the ejecta (see §\ref{sec:para ejecta}; §\ref{sec:para}; and Table \ref{tab:models} for the typical values).
The breakout radius can be found [using equation (\ref{eq:r=vt})] as:
\begin{equation}
    r_b = c\beta_m t_b .
    \label{eq:rb}
\end{equation}
$r_b/2$ is the semi-major axis of the ellipsoidally shaped cocoon.
The semi-minor axis $r_c$ can also be found (see §\ref{ap:jet}).
Hence the cocoon volume at the breakout time can be found as $V_c =\frac{4\pi}{3}{r_c^2r_b}$ (for both hemispheres).
For more details see §\ref{ap:jet}.

For late time jets, where $t_b\gg t_0$, the breakout time can approximately be found as 
$t_b \propto N_s^{-4}\chi^2{\eta\prime}^{-1}\beta_m^2 M_e \theta_0^2 L_{iso,0}^{-1}$.
With the typical jet opening angle in the extended emission phase as a reference (model EX-T in Table \ref{tab:models}), the parameter dependency of the breakout time can be found as:
\begin{equation}
t_b \sim 18 \text{ s} \left(\frac{\beta_m}{0.35}\right)^{2}\left(\frac{M_e}{0.05\msun}\right)^{}\left(\frac{\theta_0}{6^\circ}\right)^{2}\left(\frac{L_{iso,0}}{10^{49.5}}\right)^{-1} .
\label{eq:tb simple}
\end{equation}
It should be noted that this is a simplified expression that does not include all dependencies (e.g., $\eta'\propto 1-\langle{\beta_h}\rangle$, and $\chi$) and is expected to result in a factor of $\sim 2$ error; 
compared to the accurate expression in equation (\ref{eq:tb}) that is used in the following.

\subsubsection{The pre-breakout cocoon} 
The cocoon is distinguished from the relativistic jet as being slower, having $\Gamma < 10$, 
and from the ejecta as being heated by the jet shock (\citealt{2023MNRAS.520.1111H}).
As explained, before the breakout, the cocoon is assumed to take an ellipsoidal shape. 
Also, the pressure inside the cocoon $P_c$ is assumed to be dominated by radiation pressure (adiabatic index of $4/3$), and roughly constant throughout the cocoon's volume.

With these assumptions, the cocoon mass ($M_c$), energy ($E_c$), and internal energy ($E_{c,i}$) can be found until the breakout following the analytic model in \cite{2023MNRAS.520.1111H}.
Details are given in §\ref{ap:es}.

\subsection{Analytic model: post-breakout}
\label{sec:post-breakout}

We follow the same model in \cite{2023MNRAS.520.1111H} to determine the fraction of the escaped cocoon mass ($M_c^{es}/M_c$) and internal energy ($E_{c,i}^{es}/E_{c,i}$) analytically.
The main argument is that, after the jet breakout, the cocoon fluid accelerates (due to its highly thermalized energy content),
until it reaches its terminal velocity determined by Bernoulli equation as $\beta_{}=\sqrt{1-\Gamma_{inf}^2}$, where $\Gamma_{inf}\approx h\Gamma$ ($h$ is enthalpy). 
This terminal velocity is assumed to be reached roughly at $t_1\sim 2 t_b$ ($\sim 1.5t_b$ or shorter, for the plateau jet case where $t_b$ is significantly long).
Then, the cocoon fluid is designated as escaped or trapped inside the ejecta based on its velocity, so that:
\begin{equation}
  \beta_{} 
    \begin{cases}
      >\beta_m & \text{(Escaped cocoon)} ,\\
      \leqslant \beta_m & \text{(Trapped cocoon)} .
    \end{cases}       
    \label{eq:beta_inf cases}
\end{equation}
As the cocoon expansion converges to homologous (reached at $t_1$), the escaped cocoon's mass and energy can be found using the above condition (see \citealt{2023MNRAS.520.1111H}).
Analytic results are given in §\ref{ap:es} (also, for more details see \citealt{2023MNRAS.520.1111H}).

The escaped cocoon mass ($M_c^{es}$ and energy) distribution is approximated as conical (see Figure 4 in \citealt{2023MNRAS.520.1111H}), with an opening angle $\theta_c^{es}$ [see equation (\ref{eq:theta c es})], and roughly with an axial symmetric distribution.

The escaped cocoon is separated into a relativistic part and a non-relativistic part, and each is modeled (in terms of mass-density and internal energy density) with a single power-law function (see §\ref{ap:es}; and \citealt{2023MNRAS.524.4841H}).

For the trapped cocoon ($M_c^{tr}\sim M_c \gg M_c^{es}$), the mass-density is approximately considered as the same as that of the ejecta.


\begin{table*}
\centering
\caption{
Typical parameters of prompt emission's, extended emission's, and plateau emission's jets, based on observations (see Figure \ref{fig:intro}).
In total, nine models are considered, with three jet models for each of the three phases (for more information see §\ref{sec:para}).
Breakout times and escaped cocoon mass fractions, calculated using our analytic model, are also shown (see §\ref{ap:jet}; and \ref{ap:es}).
}
\label{tab:models}
\begin{tabular}{l|l|l|ccccc|cc} 
\hline
\hline
 &  &  & Jet opening & Initial jet  & Jet launch & Jet  &  Ejecta  & Breakout & Escaped cocoon\\
 & & Jet & angle & luminosity & time & duration & mass & time & mass fraction\\
Model & Phase & type &  $\theta_0$ [$^\circ$] & $L_{iso,0}$ [erg s$^{-1}$] & $t_0-t_m$ [s] & $t_e-t_0$ [s] & $M_e$ [$M_\odot$] & $t_b-t_m$ [s] & $\frac{M_c^{es}}{M_c}$ [\%] \\
\hline
\hline
PR-N & \multirow{3}{4em}{Prompt} & Narrow & $3$ & \multirow{3}{4em}{$10^{52}$} & \multirow{3}{4em}{$1$} & \multirow{3}{1em}{$2$} & \multirow{3}{4em}{$0.01$} & 1.6 & 0.7\\ 
PR-T &  & Typical & $6$ & & & &  &1.7 & 3\\ 
PR-W &  & Wide & $15$& & & &  &2.0 & 9.8\\ 
\hline
EX-N & \multirow{3}{4em}{Extended} & Narrow & $3$ & \multirow{3}{4em}{$10^{49.5}$} & \multirow{3}{4em}{$10^{0.5}$} & \multirow{3}{1em}{$10^{2.5}$} & \multirow{3}{4em}{$0.05$} &11.1 & $1.4\times 10^{-2}$\\ 
EX-T &  & Typical & $6$ & & & & &18.0 & $6.1\times 10^{-2}$\\ 
EX-W &  & Wide & $15$& & & & &49.6 & 0.4\\ 
\hline
PL-N & \multirow{3}{4em}{Plateau} & Narrow & $3$ & \multirow{3}{4em}{$10^{47}$} & \multirow{3}{4em}{$10^{2.5}$} & \multirow{3}{1em}{$10^{4}$} & \multirow{3}{4em}{$0.05$} &$1.6\times 10^3$ & $5.2\times 10^{-3}$\\ 
PL-T &  & Typical & $6$ & & & &  &$3.2\times 10^3$ & $2.4\times 10^{-2}$\\ 
PL-W &  & Wide & $15$ & & & & &$1.2\times 10^4$ & 0.3\\ 
\hline
\hline
\end{tabular}
\end{table*}

\section{Emission}
\label{sec:3}

\subsection{Main approximations}
\label{sec:approx emission}
The cocoon is assumed to radiate as a blackbody in thermal equilibrium. 
We employ the diffusion approximation, where, for a given shell, once the diffusion time becomes comparable to the dynamical time, the internal energy is approximated as instantly transforming into photons that are emitted (as in \citealt{2012ApJ...747...88N}; and similarly in \citealt{2015ApJ...802..119K}).
Radiation behind this shell is considered as unable to diffuse out.

We use the grey opacity to estimate the optical depth.
For a typical range of density and temperature of the considered cocoon fluid (in the comoving frame), we solved the ionization degrees of the fluid under the assumption of local thermodynamic equilibrium.
We confirmed that the ionization edge is always located at an energy higher
than $h \nu_{\rm peak}$ by at least a factor of 3; 
where $h$ here is Planck's constant and $\nu_{\rm peak}$ is the frequency at which the spectrum peaks following Wien's displacement law.
Hence, the strong attenuation by the bound-free process (e.g, see Figure 8 in \citealt{2019LRR....23....1M}) is not expected to affect the bulk of the emission discussed in this paper.
Thus, we adopt the same (grey) opacity both in UV and soft X-ray range,
as $\kappa = 1 \ {\rm cm^2 \ g^{-1}}$ (as for lanthanide-free material because the cocoon outflow mostly originates from the polar region; see \citealt{2017PhRvD..96l3012S,2018ApJ...860...64F}), which is derived from the detailed calculations of bound-bound opacities (\citealt{2020MNRAS.496.1369T,2020ApJ...901...29B,2023arXiv230405810B}).

For late jets, the mass-density at the jet shock is significantly lower compared to the case of prompt jets.
Therefore, we estimated the time required to realize the photon blackbody spectrum from the thermalized cocoon fluid.
Considering free-free process for electrons, and a composition of fully ionized iron-peak elements ($Z\sim 30$),
this time can be estimated as the ratio of the internal energy density of the fluid $aT^4$ over the free-free cooling rate $\varepsilon^{f f}\sim 1.7 \times 10^{-27}$ erg s$^{-1}$ cm$^{-3}$ $ T^{1 / 2} n_e n_i Z^2$ (\citealt{1979rpa..book.....R}). 
This timescale is found, at most, as comparable to the breakout time.
Therefore, our assumption of blackbody spectrum is reasonable; 
especially considering that bound-free (and bound-bound) process is even more effective at thermalizing the shocked cocoon fluid.

Finally, as suggested in \cite{2023arXiv230508575G}, radiation-mediated shocks can affect the chemical composition (via fusion or fission) and the opacity in the jet-shocked cocoon. 
However, the significance of this effect is not fully understood.

\subsection{Cooling emission from the escaped cocoon}
\label{sec:cooling}
We use an updated version of the model presented in \cite{2023MNRAS.524.4841H}, which was intended for the case of prompt jets.
In the case of late jets, the breakout time (and radius) can be large and non-negligible (e.g., compared to the photospheric radius of the emitting shell).

Therefore, in the following, quantities are expressed using alternative coordinates for laboratory time, 
radius, 
and observed time, 
with the breakout time ($t_b$), breakout radius ($r_b$), and observed breakout time [$t_{b,obs}=(1-\beta_m)t_b$] as their reference, so that: $\Delta t = t-t_b$, $\Delta r = r -r_b$, and $\Delta t_{obs} =t_{obs} -t_{b,obs}$ (respectively) [see §\ref{ap:cooling}].

With the cocoon mass determined (see §\ref{sec:post-breakout} and §\ref{ap:es}), the optical depth and the velocity of the diffusion shell as a function of the observed time can be calculated [see equation (\ref{eq:tobs R NR})].
Also, with the total internal energy determined (see §\ref{ap:es}), the internal energy density can be calculated.

Then, cooling emission can be found as the product of i) the leaking internal energy over time (i.e., power due to the change of the diffusion shell's velocity), ii) the effect of the observer's time, and iii) the effect of relativistic beaming as:
\begin{equation}
L_{bol}^j(\Delta t_{obs})=
\frac{\partial E_{c,i}^j(>\Gamma_d\beta_d,\Delta t)}{\partial \Gamma_d\beta_d }\frac{\partial \Gamma_d\beta_d }{\partial \Delta t}\times \frac{d\Delta t_{}}{d\Delta t_{obs}}\times\frac{4\pi}{\Omega_{EM}} ,
\label{eq:Lbl j}
\end{equation}
where $\Omega_{EM} \approx 4\pi(1-\cos[\max\{\arcsin[{\Gamma_d}^{-1}],\theta_{c}^{es}\}])$ is the solid angle of the emission coming from the shell moving with $\Gamma_d\beta_d$.

With the photospheric radius found as:
\begin{equation}
\Delta r_{ph}(\Delta t_{obs}) \approx c\Delta t \beta_{ph}\approx c \Delta t_{obs}\frac{\beta_{ph}}{1-\beta_{ph}},
\label{eq:rph}
\end{equation}
the observed temperature can be found using Stefan–Boltzmann law (in the relativistic limit) as
\begin{equation}
T_{obs}(\Delta t_{obs})  \approx 
      \left\{\frac{L_{bol}^{j}}{4\pi\sigma [r_{ph}/\Gamma_{ph}]^2 }\left(\frac{\Omega_{EM}}{\Omega}\right)\right\}^{\frac{1}{4}},
    \label{eq:Teff}
\end{equation}
where $\sigma$ is the Stefan-Boltzmann constant.

It should be noted that, here, by cooling emission we refer exclusively to jet heating as the source of internal energy. 
More details are given in §\ref{ap:cooling} (for details on other types of emission, see \citealt{2020MNRAS.493.1753G,2023MNRAS.524.4841H}).

\subsection{Cooling emission from the trapped cocoon}
\label{sec:trapped}
As discussed in \cite{2023MNRAS.524.4841H}, for the prompt jet case, the trapped cocoon [defined as having $\beta\leqslant \beta_m$; see equation (\ref{eq:beta_inf cases})] is not a relevant component for the cocoon cooling emission.
This is because 
i) internal energy is deposited at an early times ($\sim t_b\sim 1$ s after the merger), 
and ii) as it is concealed by both the escaped cocoon and the ejecta, by the time photons can escape out of the trapped cocoon (usually at $\sim 10^3$ s; as the escaped cocoon becomes optically thin), 
its internal energy would have been significantly reduced due to the adiabatic expansion; 
roughly by a factor of $\sim t_b/t\sim 10^{-3}$.
However, the situation is fundamentally different for the late jet case; as the jet launch and breakout is expected at much later time ($t_b\gg 1$ s), and the reduction of internal energy due to adiabatic cooling is much less severe.

The same method used for the cooling emission of the escaped cocoon is followed here.
Notable differences are: i) the power-law index of the mass-density profile is approximated here to the initial mass-density profile of the ejecta ($n=2$); ii) as the velocity of the trapped cocoon (and the ejecta) is significantly smaller ($\beta<\beta_m$), the approximation $\beta\ll 1$ is used, allowing us to express the observed time and the laboratory time much simply as $t_{obs}\sim t$; and iii) the trapped cocoon's structure is assumed as conical for simplicity.
Assuming a simplified conical shape for the trapped cocoon,
we found that its overall opening angle [defined by the mass fraction; see equation (\ref{eq:Omegac})] is larger than that of the escaped cocoon (due to the large mass fraction), particularly for wide jets, making it easier to detect.
The detailed analytic model is presented in §\ref{ap:trapped}.

\subsection{R-process powered KN emission}
\label{sec:KN}
Here we refer to all emission powered by r-process elements' radioactive heating as ``KN emission" (regardless of where it is taking place; e.g., even if it is within the cocoon).

In short, the ejecta, expanding homologously, is defined with its mass $M_e$, power-law index of its density profile $n$, its maximum velocity $\beta_m$, and its minimum velocity $\beta_0$.
The optical depth can then be calculated from the shell moving with $\beta_m$ and inward (assuming grey opacity and neglecting the escaped cocoon's contribution).
Then, as the r-process heating can be modeled with a power-law function, the energy deposition is found, allowing us to determine the diffused internal energy (i.e., luminosity) in the same manner as in §\ref{sec:cooling} and §\ref{sec:trapped} (also see \citealt{2015ApJ...802..119K,2023MNRAS.524.4841H}).
The full analytic model for the KN emission is presented in §\ref{ap:KN}.


\section{Results}
\label{sec:4}
\subsection{Dynamics}
\label{sec:results dyn}
In Table \ref{tab:models} we show the jet breakout time for our sample of models.
We found that, typically, the jet breakout time is of the order of $\sim 1$ s (prompt jets), $\sim 10$ s (extended emission's jets), and $\sim 10^3$ s (plateau emission's jets).
These breakout time are roughly consistent with the observed timing of late emission, considering that the actual ``observed" delay is shorter by a factor of $1-\beta_m$ ($\sim 0.65$ here).
This delay can easily be evaluated from observation of \sgrbs based on the start of the late emission phase (or based on the start of the cocoon emission; see §\ref{sec:cocoon parts}) and could be useful to evaluate the parameters of the jet and the ejecta [see equation (\ref{eq:tb simple})].

We find that plateau jets (if they exist) are inherently difficult to break out.
This is different from the results of \cite{2018ApJ...866L..16M}, where jet propagation did not take into account the effect of the expanding ejecta on the cocoon and on the jet collimation (see \citealt{2021MNRAS.500..627H}).
In one case (model PL-W), the jet failed to breakout (i.e., choked).
Therefore, it is difficult to explain the observed fainter plateau emissions ($\ll 10^{46}$ erg s$^{-1}$; see right panel in Figure \ref{fig:intro}) with this scenario, unless the ejecta mass is somehow significantly smaller than what has been suggested ($\sim 0.05 \msun$) for AT2017gfo (e.g., in cases of equal mass BNS where the disk/ejecta mass is small; see \citealt{2019ApJ...876L..31K}); 
or the prompt/extended jet is wide and strong enough to permanently disrupt the ejecta, leaving a cavity for the plateau jet to propagate through (although this is not trivial due to the trapped cocoon; see §\ref{sec:late jet approx}).

Table \ref{tab:models} also shows the fraction of the escaped cocoon mass for each model is shown.
These mass fractions have been calculated using equation (\ref{eq:es mass}) [also see equations (\ref{eq:tb}) and (\ref{eq:alpha tb})]. 
The larger $\alpha$ (= Energy deposited by the jet/kinetic energy of the ejecta) is, the larger the fraction of the escaped cocoon should be.

Two trends are apparent. 
First, the wider the jet opening angle, the larger the mass fraction of the escaped cocoon. 
This is due to the larger jet energy ($\propto \theta_0^2$), longer breakout times (also $\propto \theta_0^2$), resulting in a larger jet energy-density deposited into the cocoon, boosting its velocity (more details in \citealt{2023MNRAS.520.1111H}).
Second, the later the jet phase is, the much smaller the mass fraction of the escaped cocoon is.
This trend has not been discussed before, and is mainly explained by the smaller energy budget [$L_{iso,0}(t_e-t_0)$] of late jets (1/2, and 1/20 that of the prompt jet, for the extended and plateau jet, respectively), while the kinetic mass of the ejecta ($\propto M_e$) is larger at later times due to post-merger mass ejection.
Hence, for late jets, most of the cocoon mass (and energy) is concentrated in the trapped part, making its contribution to the cocoon emission more significant.

\subsection{Components of the cocoon's cooling emission}
\label{sec:cocoon parts}
In Figure \ref{fig:Lbl}, cocoon cooling emission calculated using our analytic model for a late jet model (EX-W in Table \ref{tab:models}) is presented for reference.
There are three cocoon components to consider: relativistic (escaped), non-relativistic (escaped), and trapped.

Bolometric luminosities have been calculated for each of the three cocoon components separately (see §\ref{ap:cooling} and §\ref{ap:trapped}, respectively, for more details), as shown in Figure \ref{fig:Lbl}.
Then, contributing components are connected to produce the overall light curve.

Transition time from the relativistic cocoon to the non-relativistic cocoon is set with the optical depth.
Transition time from the non-relativistic to the trapped cocoon is set with the luminosity; 
i.e., once the trapped cocoon's luminosity dominates over the non-relativistic cocoon's luminosity, we assume that transition takes place.
This is a crude simplification.
One alternative is to rather use the optical depth to determine the time of this transition.
This alternative procedure would produce a bump feature in luminosity at the transition time (sudden rise in luminosity due to the trapped cocoon's contribution; e.g., at $t_{obs}\sim 3000$ s in Figure \ref{fig:Lbl}).
However, this choice overlooks the pressure gradient at $\beta_m$ that may result in internal energy transferred from the outer trapped cocoon to the inner non-relativistic cocoon.
Future numerical simulations are needed to clarify this point.

As pointed out in §\ref{sec:results dyn}, the start of the cocoon cooling emission from a late jet is delayed from the merger time and from the prompt phase.
This is one major difference compared to cocoon emission from prompt jets (also see Figures \ref{fig:LVT}, \ref{fig:mag} and \ref{fig:X-ray}).

The very early emission comes from the escaped cocoon (relativistic or non-relativistic).
It is the brightest part of the emission, although quite short; $\sim 10^3$ s.
As explained in \cite{2023MNRAS.524.4841H} the luminosity, then, drops roughly as $\propto t_{obs}^{-2}$ for prompt jets.
But considering the late start of the emission here, time dependency of this early emission can be generalized rather as $\propto \Delta t_{obs}^{-2}$.

Internal energy available for cooling emission can be found as $\propto  L_j t_b(1-\langle{\beta_h}\rangle)$ [see equation (\ref{eq:S3})].
There are two important effects to highlight: 
i) for late jets the average jet head velocity $\langle{\beta_h}\rangle\approx \frac{t_b}{t_b-t_0}\beta_m$ is much smaller ($\sim \beta_m$) compared to late prompt jets ($\sim 2\beta_m$); \citealt{2018PTEP.2018d3E02I,2020MNRAS.491.3192H}) [see Table \ref{tab:models}] making late jets more efficient at heating the ejecta; 
and ii) the energy (and mass; as explained in §\ref{sec:results dyn}) fraction of the trapped cocoon is significantly larger for late jets compared to prompt jets [by a factor of $\sim 3$ for typical jet models; see equation (\ref{eq:es all})].
The combination of i) and ii) boosts the internal energy of late jet's cocoon (e.g., by $\sim 2\times 3$ for the extended phase). 
Hence, the internal energy of the trapped cocoon at the breakout time for the prompt jet case is roughly comparable to that of the late jets, despite the large difference in jet luminosity. 

Not all this energy is converted to photons.
One crucial point to highlight is that, in the cocoon, there are two competing processes for internal energy: i) adiabatic cooling and ii) cooling emission.
Due to the late breakout time, adiabatic cooling $\propto \frac{t_b}{t}$ [see equation (\ref{eq:adia})] is much less efficient (than in the case of prompt jet's cocoon) making the overall cocoon emission more luminous for later jets.

As explained in \cite{2023MNRAS.524.4841H}, the trapped cocoon is not relevant for prompt jets; as by the time the escaped cocoon is optically thin (at $\sim 10^3$ s) only a small fraction ($\sim \frac{t_b}{t}\sim \frac{1}{1000}$) is left to power the cooling emission.
This is much different for late jets as the adiabatic loss ($\sim \frac{t_b}{\Delta t}$) is orders of magnitude less severe [see equation (\ref{eq:adia})].

Also, the cocoon emission from late jets (trapped in particular) is released at times when the r-process powered KN emission is dimmer (compared to prompt jets).
The bright early afterglow emission (from the prompt jet) is also dimmer at late times.
Hence, the late-jet's cocoon suffers less from contamination and is easy to identify (compared to prompt jets).

Cooling emission from the trapped cocoon dominates in terms of bolometric luminosity at later times. 
As shown in Figure \ref{fig:Lbl}, for EX-W, the trapped cocoon shines at decent luminosities ($\sim 10^{43}$ erg s$^{-1}$ at $\gtrsim 10^3$ s), much higher than the early KN.
Considering that the trapped cocoon emission is blue (see §\ref{sec:Light curves}), this model offers an alternative explanation to the origin of the early blue KN in AT2017gfo (\citealt{2017Sci...358.1570D}; \citealt{2017ApJ...851L..21V};
\citealt{2018MNRAS.481.3423W}).

\subsection{Observational properties}
\label{sec:Light curves}

Figure \ref{fig:LVT} presents our analytic results for the jet models in Table \ref{tab:models}.
Cocoon emission from prompt jets, the r-process powered KN emission (see §\ref{ap:KN}), and observations of the early KN in AT2017gfo (\citealt{2017ApJ...851L..21V,2018MNRAS.481.3423W}) are shown for comparison (in grey solid lines, grey dotted lines, and grey symbols, respectively).

\subsubsection{Bolometric luminosity}
\label{sec:bolo}

Bolometric luminosities in Figure \ref{fig:LVT} (top) show two general trends in regards to the cocoon cooling emission: 
i) the later the jet (launch) phase is, the brighter the emission; and ii) the wider the jet opening angle is, the brighter the emission [assuming constant observed isotropic equivalent luminosities; see equation (\ref{eq:Liso0}) and Table \ref{tab:models}].
To our best knowledge, this is the first time that the first trend (i) has been found.
The second point (ii) has already been discussed [see equations (59) and (60) in \citealt{2023MNRAS.524.4841H}].

The delay between the merger time and the start of the cocoon cooling emission is $t_{b,obs}=(1-\beta_m)t_b$;
it is a simple function of the outer ejecta velocity $\beta_m$, and the jet breakout $t_b$;
also a function of the ejecta and the jet parameters [see equation (\ref{eq:tb simple})].
Since it is easy to evaluate such a timescale from observations, it can prove useful to evaluate the time it took the jet to break out ($t_b$) and the ejecta velocity ($\beta_m$). 
Hence, a combination of the parameters of the jet and the ejecta can also be constrained with this timescale [e.g., using equation (\ref{eq:tb simple})].

The early luminosity decreases rapidly.
This is due to its dependency on the time since the breakout as $\propto (t_{obs}-t_{b,obs})^{-2}= \Delta t_{obs}^{-2}$; this is because $E_{c,i}\propto 1/\Delta t$ due to adiabatic cooling and $L\propto E_{c,i}/\Delta t_{obs}$ [see §\ref{ap:cooling} and equation (\ref{eq:Lbl full})].
This dependency is consistent with \cite{2023MNRAS.524.4841H} [for prompt jets, where $t_{obs} - t_{b,obs}\sim t_{obs}$].

For late jets, the cocoon luminosity shows a late time excess, powered by the trapped cocoon.
Contribution of this component is much less significant in prompt jets (due to the reasons mentioned in §\ref{sec:cocoon parts}).
This component is not as bright as the early cocoon emission, coming from the escaped cocoon, but it lasts much longer thanks to the larger trapped cocoon mass.
This feature would make it easier to identify (see Figure \ref{fig:mag}) especially in wide jet cases, where it is more luminous than the KN; 
this is particularly helpful in future GW170817-like events where follow-up observations are expected to be delayed due to the time it takes to cover the region of the sky determined by the GW signal.
For plateau jets, this component can be extremely bright, even overshadowing the r-process powered KN.
Therefore, this component can offer a natural explanation to the blue KN in AT2017gfo (\citealt{2017Sci...358.1570D}; \citealt{2017ApJ...851L..21V};
\citealt{2018MNRAS.481.3423W}). 
It can also offer an explanation to the KN candidates, associated with certain \textit{s}GRBs, that somehow seem to have a luminous blue KN component (compared to AT2017gfo) and a relatively typical red KN component 
[e.g., GRB 050724 (\citealt{2017ApJ...837...50G}), GRB 060614 (\citealt{2015ApJ...811L..22J,2015NatCo...6.7323Y,2020MNRAS.493.3379R}), GRB 070714B (\citealt{2017ApJ...837...50G}), GRB 070809 (\citealt{2020NatAs...4...77J}), and GRB 150101B (\citealt{2018NatCo...9.4089T}); also see (\citealt{2018ApJ...860...62G}), and Table 2 in (\citealt{2020MNRAS.493.3379R}) for a summary].

Emission from the trapped cocoon, with its delayed start and long timescale, contributes to making the cocoon emission from late jets easier to detect and to exploit, as a tool to probe BNS and \textit{s}GRBs (also its color; see §\ref{sec:temp}).
Also, the trapped cocoon is wider than the escaped cocoon, and can be as wide as $\sim 60^\circ-70^\circ$ (for wide jets), making it even more accessible (see §\ref{sec:rate X} and §\ref{ap:trapped}).
Considering the model PL-T (see Table \ref{tab:models}), the parameter dependency of this trapped cocoon emission can be found as:
\begin{equation}
\begin{split}
&L_{bol}^{tr}\approx 5\times10^{42} \text{ erg s$^{-1}$}\\
&\left(\frac{\theta_0}{6^{\circ}}\right)^{2} 
\left(\frac{L_{iso,0}}{10^{47}\text{ erg s$^{-1}$}}\right)
\left(\frac{t_b}{3\times 10^3 \text{ s}}\right)\left(\frac{\kappa}{\text{$1$ cm$^{2}$ g$^{-1}$}}\right)^{\frac{p-2}{2}}\\
&\left(\frac{M_c}{4\times 10^{-3}\msun}\right)^{\frac{p-2}{2}}
\left(\frac{t_{obs}}{6\times 10^3 \text{ s}}\right)^{-p}, \\
\end{split}
\label{eq:Lbl PL-T}
\end{equation}
where $t_{obs}\propto (\kappa M_c)^{\frac{1}{2}}$ [see equation (\ref{eq:time tr})], 
$t_{obs}\sim 2t_b$ is roughly considered as the starting time,
and $p$ is the power-law index of the time dependency of this luminosity, with $p\sim 1.1 -1.3$ at the early phase of this emission; 
roughly between $t_{obs}\sim 2t_b$ and $20\times t_{b}$.
This shallow power-law index $p$ ($\sim 2$ for the escaped cocoon; see \citealt{2023MNRAS.524.4841H}) is due to the less severe adiabatic cooling at early times (as $\frac{\Delta t_1}{\Delta t}\sim 1$ for $t\gtrsim t_1\sim 2t_b$).
For more details see §\ref{ap:trapped} and equation (\ref{eq:Lbl tr}) in particular.
Dependency of this luminosity can be used to estimate the combination $L_{iso,0}\theta_0^2\propto L_j$ (i.e., jet power), if this emission is observed (i.e., both $L_{bol}$ and $t_b$).

\begin{figure}
    \centering
    \includegraphics[width=0.99\linewidth]{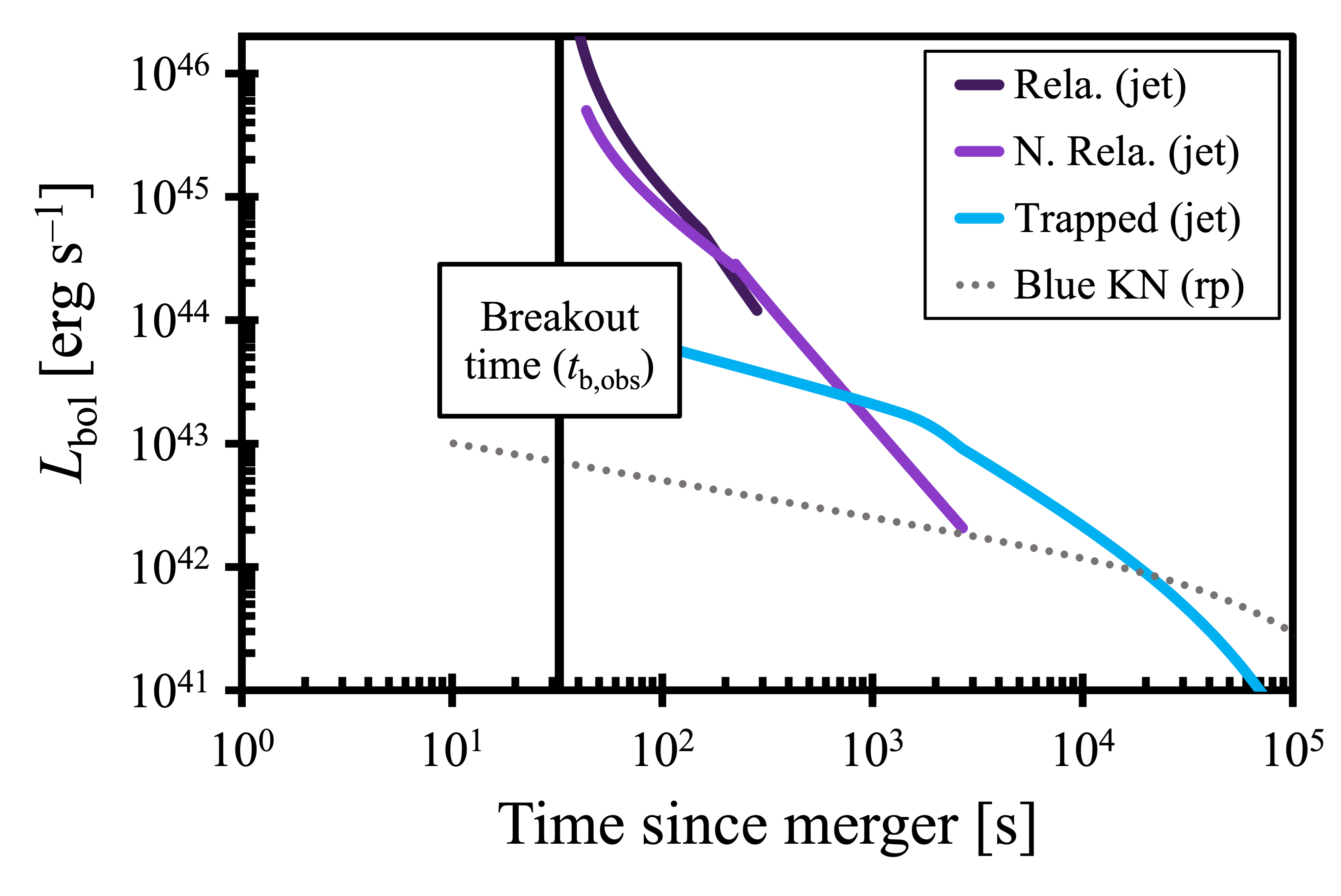} 
  \caption{Cooling emission from the jet-shock-heated cocoon, for the model EX-W.
  Three cocoon components are shown: relativistic cocoon (dark purple), non-relativistic cocoon (light purple), and trapped cocoon (light blue) [using equations (\ref{eq:Lbl full}) and (\ref{eq:Lbl tr})].
  The vertical (black) line indicates the breakout time in the observer's frame [$t_{b,obs}=(1-\beta_m)t_b$; see equation (\ref{eq:t obs lab})].
  Analytic estimation of the r-process powered KN emission is also shown (dotted grey line) [calculated using equation (\ref{eq:Lbl KN})].
  The starting time of light curves is set as the start of the free-expansion phase $t_1\sim 2t_b$ (see §\ref{ap:cooling}).
  The transition of emission from the non-relativistic cocoon to the trapped cocoon is set to take place once the trapped cocoon's luminosity dominates over the non-relativistic cocoon's luminosity (here at $\sim 10^3$ s; see §\ref{sec:cocoon parts}).
  }
  \label{fig:Lbl} 
\end{figure}
\begin{figure*}
    \centering
    \begin{subfigure}
    \centering
    \includegraphics[width=0.49\linewidth]{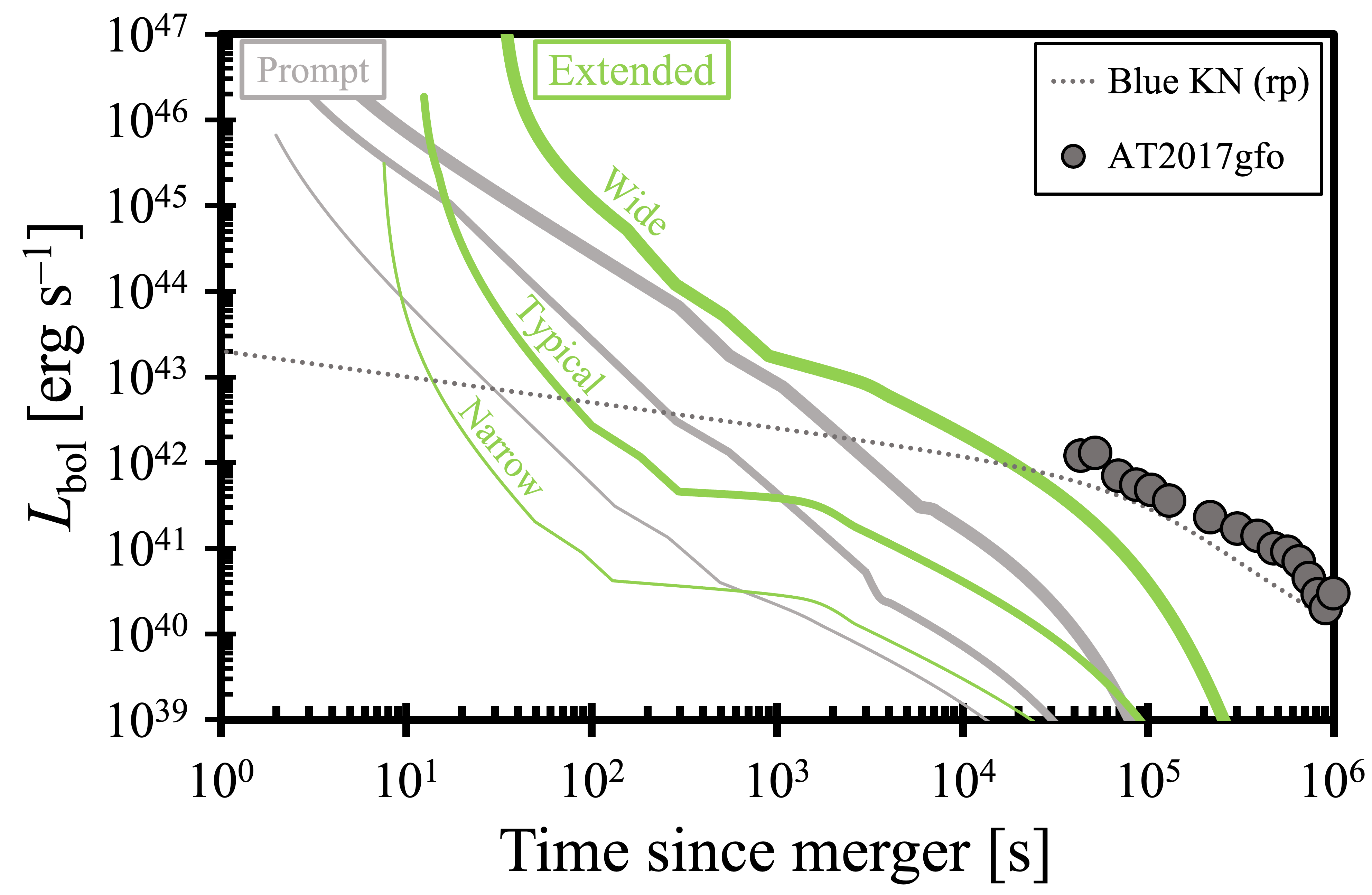}
    \includegraphics[width=0.49\linewidth]{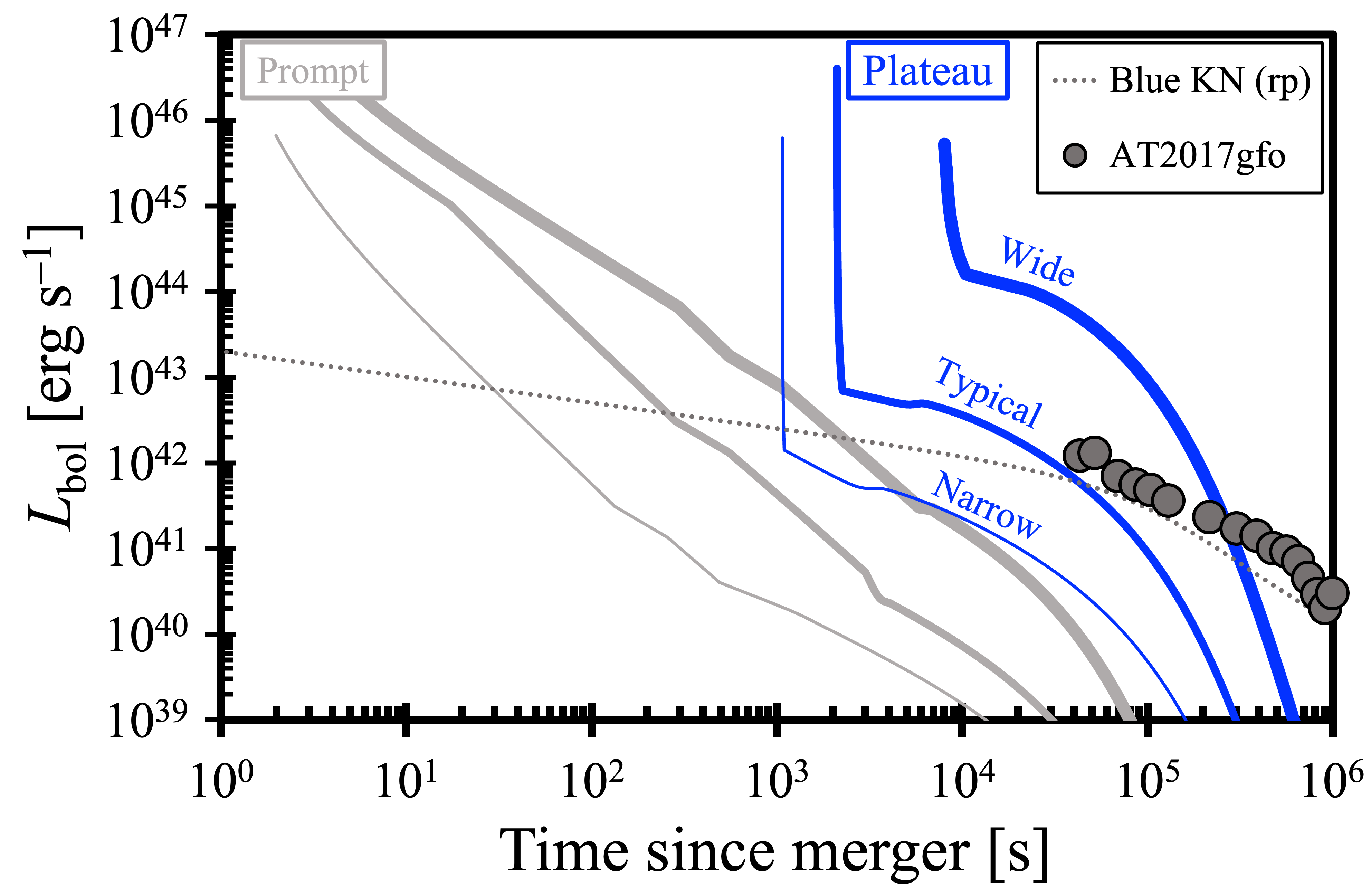}
    \centering
    \begin{subfigure}
    \centering
    \includegraphics[width=0.49\linewidth]{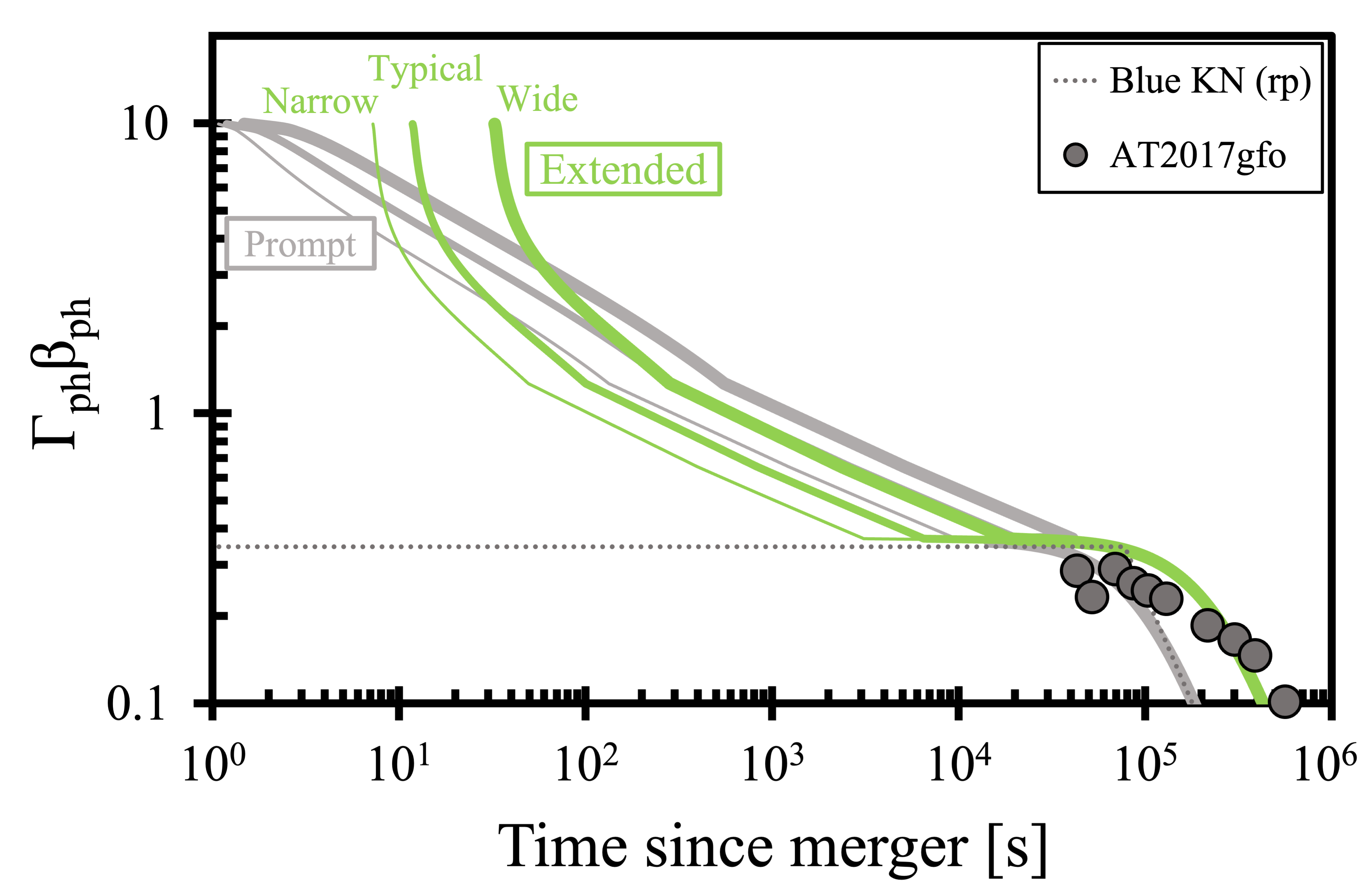}
    \includegraphics[width=0.49\linewidth]{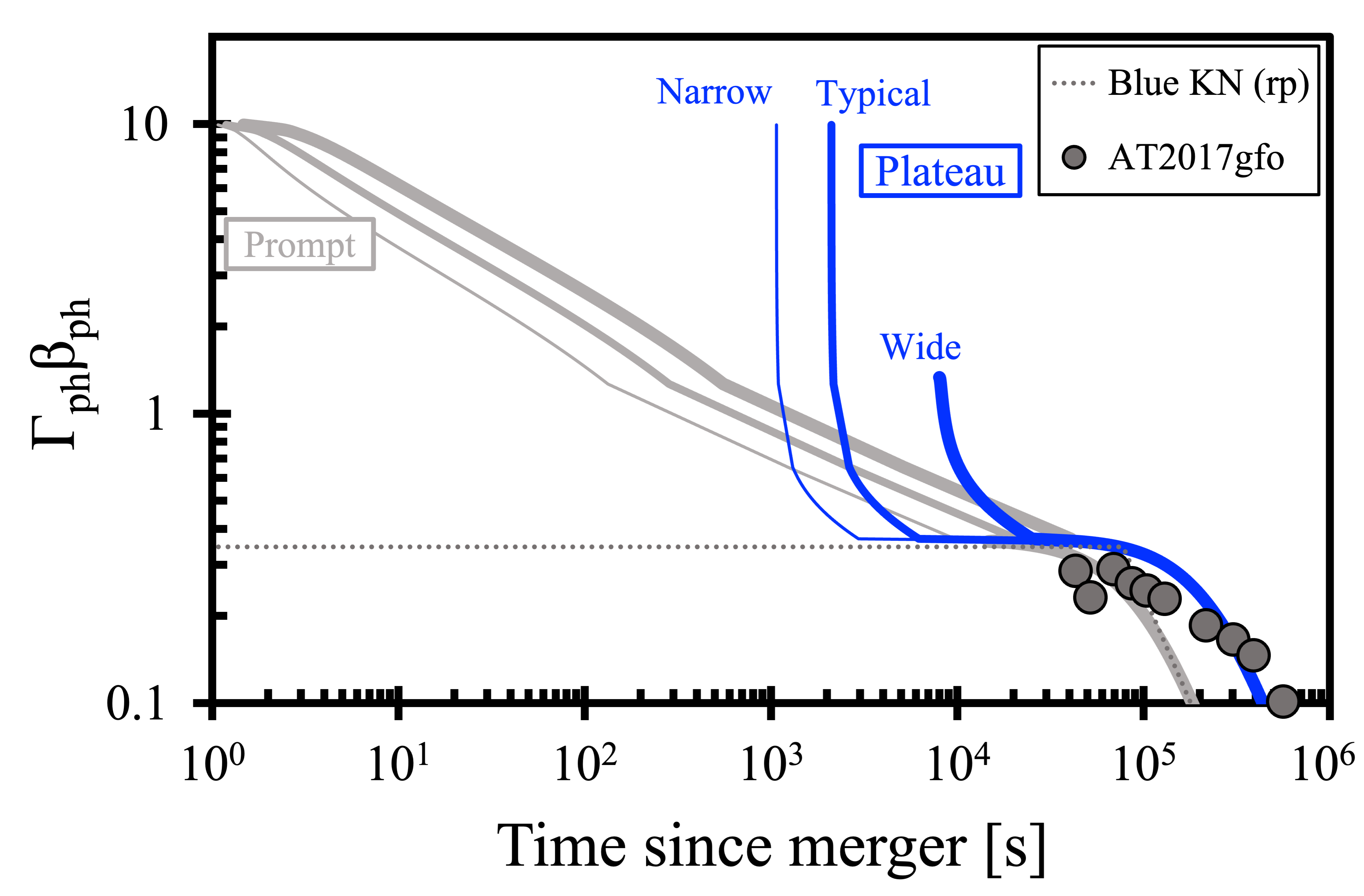}
  \end{subfigure}
        \end{subfigure}
    \centering
    \begin{subfigure}
    \centering
    \includegraphics[width=0.49\linewidth]{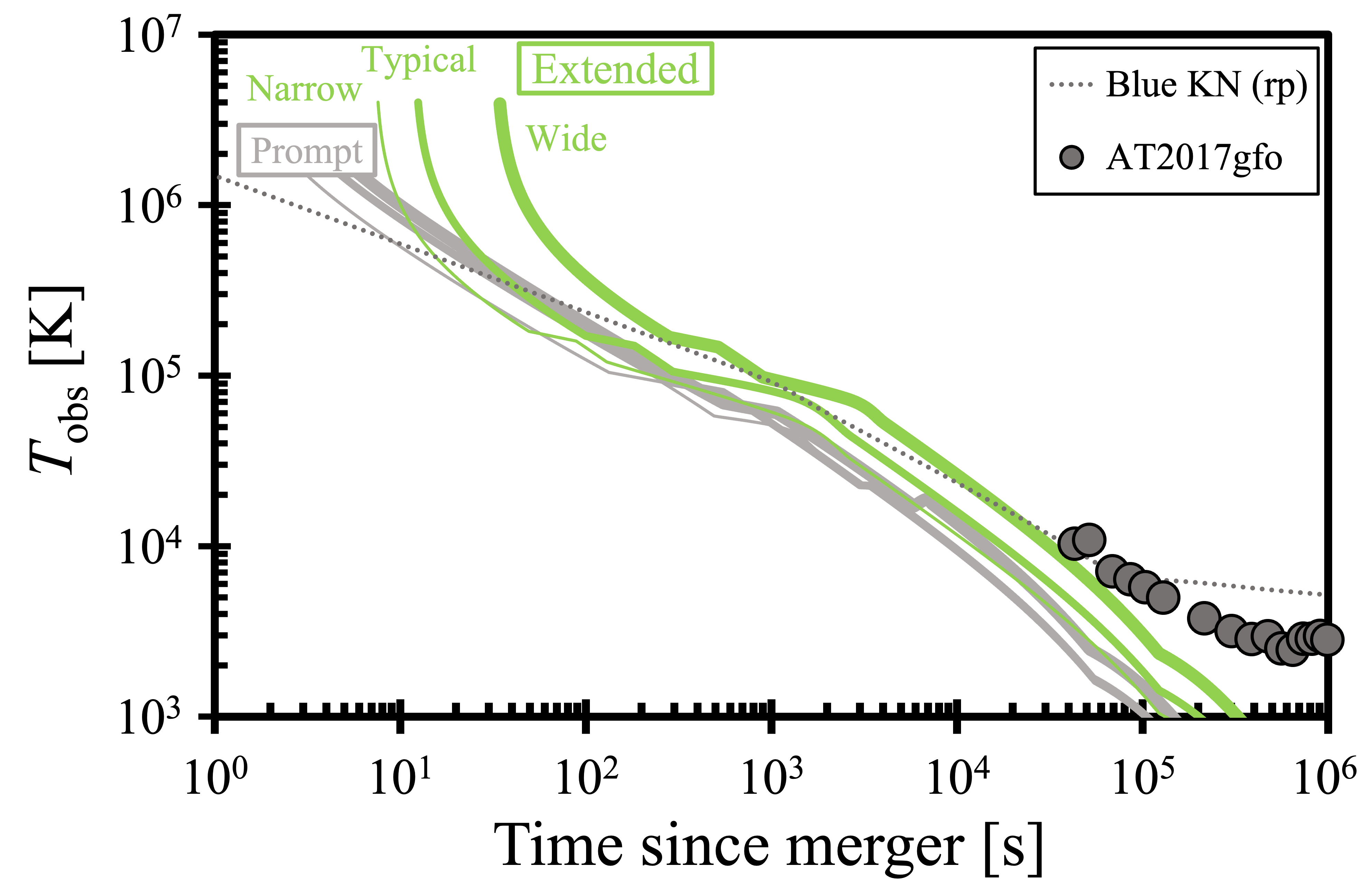}
    \includegraphics[width=0.49\linewidth]{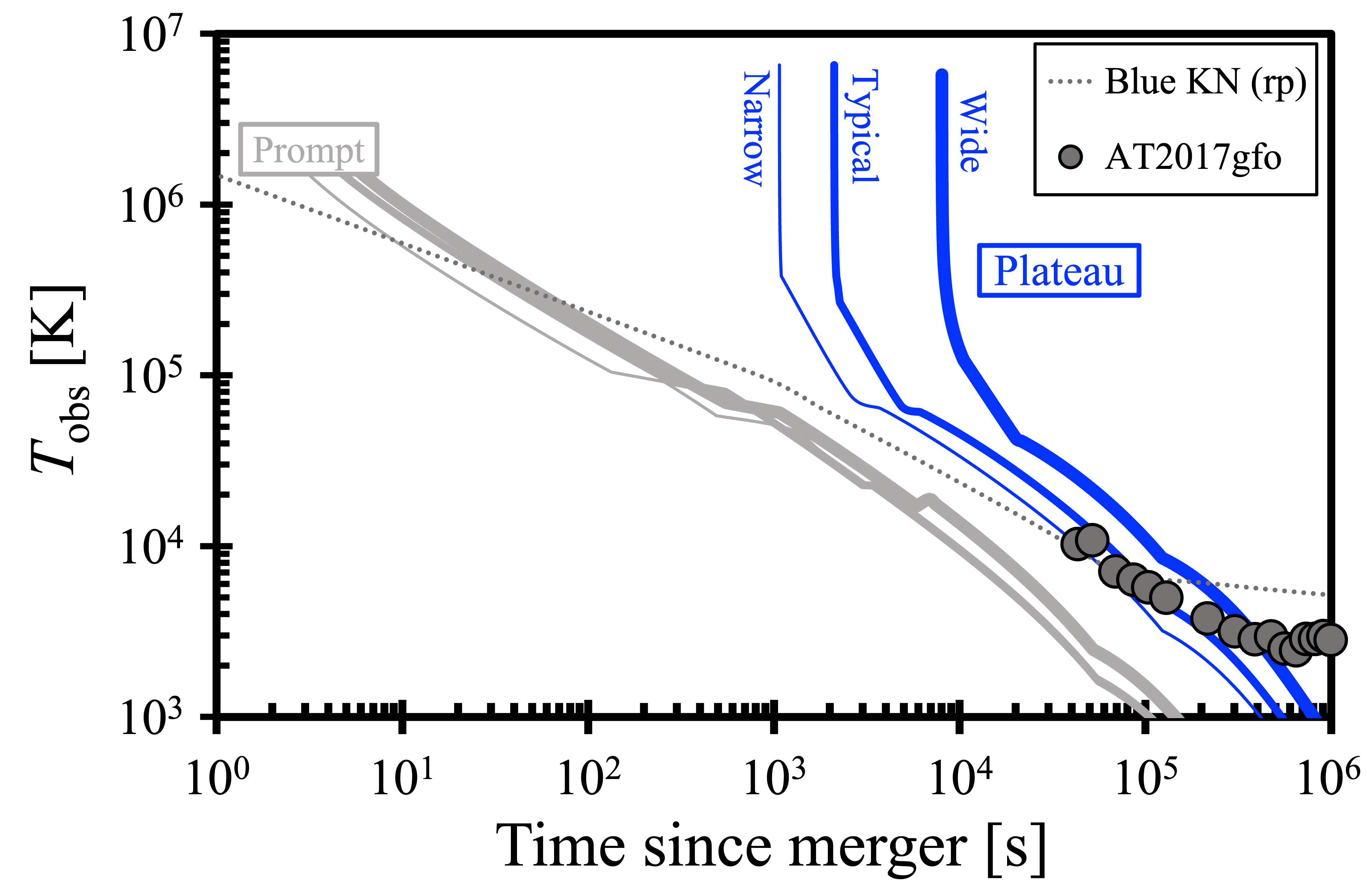}
  \end{subfigure}
  \caption{
  Bolometric isotropic luminosity (top), photospheric four-velocity (middle), and observed temperature (bottom), for different jet models [prompt (grey), extended (green; left panels), and plateau (blue; right panels)] with different opening angle [narrow (thin), typical (medium), and wide (thick)], as a function of the observed time since the merger.
  Our analytic model of the r-process powered early KN (grey dotted line) is shown (see §\ref{ap:KN}) for comparison, as well as the observations of the KN in AT2017gfo (grey circles; \citealt{2017ApJ...851L..21V,2018MNRAS.481.3423W}).
  The starting time of the cocoon emission is set as the start of the free-expansion phase $t_1$ (see §\ref{ap:cooling}).
  For narrow and typical prompt jet models, $T_{obs}$ evolution at later times is nearly identical.
  }
  \label{fig:LVT} 
\end{figure*}

\subsubsection{Photospheric four-velocity}
\label{sec:photo}
Photospheric four-velocity is shown in Figure \ref{fig:LVT} (middle panels).
The late time evolution is similar for all jet models, as the photosphere moves inward, across the escaped cocoon, and to the trapped cocoon.
As explained in \cite{2023MNRAS.524.4841H}, the photospheric velocity can be inferred from observations, and is potentially a powerful tool to reveal the relativistic nature of the cocoon.
Also, it can be used to, in theory, measure the product $(\kappa M_e)^{\frac{1}{7}}$.
Although it is challenging to break the apparent model degeneracy by only relaying on the observed photospheric velocity at later times, as it can be seen from Figure \ref{fig:LVT}; 
distinguishing between prompt, extended, and plateau jets is feasible provided that early observations ($\lesssim 10^2$ s after the jet breakout) of the photospheric velocity are available\footnote{It should be stressed that, in the early part, measuring this photospheric velocity from observation requires extra attention to relativistic effects (see §4.6.2 in \citealt{2023MNRAS.524.4841H} for a robust approach).}.

At early times, and immediately after the breakout, the photospheric velocity drops from $\Gamma_{ph}\beta_{ph}\sim 10$ to $\sim 1$ in $\sim 10^2-10^3$ s.
This unique evolution of $\Gamma_{ph}\beta_{ph}$ is a very useful quantity to monitor in order to identify the cocoon and extract information about its environment (e.g., parameters of the engine/jet and the ejecta).
High cadence observations are ideal to monitor this fast evolution.

\subsubsection{Observed temperature}
\label{sec:temp}
With our blackbody approximation, the color of the cocoon emission is determined by the observed temperature.
In Figure \ref{fig:LVT} (bottom panel), $T_{obs}$ is plotted as a function of the observed time since the merger $t_{obs}-t_{m,obs}$.
The evolution of $T_{obs}$ is determined by the time since the breakout $t_{obs}-t_{b,obs}$. 
The difference in $t_{obs}-t_{b,obs}$ causes an apparent difference between $T_{obs}$ for prompt and for late jet models in Figure \ref{fig:LVT}. 
Apart from this, the overall color of the cocoon emission from prompt and late jets is intrinsically similar.

\begin{figure*}
    \centering
    \begin{subfigure}
    \centering
    \includegraphics[width=0.49\linewidth]{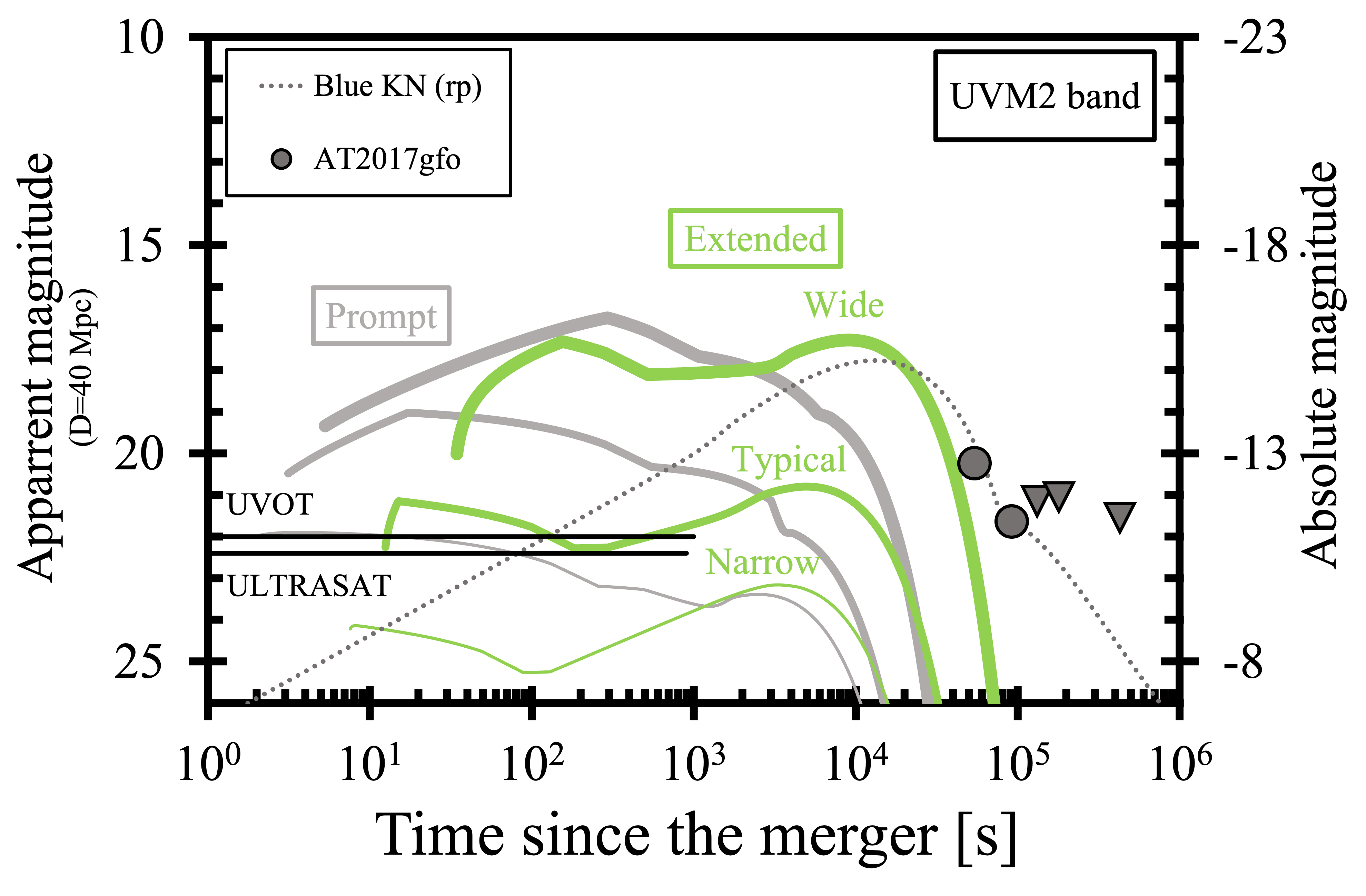}
    \includegraphics[width=0.49\linewidth]{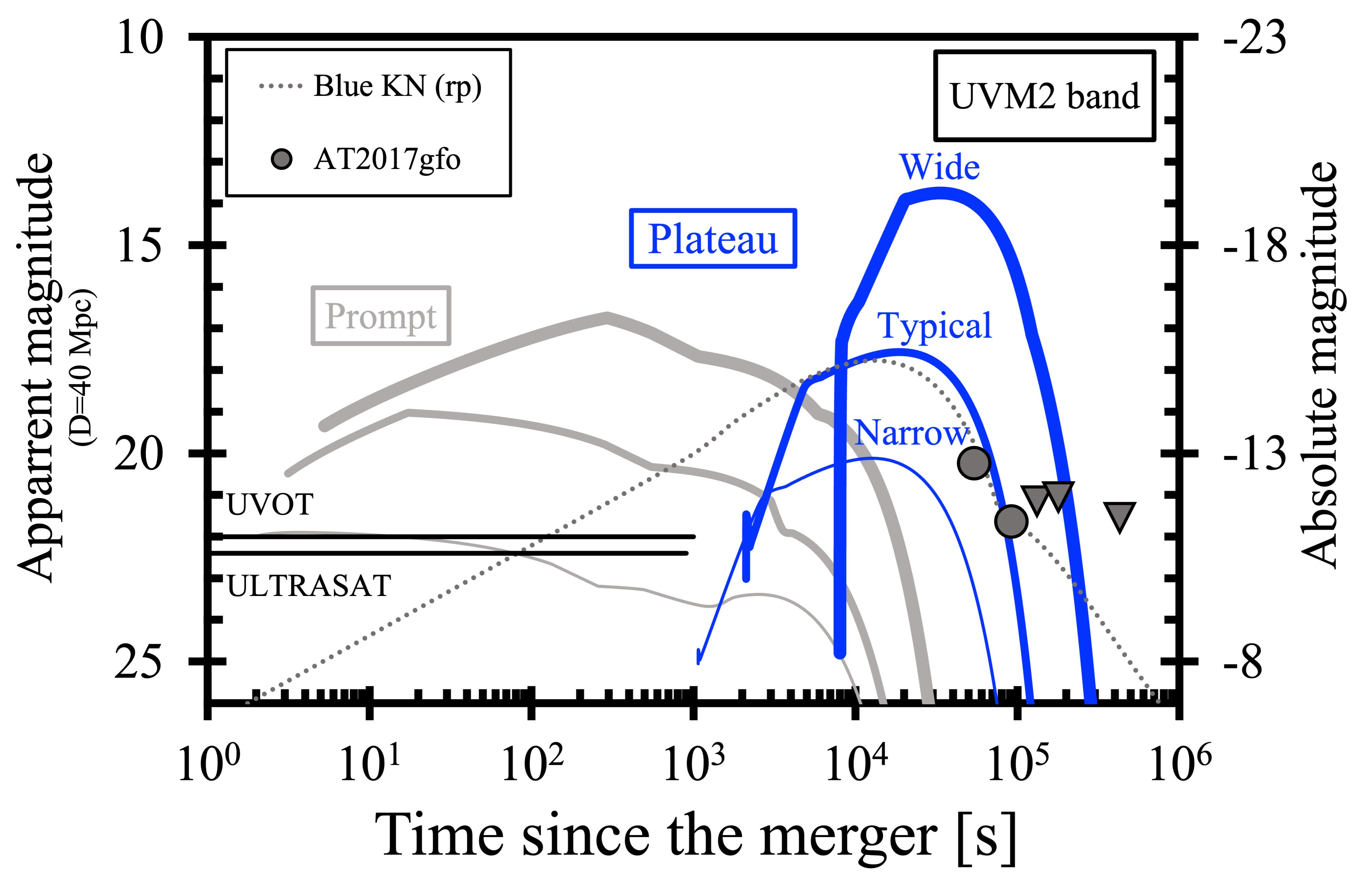}
    \centering
    \begin{subfigure}
    \centering
    \includegraphics[width=0.49\linewidth]{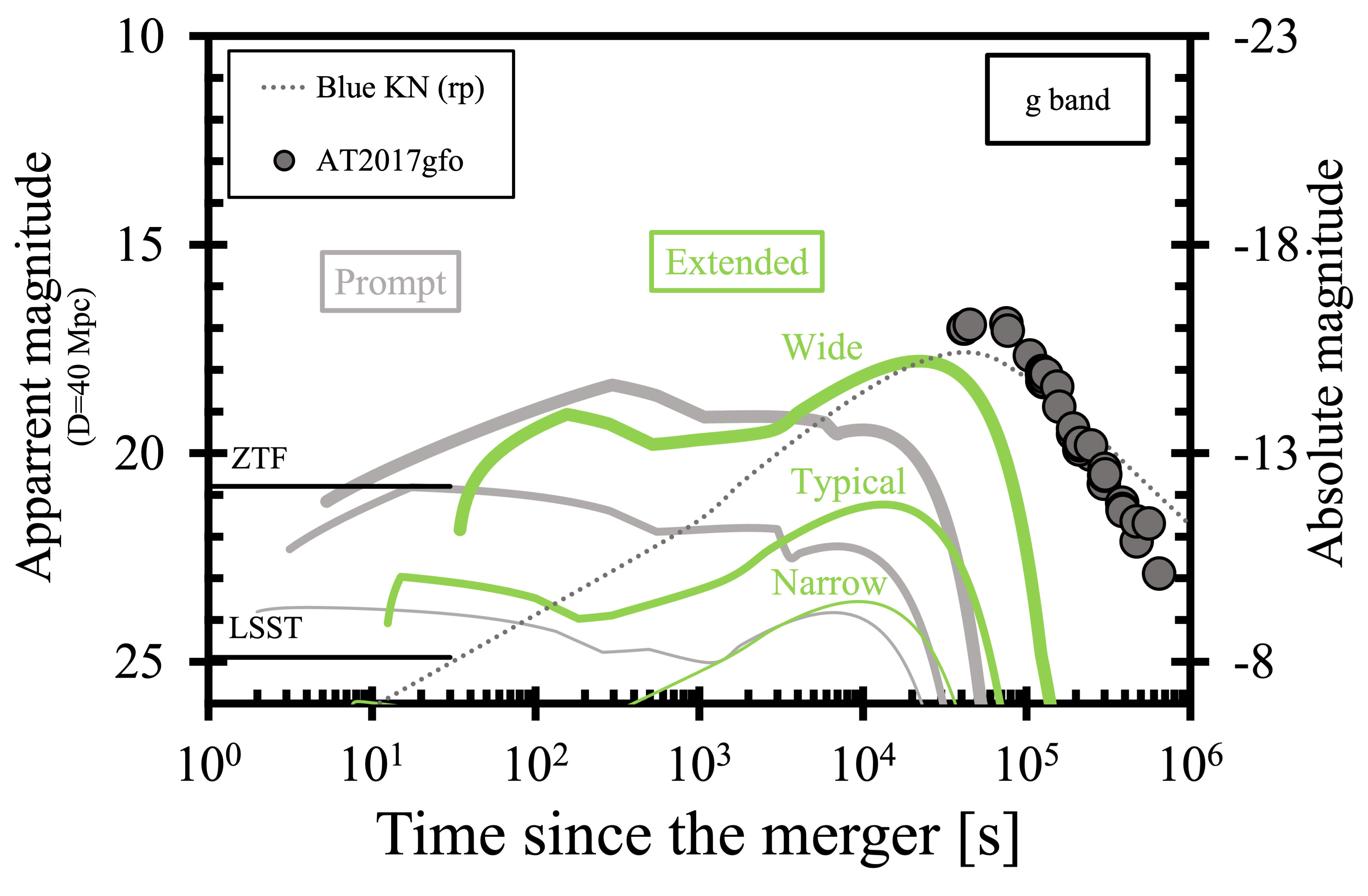}
    \includegraphics[width=0.49\linewidth]{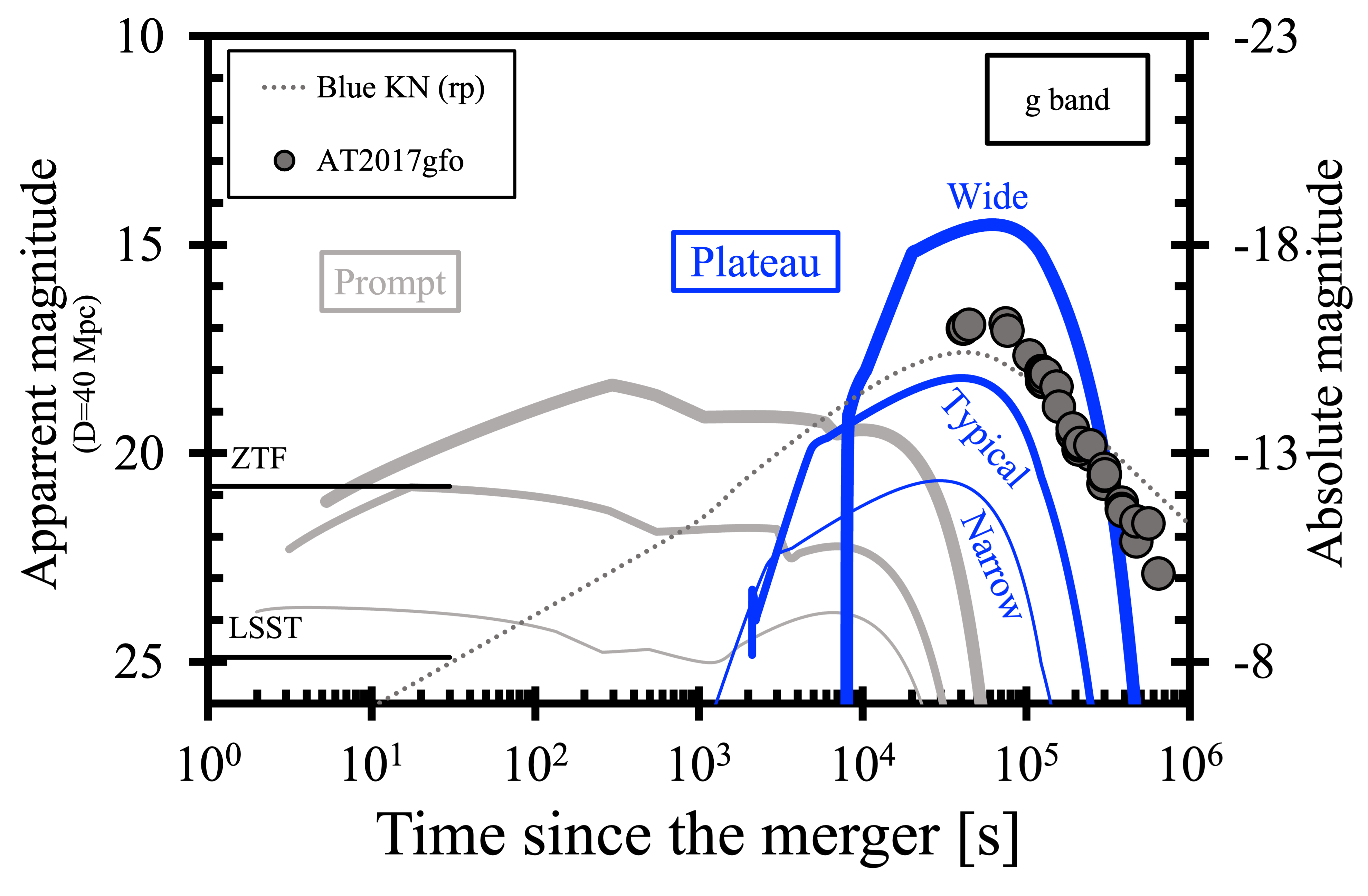}
  \end{subfigure}
        \end{subfigure}
    \centering
  \caption{Apparent (left axis; for a distance $D=40$ Mpc) and absolute (right axis) magnitudes of the cocoon emission from different jet phases [prompt (light grey), extended (green), and plateau (blue)] and different jet models [narrow (thin lines), typical (medium), and wide (thick)] as in Table \ref{tab:models}]. 
  Top and bottom panels are for UVOT's UVM2 band, and g band; respectively.
  The expected magnitude of the r-process powered KN is shown (dotted grey line) using our analytic model in §\ref{ap:KN}.
  Observation of the KN AT2017gfo are also shown (grey circles for detection, and triangles for upper limits; taken from \citealt{2017Sci...358.1559K}; \citealt{2017Sci...358.1570D}; \citealt{2017ApJ...851L..21V} and the references therein).
  Limiting magnitude of the relevant detectors is indicated [horizontal black lines; UVOT: $\sim 22$ for an exposure time of $\sim 1000$ s (\citealt{2005SSRv..120...95R}); ULTRASAT: $\sim 22.4$ for $900$ s (\citealt{2014AJ....147...79S}); ZTF: 20.8 for $30$ s (\citealt{2020ApJ...905...32B}); LSST: 24.9 for $30$ s (\citealt{2019ApJ...873..111I})].
  All models are shown in the frame of the source (no reddening applied) and AT2017gfo observations have been de-reddened [using Milky Way extinction model; $R_v=3.1$ and $E(B-V)=0.11$] using (\citealt{2010ApJ...721.1608B}; for UVM2 band) and (\citealt{1998ApJ...500..525S}; for g band).
  A jet with parameters similar to those of the typical plateau jet model (PL-T) appears to reproduce the blue kilonova (KN) of AT2017gfo effectively. There is a slight difference between the analytic models in the UV band and the g band, e.g., in comparison to AT2017gfo. This discrepancy, although the emission process is a blackbody in both cases, arises from our use of a simplified, non-wavelength-dependent mean grey opacity ($\kappa=1$ cm$^{2}$ {g}$^{-1}$; see \citealt{2023arXiv230405810B}).
  }
  \label{fig:mag} 
\end{figure*}

In all cases, immediately after the breakout, for a few tens of seconds, the observed temperature from the escaped cocoon is so high ($\gtrsim 10^6$K), that blackbody emission is expected to peak in soft X-ray. The temperature dependency, of the peak of the emission, can be expressed using Wien's displacement law as
\begin{equation}
    h \nu_{\text {peak }}\approx 0.24\, \mathrm{keV}  \left(\frac{T_{obs}}{10^6 \text{\:K}}\right) .
\end{equation}
As explained in §\ref{sec:approx emission}, strong attenuation by the bound-free process was found as located at an energy higher
than $h \nu_{\rm peak}$ (by at least a factor of 3) hence its irrelevance.
The idea that early cocoon emission can power a bright soft X-ray transients has already been presented in \cite{2023MNRAS.524.4841H}.
This X-ray emission part is discussed in more details later in §\ref{sec:X-ray} (also see Figure \ref{fig:X-ray}).
At later times, emission from the escaped cocoon is expected to peak at UV bands ($\sim 10^4-10^5$K) [see Figure \ref{fig:mag}].

In the case of prompt jets, after the escaped cocoon becomes optically thin at late times ($\gtrsim 10^3 - 10^4$ s), the trapped cocoon's emission is irrelevant due to its extreme faintness (e.g., relative to the r-process powered KN).
However, for late jets, the trapped cocoon is brighter and lasts longer (due to the arguments presented in  §\ref{sec:cocoon parts}).
Temperature of the trapped cocoon part is on the order of $\sim$ a few $10^4$ K, making it bright in optical bands too.
This is comparable to the temperature range of the blue KN (e.g., in AT2017gfo), and suggests that the trapped cocoon component, heated by late engine activity, can be one possible explanation of the blue KNe component (in AT2017gfo; \citealt{2017Sci...358.1559K}; \citealt{2017Sci...358.1570D}; \citealt{2017ApJ...851L..21V}).
Also, being bright in optical makes this cocoon emission very accessible from an observational point of view (compared to UV observations) as optical facilities are more abundant, and optical emission is less affected by extinction.

In all cases, the fast temperature evolution shows the multi-messenger and unique nature of the cocoon emission in \textit{s}GRBs; as it shifts from soft X-ray to UV, and then to optical bands in a matter of hours.
This makes it easy to identify, although high cadence observations are required.
In late jets, this emission come at later times (e.g., after the merger) making it even more accessible.

\section{Discussion}
\label{sec:5}
\subsection{UV/Optical signal}
\label{sec:mag}
In Figure \ref{fig:mag} we show the magnitudes (absolute, and apparent for a distance of $40$ Mpc as for GW170817/AT2017gfo) of the cocoon emission, in comparison to that of the r-process powered KN and observations of AT2017gfo.

The pattern of late engine-powered cocoon emission, compared to that of the prompt cocoon, show three main differences.
First, the emission is delayed due to the late jet breakout.
Second, the early part (from the mildly relativistic escaped cocoon) is slightly dimmer.
Third, the cocoon emission is still luminous at late times ($\sim 10^3-10^5$ s; $\sim$ 1 h - 1 d), thanks to the contribution of the trapped cocoon (due to the same arguments presented in §\ref{sec:cocoon parts} and §\ref{sec:bolo}).

For the plateau jet case (also for extended wide jet), in terms of magnitude, the late time peak (trapped cocoon) coincides with the peak of the r-process powered KN component (grey dotted line in Figure \ref{fig:mag}).
Again, this shows that the late-jet's trapped cocoon is capable of explaining the observed blue KN component by itself.
With the wide plateau jet, this component is far brighter, mainly due to the fact that the jet failed to breakout ($t_b>t_e$).
This case is already excluded for AT2017gfo, but could be a possible explanation of KNe candidates claimed by \cite{2020MNRAS.493.3379R} that have typical red KN components but their blue KN components are exceptionally bright (e.g., \sgrbs 050724, 060614, and 070714B; also see \citealt{2015ApJ...811L..22J,2017ApJ...837...50G,2018NatCo...9.4089T,2018ApJ...860...62G,2020NatAs...4...77J} and the references within).

Comparison with the r-process powered KN (in Figure \ref{fig:mag}), and considering limiting magnitudes of different instruments (in UV: UVOT and the upcoming ULTRASAT; in optical: ZTF and the upcoming LSST), suggest that cocoon emission from wider late jets dominates the KN for longer timescales, and can easily be detected.
For typical jet opening angles, the emission is dimmer, especially for the extended emission's jet case, as the trapped cocoon is not powerful enough to compete with the r-process powered KN.
For narrow jets, the cocoon emission seems extremely difficult to detect.

\cite{2017ApJ...834...28N}, followed by \cite{2018MNRAS.473..576G} and \cite{2023MNRAS.524.4841H}, explained that the cocoon emission is mainly a UV transient.
This is confirmed here, in more specific details;
while the escaped part of the cocoon shines in UV, in the form of a fast transient, the trapped cocoon part (only for late jets) is visible in optical at later times for longer.

These findings stress the importance of a wide field UV survey, such as the ULTRASAT mission ($\sim 200$ deg$^2$; \citealt{2014AJ....147...79S}); which could detect the cocoon emission without the need of a GW signal.
Optical wide field surveys (e.g., LSST) are also important. 
However, high cadence ($\lesssim 10^3$ s) is necessary, otherwise, the detection would come at later times when the uniqueness of the cocoon emission (e.g., in terms of the fast photospheric evolution) could be missed, making it difficult to identify due to contamination by other fast optical transients (e.g., peculiar ultra-stripped supernovae \citealt{2023ApJ...949..120H,2023A&A...675A.201A}; also, galactic dwarf novae, and variable quasars \citealt{2023arXiv231010066O}).

\subsection{X-ray cocoon signal}
\label{sec:X-ray}
\begin{figure*}
    \centering
    \begin{subfigure}
    \centering
    \includegraphics[width=0.79\linewidth]{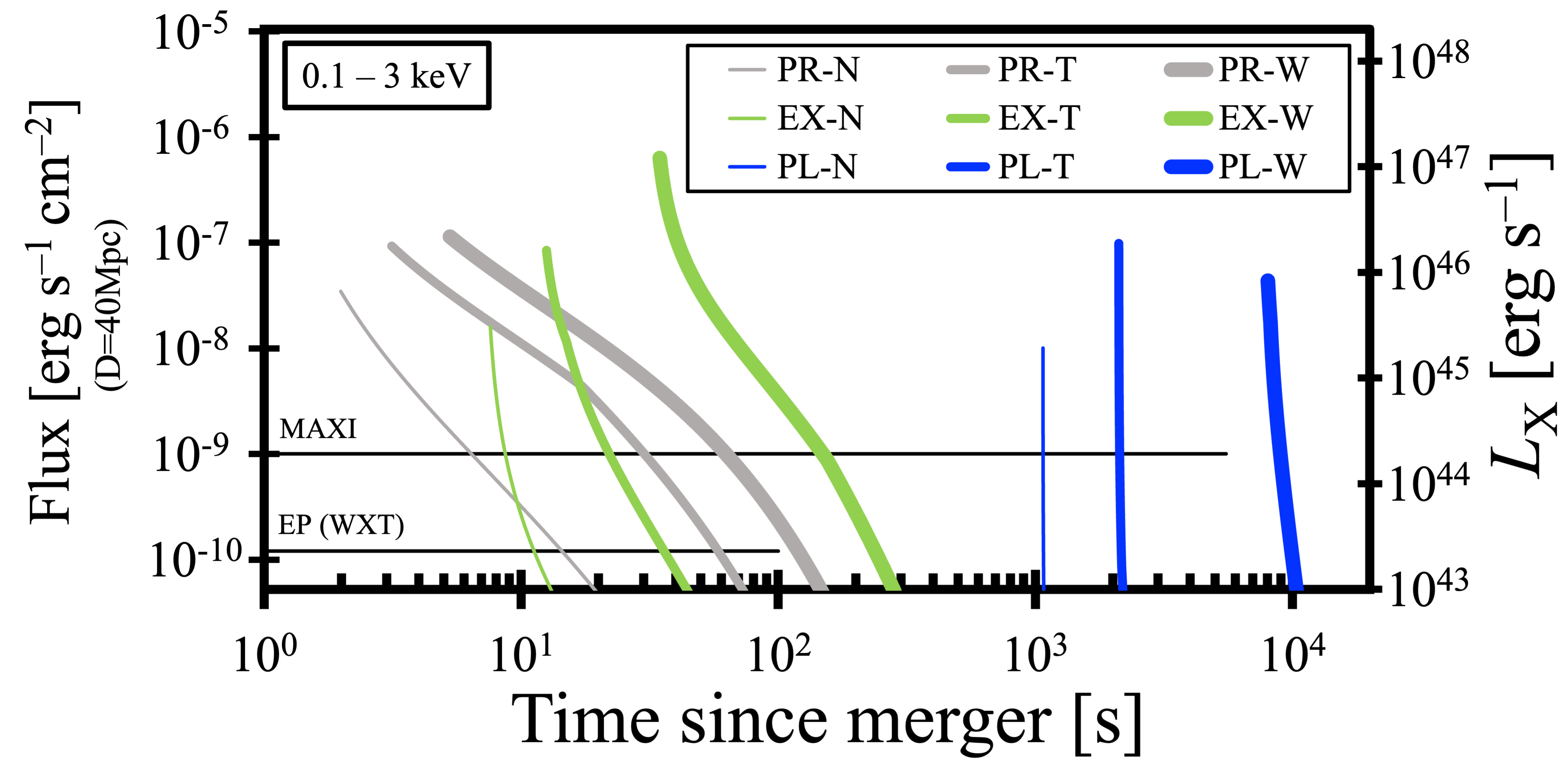}
  \end{subfigure}
    \centering
    \begin{subfigure}
    \centering
    \includegraphics[width=0.79\linewidth]{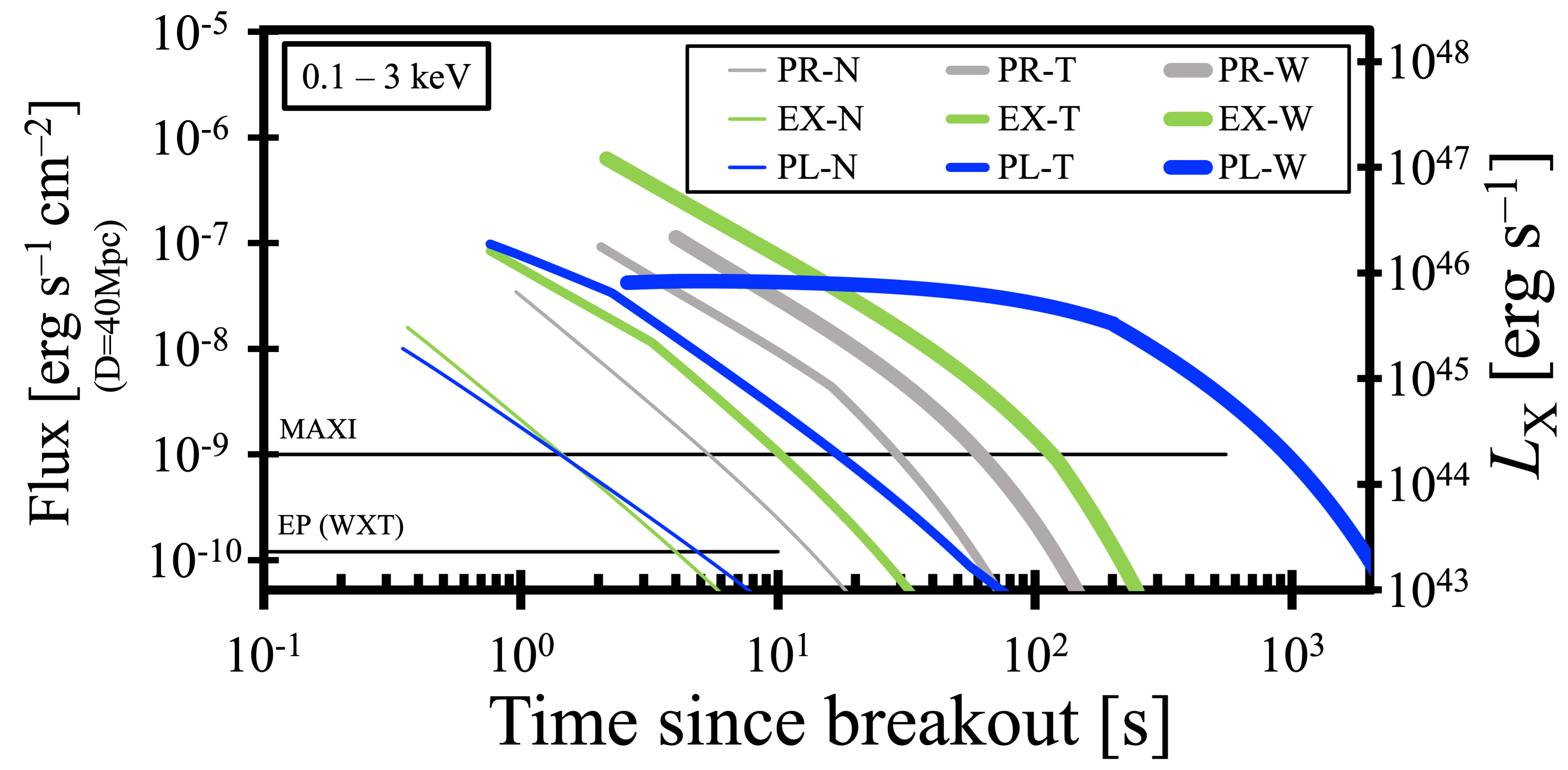}
  \end{subfigure}
  \caption{Cocoon emission in soft X-ray (0.1-1 keV) as a function of time since the merger (top) and time since the jet breakout (bottom).
  Both the flux (left, for $D=40$ Mpc) and the luminosity (right) are shown.
  Prompt (grey), extended (green), and plateau (blue) jet models with narrow (thin), typical (medium) and wide (thick) opening angles are shown (see Table \ref{tab:models}).
  Sensitivity of MAXI ($\sim 10^{-9}$ erg cm$^{-2}$ s$^{-1}$ per orbit in $2-10$ keV) and Einstein probe's WXT ($\sim 1.2\times10^{-10}$ erg cm$^{-2}$ s$^{-1}$ for 100 s in $0.5-4$ keV) instruments are shown (\citealt{2009PASJ...61..999M,2022hxga.book...86Y}).
  The starting time of this cocoon emission is set as the start of the free-expansion phase $t_1$ (see §\ref{ap:cooling}).
  }
  \label{fig:X-ray} 
\end{figure*}

In Figure \ref{fig:X-ray} we present the cocoon emission in soft X-ray.
This X-ray luminosity has been calculated by integrating Plank law in the range $0.1 - 3$ keV.
Note that the possibility of strong attenuation by the bound-free process has been considered, but we found that the effect is weak in the escaped cocoon at the considered soft X-ray range (as discussed in §\ref{sec:approx emission}).

Depending on the engine phase, the onset of the X-ray emission is different relative to the merger time (top) [or to the prompt emission time].
Emission since the jet breakout time (bottom) show that the X-ray emission is similar for the different models; 
with two exceptions: the wider the jet opening angle the longer the emission (due to the large escaped cocoon mass), and the model PL-W is an outlier due to the different fate of its jet (failed; i.e., no relativistic cocoon) and the non-relativistic cocoon being the emitting component, making its time evolution slower.

This shows that the cocoon breakout can trigger a soft X-ray transient (in addition to emission in a multitude of other bands; see Figure \ref{fig:mag}).
This X-ray emission is different from other discussed emissions that can be observed in X-ray, such as shock breakout, and cocoon afterglow (\citealt{2010ApJ...725..904N,2017ApJ...834...28N}).
Also, this emission is similar in nature to that expected in the case of long GRBs (\citealt{2018MNRAS.478.4553D}), although here it is fainter (and shorter) by about one order of magnitude (each) due to the weaker jet and the less massive ejecta.

The upcoming Einstein Probe (EP) mission (expected in 2024; \citealt{2022arXiv220909763Y}) in particular, with its very wide field of view ($\sim 3600$ deg$^2$) would be ideal to detect such transient, independently of GW/GRB signals (also the HiZ-GUNDAM mission planned for the next decade, \citealt{2020SPIE11444E..2ZY}; SVOM telescope with a FoV of $\sim$1 degree (\citealt{2022IJMPD..3130008A}); 
and KOYOH, small satellites with wide-field X-ray launched recently\footnote{\url{https://space.skyrocket.de/doc_sdat/koyoh.htm}}).
With EP's deep sensitivity, detection could be possible for distances much larger than $40$ Mpc.
Spectroscopic analysis will allow us to identify it as cocoon emission; 
given that the spectrum is blackbody with its peak shifting to lower frequencies with time (see Figure \ref{fig:LVT}; for more details see the dependency of $T_{obs}$ in \citealt{2023MNRAS.524.4841H}).

Another relevant soft X-ray detector for this emission is the all sky monitor MAXI (\citealt{2009PASJ...61..999M}).
In fact, as argued in \cite{2023MNRAS.524.4841H}, such transients might have already been detected by MAXI in the past but remained unidentified (MAXI Unidentified Short Soft Transient; e.g., \citealt{2016GCN.20206....1D}; \citealt{2015ATel.7960....1M}; \citealt{2015GCN.17772....1H}; \citealt{2014GCN.16686....1U}; etc.) detected by MAXI (in $2-10$ keV), but most likely missed by Swift/BAT (only sensitive to harder X-ray, $15-50$ keV; same for Fermi/GBM) and also missed by late Swift/XRT follow-ups due to their fast decay.
This is supported by data indicating that the rate of MAXI's X-ray transients is three times higher than the GRB rate, which might suggest that contribution from off-axis GRB events have been captured (although contamination from galactic transients is another possibility; \citealt{2014PASJ...66...87S}).
As a note, with MAXI's X-ray band and sensitivity, the duration of this emission is expected to be shorter than in Figure \ref{fig:X-ray}.
Also, this X-ray emission could be present in the sample of fast X-ray transient candidates observed by Chandra in $0.5-7.0$ keV (\citealt{2017MNRAS.467.4841B,2022A&A...663A.168Q,2023A&A...675A..44Q}) although the duration of most of the candidates is much longer ($\sim 10^3 -10^4$ s; with a few exceptions, e.g., XRT 191127).
Also, it is worth noting that this X-ray emission can also explain some X-Ray Flashes (XRFs; \citealt{2002ApJ...571L..31Y,2005ApJ...629..311S,2008ApJ...679..570S}).

Our modeling of this X-ray emission is simplistic, assuming LTE, grey opacity, etc.
These are crude but fairly reasonable considerations (as discussed in §\ref{sec:approx emission}).
Also, extinction in the host galaxy has not been taken into account and is assumed to have a limited effect on this emission (mostly Galactic as BNS mergers are expected to occur at the outskirts of their host galaxies).
Lastly, one challenge related to detecting this X-ray emission is contamination by early jet afterglow (non-thermal) emission.
Therefore, ideal events are \sgrbs seen with a LOS slightly outside the jet core.

\subsection{Future detection}
\label{sec:rate}
Here we estimate the number of cocoon-powered events that could be detected in the near future.
For the rate of BNS mergers, we take $\sim 320^{+490}_{-240}$ Gpc$^{-3}$ yr$^{-1}$ as our fiducial value (\citealt{2021ApJ...913L...7A}; based on the first half of O3; also see \citealt{2022LRR....25....1M}).
We roughly assume that every BNS merger will launch a jet (as in GW170817/GRB 170817A; \citealt{2017ApJ...848L..13A}).
We consider the cocoon from the wide jet, as our model (with $\theta_0\sim \theta_j\sim 15^\circ$); and we take the probability of such wide jet among \textit{s}GRBs as $\sim 0.2$  (see Figure 4 in \citealt{2023ApJ...959...13R}; also see \citealt{2015ApJ...815..102F}).
We consider the case of late jet launch after the prompt jet; we assume that there is a probability of $\sim 0.7$ for this to happen (following \citealt{2017ApJ...846..142K} results suggesting that $\sim 70\%$ of Swift-\textit{s}GRBs show signs of late emission).

\subsubsection{Detection in soft X-ray}
\label{sec:rate X}
Let us estimate the detection rate of the cocoon emission by Einstein Probe (EP) mission.
The opening angle of the escaped cocoon is on the order of $\theta_c^{es}\sim 30^\circ$ for the wide jet model [see equation (\ref{eq:theta c es}); also see simulation results in \citealt{2023MNRAS.520.1111H}]. 
This is also the expected opening angle for the soft X-ray emission (as $\theta_c> 1/\Gamma$).
This reduces the number of detectable events by a factor $[{1-\cos(\theta_c^{es})}]^{}\sim 0.13$.
Then, considering the sensitivity of EP (see Figure \ref{fig:X-ray}), and considering late jet models (for $D=40$ Mpc),
the cocoon emission can be detected up to $\sim 25$ times further away (up to $\sim 1$ Gpc away in an optimistic scenario)\footnote{This is reasonable considering the early part of the X-ray emission that is expected (roughly with extrapolation) to be even brighter; although our analytic model is unable to solve this at very early times (as $t\sim t_{b}$).}.
EP's field of view as a fraction of all sky reduces the number of events by the factor $\sim \frac{3600}{41253}\sim 0.087$.
The yearly number of events to be expected can be found as the product of, 
BNS merger rate ($\sim 320^{+490}_{-240}$ Gpc$^{-3}$ yr$^{-1}$),
volume ($4\pi/3$ Gpc$^3$),
the probability of \textit{s}GRB-jet launch in a BNS merger (taken as 1), 
the probability of a wide jet launch ($\sim 0.2$), 
the probability of a late jet launch ($\sim 0.7$),
the probability of the cocoon emission in the line-of-sight ($\sim0.13$), and
the probability of the emission being in the field of view of EP's WXT instrument ($\sim0.087$).
Hence, the rate of yearly detection of this soft X-ray emission can be found as:
\begin{equation}
    R_{X, cocoon}\sim 2.1^{+3.2}_{-1.6}\times \left[\frac{D}{1\rm{ \:Gpc}}\right]^3 \rm{yr}^{-1}.
    \label{eq:rateX}
\end{equation}
This detection rate is equivalent to $\sim 1/600$ BNS mergers within 1 Gpc being detected.
Considering the more recent, but much less constrained, estimate of BNS event rate (as $10-1700$ Gpc$^{-3}$ yr$^{-1}$; \citealt{2023PhRvX..13a1048A}), the yearly detection rate will take a wider range of values as $0.07-11\rm~yr^{-1}$.
The BNS event rate has also been estimated using knowledge of the jet structure and GRB emission as  $2-680$ Gpc$^{-3}$ yr$^{-1}$ (\citealt{2023ApJ...954...92H}), but more observations of BNS mergers (and \textit{s}GRBs) are needed to further constrain it.

Hence, we expect that EP will be able to independently detect cocoon cooling emission in soft X-ray (also HiZGUNDAM and other soft X-ray instruments).
With the properties of this emission known (see §\ref{sec:X-ray}), identification will be possible.
Such detection will have several important uses, such as constraining the rate of BNS merger and the fraction of \textit{s}GRB-jets that are choked (failed).
Such detection could also be used to trigger follow-up campaigns and identify KNe independently.
Therefore, this detection process is important as it offers us a new channel to probe BNS mergers in the Universe, independently of \sgrbs and GWs.

As the opening angle of the escaped cocoon ($\theta_c^{es}$) [see equation (\ref{eq:theta c es})] is $\sim 2$ times larger than the jet opening angle ($\theta_j$), 
simple estimation indicates that the expected number of these X-ray transients is about $\sim 4$ times larger than the number of \textit{s}GRBs in the local Universe ($<$ 1 Gpc).

\subsubsection{Detection as an EM counterpart to GW}
\label{sec:rate ligo}
Here, we roughly evaluate the detection rate of the cocoon emission as a counterpart of GW events, in optical or UV bands, and for BNS merger events observed with at least 2 GW detectors.
We focus on the upcoming O5 run of LIGO-VIRGO-KAGRA.
With O5 run, LIGO is expected to detect BNS mergers up to $\sim$200 Mpc (angle-averaged sensitivity; \citealt{2020LRR....23....3A}).
We assume that, at such distance, early EM follow-up and localization will be successful (e.g., with LSST's wide field of view).
The duration of O5 is set for 2 years. 
However, as each detector is expected to be active for $\sim 70-80\%$ of the time, effectively the duration will be $\sim 1$ year (\citealt{2020LRR....23....3A,2021CQGra..38m5014D}).

We take into account contribution from the trapped cocoon, as its emission has the widest opening angle and can be as large as $\theta_c\sim 70^\circ$ (see §\ref{ap:trapped}) [considering that it is concealed by the ejecta on the sides making it dimmer].
Hence, considering that we are using an angle-averaged sensitivity, the probability of detecting BNS with a viewing angle within $\theta_c\sim 70^\circ$ of the polar axis\footnote{Here, it has been assumed that the \sgrb jet is launched perfectly aligned with the polar direction, which is seemingly the case for GW170817/GRB 170817A (see \citealt{2019MNRAS.489.1919T}).} can be found as $1-\cos\theta_c\approx 0.66$.

The number of events to be expected in the O5 run can be found as the product of, 
BNS merger rate ($\sim 320^{+490}_{-240}$ Gpc$^{-3}$ yr$^{-1}$),
observable volume ($\sim 0.03$ Mpc$^3$),
the probability of \textit{s}GRB-jet launch (taken as 1), 
the probability of a wide jet launch ($\sim 0.2$), 
the probability of a late jet launch ($\sim 0.7$),
the probability of BNS merging so that an on-axis line-of-sight of the trapped cocoon, within $\theta_c$, is possible ($\sim 0.66$), and
the time of O5 multiplied by the duty cycle ($\sim 2$ yr$\times 0.5$).
This gives:
\begin{equation}
    N_{cocoon,O5}\sim 1.0_{-0.7}^{+1.5} \times \left[\frac{D}{200\:\rm{Mpc}}\right]^3\left[\frac{t_{O5}}{2\:\rm{yr}}\right] .
    \label{eq:rateO5}
\end{equation}
This is equivalent to $\sim 1/20$ of every BNS mergers within 200 Mpc being detected.
This rate of detection is slightly lower than that found in equation (\ref{eq:rateX}) [due to the limitation in distance $D$] but overall similar remarks can be made.
Considering recent analysis by \cite{2023PhRvX..13a1048A} suggesting a BNS merger rate of $10-1700$ Gpc$^{-3}$ yr$^{-1}$, the above number of events (for 2 years) is found within the range $\sim 0.03-5.3$ Gpc$^{-3}$.

Therefore, although these estimations are based on preliminary parameters (i.e., of O5 that will most likely be subject to changes), we estimate that the odds are favorable for the cocoon emission to be detected and well monitored in O5 (as an EM counterpart to GWs) for the first time.
Hence, O5 will play an important role at answering questions related to GW170817/AT2017gfo, \textit{s}GRBs, and late engine activity.

For the LIGO-VIRGO-KAGRA O4B run, at the time of writing, VIRGO's sensitivity is planned to reach 80 Mpc for the last 3 quarters of 2024 (although this is constantly subject to change).
If so, the expected detection rate in O4B would be much lower than that found for O5 [see equation (\ref{eq:rateO5})], by a factor of $\sim(80/200)^3(0.75/2)\sim 1/40$; i.e., $\sim 0.024\ll 1$ event in O4B.

It should be noted that combining these detection channels with targets of opportunity from GRB detectors, is expected to increase even further the number of events.
Also, as the detection rate is $\propto D^3$, future 3rd generation of GW detectors (Cosmic Explorer \citealt{2019BAAS...51g..35R}; and Einstein Telescope \citealt{2010CQGra..27h4007P,2010CQGra..27s4002P,2020JCAP...03..050M}) expected in the coming decade and with future wide field detectors (e.g., LSST; \citealt{2019ApJ...873..111I})
will dramatically increase the detection rate by $\sim 3$ orders of magnitude (e.g., \citealt{2023ApJ...942...88Z}).



\section{Summary \& Conclusion}
\label{sec:6}
Considering the observational evidence of late extended/plateau emission phases after the prompt emission in most \textit{s}GRBs (e.g., see \citealt{2017ApJ...846..142K}), 
and that these late emission phases are most likely due to late engine activity,
we explored the scenario of propagation of late jet outflow in BNS merger ejecta (see Figure \ref{fig:f0}).
Following observational analysis of \sgrbs (\citealt{2015ApJ...815..102F,2017ApJ...846..142K,2023ApJ...959...13R}),
we estimated the typical parameters of prompt jets and late jets 
in consistency with the observed prompt, extended, and plateau emission phases (see §\ref{sec:para}; and Figure \ref{fig:intro}).
For each of these three phases, we considered three types of jets according to their width: narrow ($3^\circ$), typical ($6^\circ$), and wide ($15^\circ$) [see Table \ref{tab:models}].

Based on the analytic model for prompt jet propagation in expanding medium \cite{2020MNRAS.491.3192H,2021MNRAS.500..627H}, we analytically solved late jets' propagation and breakout [see equation (\ref{eq:tb simple}); §\ref{ap:jet}].
Following the same arguments as in \cite{2023MNRAS.520.1111H}, we estimated the properties of the jet-shock heated cocoon before and after the breakout (see §\ref{ap:es}).
Then, in §\ref{sec:3}, we presented an analytic model to estimate the cocoon cooling emission. 
Our model was based on \cite{2023MNRAS.524.4841H} with additional improvements as required for late jets (see §\ref{ap:cooling}).
To our best knowledge, this is the first time that the cocoon of late jets have been considered, and its emission have been calculated.

Our results indicate that typical plateau jets can barely breakout (Table \ref{tab:models}).
This indicates that, although it is possible to explain plateau emission of $\sim 10^{46}$ erg s$^{-1}$ with jet luminosities $\sim 10^{47}$ erg s$^{-1}$ and radiation efficiency $\eta_{\gamma}\sim 0.1$,
it would be difficult to explain fainter plateau emissions that have been observed (see Figure \ref{fig:intro}), unless the ejecta mass is much smaller than what has been observed in AT2017gfo ($\sim 0.05 \msun$ or prompt/extended jets are strong enough to permanently leave a cavity (see §\ref{sec:late jet approx}).
Such low mass is plausible in systems where the disk/ejecta mass is small, which is favored by mergers of BNS of equal masses (see \citealt{2019ApJ...876L..31K}).
Another possibility is plateau jets with $\eta_\gamma\gtrsim 0.1$.

We explained that the cocoon emission initiates upon the jet breakout. Consequently, the delay between the merger and the onset of the cocoon emission can be easily measured, rendering it a useful quantity as a proxy for the jet breakout time [see equation (\ref{eq:tb simple})].

We found that for later jets, 
the cocoon emission is more luminous and lasts longer, compared to prompt jets (see top panels in Figure \ref{fig:LVT}).
We showed that late jet head velocity ($\langle{\beta_h}\rangle\approx \frac{t_b}{t_b-t_0}\beta_m\sim \beta_m$) is smaller compared to prompt jets ($\langle{\beta_h}\rangle\sim 2\beta_m$; \citealt{2018PTEP.2018d3E02I,2020MNRAS.491.3192H}) making late jets more efficient at depositing energy into the cocoon.
Furthermore, because breakout times are longer in late jets, adiabatic cooling ($\propto t_b/t$) is less efficient, and more internal energy is available for the cooling emission. 
In prompt jets, compared to late jets, shock heating takes place much earlier ($\sim 1$ s; compared $\sim 10-10^4$ s), and adiabatic cooling consumes much of the internal energy before cooling emission start to diffuse out.
Consequently, for certain late jets, the trapped cocoon (irrelevant in prompt jets) can be brighter than the expected r-process powered KN at late times ($\sim 10^4-10^5$ s) [see Figures \ref{fig:Lbl} and \ref{fig:LVT} (top panels); also see equation (\ref{eq:Lbl PL-T})].

We also showed that the brightness of the cocoon emission is strongly dependent on the jet opening angle, in agreement with \cite{2023MNRAS.524.4841H} [see their equations (59) and (60)].

The cocoon emission was found to peak in the UV during most of its emission, $\sim 10^2-10^5$ s after the merger (see bottom panels in Figure \ref{fig:LVT} and top panels in Figure \ref{fig:mag}), making it an excellent target for the upcoming ULTRAST mission.
The trapped cocoon (of late jets) is found to peak with $\sim 10^4-10^3$ K, at $\sim 10^3-10^5$ s after the merger, making it bright in optical bands as well.
This makes it easier to identify due to the less extinction in optical, and the large number of facilities/surveys in optical bands with more depth in their observations (e.g., LSST).

At late times ($\sim 10^4-10^5$ s), we found that late jets (plateau in particular) with their cocoons (e.g., model PL-T) can produce a blue KN-like emission, similar to the one in AT2017gfo (see Figures \ref{fig:LVT} and \ref{fig:mag}).
For wider jets (e.g., model PL-W), an extremely bright blue KN can be produced (see Figures \ref{fig:LVT} and \ref{fig:mag}).
This might be one possible scenario for the exceptionally bright blue KNe identified in some \sgrbs (\citealt{2020MNRAS.493.3379R}; also see \citealt{2013Natur.500..547T,2015ApJ...811L..22J,2017ApJ...837...50G,2018NatCo...9.4089T,2018ApJ...860...62G,2020NatAs...4...77J} and the references therein).
In this scenario the brightness of the red KN is determined by the ejecta mass, while the brightness of the blue KN is determined by energy deposition into the ejecta by the late jet (Hamidani et al. in preparation).

Previously, it has been suggested that AT2017gfo's early data can be explained with the prompt jet's cocoon emission \citep{2018ApJ...855..103P}.
However, we have clarified that the prompt jet's cocoon, given its parameters, is not able to explain AT2017gfo's data (see grey lines Figures \ref{fig:LVT} and \ref{fig:mag}; also see arguments in \citealt{2023MNRAS.520.1111H,2023MNRAS.524.4841H}).
Instead, we showed that late jets (plateau in particular) with their cocoon present a viable alternative to explain the early AT2017gfo data and the blue KN (see blue lines Figures \ref{fig:LVT}  and \ref{fig:mag}).

We also showed that the cocoon is bright in soft X-ray $10-100$ s after its breakout (see Figure \ref{fig:X-ray}).
This confirmed previous claims in \cite{2023MNRAS.524.4841H}.
We explained that such soft X-ray emission is within reach of the upcoming Einstein Probe mission.

We estimated the rate of detection of the cocoon emission by Einstein Probe, or as an EM counterpart to GWs in the O5 run of LIGO-VIRGO-KAGRA, and found that it can be detected on a yearly basis [see equations (\ref{eq:rateX}) and (\ref{eq:rateO5})].
This detection rate is expected to increase by $\sim 2-3$ orders of magnitude with future 3rd generation of GW detectors (Cosmic Explorer \citealt{2019BAAS...51g..35R}; and Einstein Telescope \citealt{2010CQGra..27h4007P,2010CQGra..27s4002P,2020JCAP...03..050M}) expected in the coming decade and with future wide field detectors such as LSST (\citealt{2019ApJ...873..111I}).

Traditionally BNS mergers have been tracked indirectly through the prompt emission of \textit{s}GRBs.
However, \sgrbs jets are beamed, and considering that their typical opening angle is $\theta_j\sim 6^\circ$ (\citealt{2023ApJ...959...13R}), only $\sim 1/200$ \sgrbs (or BNS mergers) is on-axis and can theoretically be detected.
This is without considering that an unknown fraction \textit{s}GRB-jets are choked and fail to produce a \textit{s}GRB.
Also, for instruments like Swift/BAT, this number is reduced by about one order of magnitude due to the limited field of view.
This has resulted in a bias toward intrinsically bright, on-axis, and distant \textit{s}GRBs, rendering them challenging to localize, follow up (e.g., in optical), and exploit.

GWs is another tool to detect BNS mergers.
GWs offer great advantages, as they are less sensitive to the viewing angle, and give the possibility of detecting BNS even if the \sgrb jet is off-axis.
This feature favors the detection of off-axis nearby events.
Closer distances for GW detected BNS events can make push these otherwise faint transients within the range of ground-based telescope facilities.
GWs also can detect BNS source a few minutes before the actual merger.
However, there are two disadvantages associated with GW detection of BNS: i) the spatial localization is poor (although Fermi GBM's localization can be poorer) requiring more GW detectors, and ii) the current limited detectors' horizon making BNS detections with GWs quite rare.

Here, we showed that the cocoon of \textit{s}GRB-jets can be another channel to detect BNS mergers, in particular in soft X-ray (also in UV), with wide field and high cadence observations.
The upcoming EP and ULTRASAT missions are ideal for detecting this emission (also the upcoming UVEX mission; \citealt{2021arXiv211115608K}).
With EP, we estimate that $\sim 1/600$ BNS mergers within $\sim 1$ Gpc could be detected [see equation (\ref{eq:rateX})]. 

The cocoon emission has its advantages for identifying BNS mergers, compared to \textit{s}GRBs, as it is not biased to the jet; 
it may even accompany BNS mergers that are off-axis or with failed jets (also low-$\Gamma$ jets; \citealt{2016ApJ...829..112L}); 
and is good at detecting more local BNS mergers. 
The cocoon emission (e.g., with EP) has also its advantages compared to current GW detectors in terms of the horizon of the detection, exposure time, sky localization, and number of events [see equations (\ref{eq:rateX}) and (\ref{eq:rateO5})].
Although off-axis afterglow emission (non-thermal) from structured \textit{s}GRB-jet could contaminate this cooling emission (thermal), 
difference in timescales is expected.

Hence, the cocoon and its emission can be a useful tool.
In the era of multi-messenger astronomy, understanding and tracking this cocoon emission, in coordination with \sgrb and GW observations, could help make breakthroughs regarding the central engine of \textit{s}GRBs, their jets, and the origin of heavy elements.


\section{Data availability}
The data underlying this article will be shared on reasonable request to the corresponding author.

\begin{acknowledgements}
We thank
    Amir Levinson, 
    Asano Katsuaki, 
    Banerjee Smaranika, 
    Daisuke Yonetoku,
    Ehud Nakar,
    Gavin Lamb,
    Hirotaka Ito,
    Kazumi Kashiyama, 
    Kenta Hotokezaka,
    Koutarou Kyutoku, 
    Kota Hayashi,
    Ore Gottlieb, 
    Ryo Yamazaki,
    Shigeru Yoshida,
    Shinya Wanajo,
    Shoichi Yamada,
    Shota Kisaka, 
    Matsumoto Tatsuya, 
    and Yutaka Ohira,
    for their fruitful discussions and comments. 

    This research was supported by Japan Science and Technology Agency (JST) FOREST Program (Grant Number JPMJFR212Y),
    the Japan Society for the Promotion of Science (JSPS) Grant-in-Aid for Scientific Research (19H00694, 20H00158, 20H00179, 21H04997, 23H00127, 23H04894, 23H04891, 23H05432), 
    JSPS Bilateral Joint Research Project, and National Institute for Fusion Science (NIFS) Collaborative Research Program (NIFS22KIIF005).
  
    This work was partly supported by 
    JSPS KAKENHI nos. 20H00158, 20H01904, 20H01901, 22H00130, 23H04900, 23H05430, 23H01172 (KI), 22K14028, 21H04487, 23H04899 (S.S.K.), 23K19059. 
    S.S.K. acknowledges the support by  the Tohoku Initiative for Fostering Global Researchers for Interdisciplinary Sciences (TI-FRIS) of MEXT’s Strategic Professional Development Program for Young Researchers.

    Numerical computations were achieved thanks to the following: Cray XC50 of the Center for Computational Astrophysics at the National Astronomical Observatory of Japan, and Cray XC40 at the Yukawa Institute Computer Facility.
\end{acknowledgements}

\bibliography{0-P6}

\begin{thebibliography}{}
\expandafter\ifx\csname natexlab\endcsname\relax\def\natexlab#1{#1}\fi
\providecommand{\url}[1]{\href{#1}{#1}}
\providecommand{\dodoi}[1]{doi:~\href{http://doi.org/#1}{\nolinkurl{#1}}}
\providecommand{\doeprint}[1]{\href{http://ascl.net/#1}{\nolinkurl{http://ascl.net/#1}}}
\providecommand{\doarXiv}[1]{\href{https://arxiv.org/abs/#1}{\nolinkurl{https://arxiv.org/abs/#1}}}

\bibitem[{{Abbott} {et~al.}(2017{\natexlab{a}}){Abbott}, {Abbott}, {Abbott}, {Acernese}, {Ackley}, {Adams}, {Adams}, {Addesso}, {Adhikari}, {Adya}, \& et~al.}]{2017PhRvL.119p1101A}
{Abbott}, B.~P., {Abbott}, R., {Abbott}, T.~D., {et~al.} 2017{\natexlab{a}}, Physical Review Letters, 119, 161101, \dodoi{10.1103/PhysRevLett.119.161101}

\bibitem[{{Abbott} {et~al.}(2017{\natexlab{b}}){Abbott}, {Abbott}, {Abbott}, {Acernese}, {Ackley}, {Adams}, {Adams}, {Addesso}, {Adhikari}, {Adya}, \& et~al.}]{2017ApJ...848L..13A}
---. 2017{\natexlab{b}}, \apjl, 848, L13, \dodoi{10.3847/2041-8213/aa920c}

\bibitem[{{Abbott} {et~al.}(2020){Abbott}, {Abbott}, {Abbott}, {Abraham}, {Acernese}, {Ackley}, {Adams}, {Adya}, {Affeldt}, {Agathos}, {Agatsuma}, {Aggarwal}, {Aguiar}, {Aiello}, {Ain}, {Ajith}, {Akutsu}, {Allen}, {Allocca}, {Aloy}, {Altin}, {Amato}, {Ananyeva}, {Anderson}, {Anderson}, {Ando}, {Angelova}, {Antier}, {Appert}, {Arai}, {Arai}, {Arai}, {Araki}, {Araya}, {Araya}, {Areeda}, {Ar{\`e}ne}, {Aritomi}, {Arnaud}, {Arun}, {Ascenzi}, {Ashton}, {Aso}, {Aston}, {Astone}, {Aubin}, {Aufmuth}, {Aultoneal}, {Austin}, {Avendano}, {Avila-Alvarez}, {Babak}, {Bacon}, {Badaracco}, {Bader}, {Bae}, {Bae}, {Baiotti}, {Bajpai}, {Baker}, {Baldaccini}, {Ballardin}, {Ballmer}, {Banagiri}, {Barayoga}, {Barclay}, {Barish}, {Barker}, {Barkett}, {Barnum}, {Barone}, {Barr}, {Barsotti}, {Barsuglia}, {Barta}, {Bartlett}, {Barton}, {Bartos}, {Bassiri}, {Basti}, {Bawaj}, {Bayley}, {Bazzan}, {B{\'e}csy}, {Bejger}, {Belahcene}, {Bell}, {Beniwal}, {Berger}, {Bergmann}, {Bernuzzi}, {Bero}, {Berry}, {Bersanetti}, {Bertolini},
  {Betzwieser}, {Bhandare}, {Bidler}, {Bilenko}, {Bilgili}, {Billingsley}, {Birch}, {Birney}, {Birnholtz}, {Biscans}, {Biscoveanu}, {Bisht}, {Bitossi}, {Bizouard}, {Blackburn}, {Blair}, {Blair}, {Blair}, {Bloemen}, {Bode}, {Boer}, {Boetzel}, {Bogaert}, {Bondu}, {Bonilla}, {Bonnand}, {Booker}, {Boom}, {Booth}, {Bork}, {Boschi}, {Bose}, {Bossie}, {Bossilkov}, {Bosveld}, {Bouffanais}, {Bozzi}, {Bradaschia}, {Brady}, {Bramley}, {Branchesi}, {Brau}, {Briant}, {Briggs}, {Brighenti}, {Brillet}, {Brinkmann}, {Brisson}, {Brockill}, {Brooks}, {Brown}, {Brown}, {Brunett}, {Buikema}, {Bulik}, {Bulten}, {Buonanno}, {Buskulic}, {Buy}, {Byer}, {Cabero}, {Cadonati}, {Cagnoli}, {Cahillane}, {Bustillo}, {Callister}, {Calloni}, {Camp}, {Campbell}, {Canepa}, {Cannon}, {Cannon}, {Cao}, {Cao}, {Capocasa}, {Carbognani}, {Caride}, {Carney}, {Carullo}, {Casanueva Diaz}, {Casentini}, {Caudill}, {Cavagli{\`a}}, {Cavalier}, {Cavalieri}, {Cella}, {Cerd{\'a}-Dur{\'a}n}, {Cerretani}, {Cesarini}, {Chaibi}, {Chakravarti}, {Chamberlin},
  {Chan}, {Chan}, {Chao}, {Charlton}, {Chase}, {Chassande-Mottin}, {Chatterjee}, {Chaturvedi}, {Chatziioannou}, {Cheeseboro}, {Chen}, {Chen}, {Chen}, {Chen}, {Chen}, {Chen}, {Cheng}, {Cheong}, {Chia}, {Chincarini}, {Chiummo}, {Cho}, {Cho}, {Cho}, {Christensen}, {Chu}, {Chu}, {Chu}, {Chua}, {Chung}, {Chung}, {Ciani}, {Ciobanu}, {Ciolfi}, {Cipriano}, {Cirone}, {Clara}, {Clark}, {Clearwater}, {Cleva}, {Cocchieri}, {Coccia}, {Cohadon}, {Cohen}, {Colgan}, {Colleoni}, {Collette}, {Collins}, {Cominsky}, {Constancio}, {Conti}, {Cooper}, {Corban}, {Corbitt}, {Cordero-Carri{\'o}n}, {Corley}, {Cornish}, {Corsi}, {Cortese}, {Costa}, {Cotesta}, {Coughlin}, {Coughlin}, {Coulon}, {Countryman}, {Couvares}, {Covas}, {Cowan}, {Coward}, {Cowart}, {Coyne}, {Coyne}, {Creighton}, {Creighton}, {Cripe}, {Croquette}, {Crowder}, {Cullen}, {Cumming}, {Cunningham}, {Cuoco}, {Dal Canton}, {D{\'a}lya}, {Danilishin}, {D'Antonio}, {Danzmann}, {Dasgupta}, {da Silva Costa}, {Datrier}, {Dattilo}, {Dave}, {Davier}, {Davis}, {Daw}, {Debra},
  {Deenadayalan}, {Degallaix}, {de Laurentis}, {Del{\'e}glise}, {Pozzo}, {Demarchi}, {Demos}, {Dent}, {de Pietri}, {Derby}, {De Rosa}, {de Rossi}, {Desalvo}, {de Varona}, {Dhurandhar}, {D{\'\i}az}, {Dietrich}, {di Fiore}, {di Giovanni}, {di Girolamo}, {di Lieto}, {Ding}, {di Pace}, {di Palma}, {di Renzo}, {Dmitriev}, {Doctor}, {Doi}, {Donovan}, {Dooley}, {Doravari}, {Dorrington}, {Downes}, {Drago}, {Driggers}, {Du}, {Ducoin}, {Dupej}, {Dwyer}, {Easter}, {Edo}, {Edwards}, {Effler}, {Eguchi}, {Ehrens}, {Eichholz}, {Eikenberry}, {Eisenmann}, {Eisenstein}, {Enomoto}, {Essick}, {Estelles}, {Estevez}, {Etienne}, {Etzel}, {Evans}, {Evans}, {Fafone}, {Fair}, {Fairhurst}, {Fan}, {Farinon}, {Farr}, {Farr}, {Fauchon-Jones}, {Favata}, {Fays}, {Fazio}, {Fee}, {Feicht}, {Fejer}, {Feng}, {Fernandez-Galiana}, {Ferrante}, {Ferreira}, {Ferreira}, {Ferrini}, {Fidecaro}, {Fiori}, {Fiorucci}, {Fishbach}, {Fisher}, {Fishner}, {Fitz-Axen}, {Flaminio}, {Fletcher}, {Flynn}, {Fong}, {Font}, {Forsyth}, {Fournier}, {Frasca}, {Frasconi},
  {Frei}, {Freise}, {Frey}, {Frey}, {Fritschel}, {Frolov}, {Fujii}, {Fukunaga}, {Fukushima}, {Fulda}, {Fyffe}, {Gabbard}, {Gadre}, {Gaebel}, {Gair}, {Gammaitoni}, {Ganija}, {Gaonkar}, {Garcia}, {Garc{\'\i}a-Quir{\'o}s}, {Garufi}, {Gateley}, {Gaudio}, {Gaur}, {Gayathri}, {Ge}, {Gemme}, {Genin}, {Gennai}, {George}, {George}, {Gergely}, {Germain}, {Ghonge}, {Ghosh}, {Ghosh}, {Ghosh}, {Giacomazzo}, {Giaime}, {Giardina}, {Giazotto}, {Gill}, {Giordano}, {Glover}, {Godwin}, {Goetz}, {Goetz}, {Goncharov}, {Gonz{\'a}lez}, {Gonzalez Castro}, {Gopakumar}, {Gorodetsky}, {Gossan}, {Gosselin}, {Gouaty}, {Grado}, {Graef}, {Granata}, {Grant}, {Gras}, {Grassia}, {Gray}, {Gray}, {Greco}, {Green}, {Green}, {Gretarsson}, {Groot}, {Grote}, {Grunewald}, {Gruning}, {Guidi}, {Gulati}, {Guo}, {Gupta}, {Gupta}, {Gustafson}, {Gustafson}, {Haegel}, {Hagiwara}, {Haino}, {Halim}, {Hall}, {Hall}, {Hamilton}, {Hammond}, {Haney}, {Hanke}, {Hanks}, {Hanna}, {Hannam}, {Hannuksela}, {Hanson}, {Hardwick}, {Haris}, {Harms}, {Harry}, {Harry},
  {Hasegawa}, {Haster}, {Haughian}, {Hayakawa}, {Hayama}, {Hayes}, {Healy}, {Heidmann}, {Heintze}, {Heitmann}, {Hello}, {Hemming}, {Hendry}, {Heng}, {Hennig}, {Heptonstall}, {Heurs}, {Hild}, {Himemoto}, {Hinderer}, {Hiranuma}, {Hirata}, {Hirose}, {Hoak}, {Hochheim}, {Hofman}, {Holgado}, {Holland}, {Holt}, {Holz}, {Hong}, {Hopkins}, {Horst}, {Hough}, {Howell}, {Hoy}, {Hreibi}, {Hsieh}, {Huang}, {Huang}, {Huang}, {Huerta}, {Huet}, {Hughey}, {Hulko}, {Husa}, {Huttner}, {Huynh-Dinh}, {Idzkowski}, {Iess}, {Ikenoue}, {Imam}, {Inayoshi}, {Ingram}, {Inoue}, {Inta}, {Intini}, {Ioka}, {Irwin}, {Isa}, {Isac}, {Isi}, {Itoh}, {Iyer}, {Izumi}, {Jacqmin}, {Jadhav}, {Jani}, {Janthalur}, {Jaranowski}, {Jenkins}, {Jiang}, {Johnson}, {Jones}, {Jones}, {Jones}, {Jonker}, {Ju}, {Jung}, {Jung}, {Junker}, {Kajita}, {Kalaghatgi}, {Kalogera}, {Kamai}, {Kamiizumi}, {Kanda}, {Kandhasamy}, {Kang}, {Kanner}, {Kapadia}, {Karki}, {Karvinen}, {Kashyap}, {Kasprzack}, {Katsanevas}, {Katsavounidis}, {Katzman}, {Kaufer}, {Kawabe}, {Kawaguchi},
  {Kawai}, {Kawasaki}, {Keerthana}, {K{\'e}f{\'e}lian}, {Keitel}, {Kennedy}, {Key}, {Khalili}, {Khan}, {Khan}, {Khan}, {Khan}, {Khazanov}, {Khursheed}, {Kijbunchoo}, {Kim}, {Kim}, {Kim}, {Kim}, {Kim}, {Kim}, {Kim}, {Kim}, {Kimball}, {Kimura}, {King}, {King}, {Kinley-Hanlon}, {Kirchhoff}, {Kissel}, {Kita}, {Kitazawa}, {Kleybolte}, {Klika}, {Klimenko}, {Knowles}, {Knyazev}, {Koch}, {Koehlenbeck}, {Koekoek}, {Kojima}, {Kokeyama}, {Koley}, {Komori}, {Kondrashov}, {Kong}, {Kontos}, {Koper}, {Korobko}, {Korth}, {Kotake}, {Kowalska}, {Kozak}, {Kozakai}, {Kozu}, {Kringel}, {Krishnendu}, {Kr{\'o}lak}, {Kuehn}, {Kumar}, {Kumar}, {Kumar}, {Kumar}, {Kumar}, {Kume}, {Kuo}, {Kuo}, {Kuo}, {Kuroyanagi}, {Kusayanagi}, {Kutynia}, {Kwak}, {Kwang}, {Lackey}, {Lai}, {Lam}, {Landry}, {Lane}, {Lang}, {Lange}, {Lantz}, {Lanza}, {Lartaux-Vollard}, {Lasky}, {Laxen}, {Lazzarini}, {Lazzaro}, {Leaci}, {Leavey}, {Lecoeuche}, {Lee}, {Lee}, {Lee}, {Lee}, {Lee}, {Lee}, {Lee}, {Lehmann}, {Lenon}, {Leonardi}, {Leroy}, {Letendre}, {Levin},
  {Li}, {Li}, {Li}, {Li}, {Lin}, {Lin}, {Lin}, {Lin}, {Linde}, {Linker}, {Littenberg}, {Liu}, {Liu}, {Liu}, {Lo}, {Lockerbie}, {London}, {Longo}, {Lorenzini}, {Loriette}, {Lormand}, {Losurdo}, {Lough}, {Lousto}, {Lovelace}, {Lower}, {L{\"u}ck}, {Lumaca}, {Lundgren}, {Luo}, {Lynch}, {Ma}, {Macas}, {Macfoy}, {Macinnis}, {MacLeod}, {Macquet}, {Maga{\~n}a-Sandoval}, {Zertuche}, {Magee}, {Majorana}, {Maksimovic}, {Malik}, {Man}, {Mandic}, {Mangano}, {Mansell}, {Manske}, {Mantovani}, {Marchesoni}, {Marchio}, {Marion}, {M{\'a}rka}, {M{\'a}rka}, {Markakis}, {Markosyan}, {Markowitz}, {Maros}, {Marquina}, {Marsat}, {Martelli}, {Martin}, {Martin}, {Martynov}, {Mason}, {Massera}, {Masserot}, {Massinger}, {Masso-Reid}, {Mastrogiovanni}, {Matas}, {Matichard}, {Matone}, {Mavalvala}, {Mazumder}, {McCann}, {McCarthy}, {McClelland}, {McCormick}, {McCuller}, {McGuire}, {McIver}, {McManus}, {McRae}, {McWilliams}, {Meacher}, {Meadors}, {Mehmet}, {Mehta}, {Meidam}, {Melatos}, {Mendell}, {Mercer}, {Mereni}, {Merilh}, {Merzougui},
  {Meshkov}, {Messenger}, {Messick}, {Metzdorff}, {Meyers}, {Miao}, {Michel}, {Michimura}, {Middleton}, {Mikhailov}, {Milano}, {Miller}, {Miller}, {Millhouse}, {Mills}, {Milovich-Goff}, {Minazzoli}, {Minenkov}, {Mio}, {Mishkin}, {Mishra}, {Mistry}, {Mitra}, {Mitrofanov}, {Mitselmakher}, {Mittleman}, {Miyakawa}, {Miyamoto}, {Miyazaki}, {Miyo}, {Miyoki}, {Mo}, {Moffa}, {Mogushi}, {Mohapatra}, {Montani}, {Moore}, {Moraru}, {Moreno}, {Morisaki}, {Moriwaki}, {Mours}, {Mow-Lowry}, {Mukherjee}, {Mukherjee}, {Mukherjee}, {Mukund}, {Mullavey}, {Munch}, {Mu{\~n}iz}, {Muratore}, {Murray}, {Nagano}, {Nagano}, {Nagar}, {Nakamura}, {Nakano}, {Nakano}, {Nakashima}, {Nardecchia}, {Narikawa}, {Naticchioni}, {Nayak}, {Negishi}, {Neilson}, {Nelemans}, {Nelson}, {Nery}, {Neunzert}, {Ng}, {Ng}, {Nguyen}, {Ni}, {Nichols}, {Nishizawa}, {Nissanke}, {Nocera}, {North}, {Nuttall}, {Obergaulinger}, {Oberling}, {O'Brien}, {Obuchi}, {O'Dea}, {Ogaki}, {Ogin}, {Oh}, {Oh}, {Ohashi}, {Ohishi}, {Ohkawa}, {Ohme}, {Ohta}, {Okada}, {Okutomi},
  {Oliver}, {Oohara}, {Ooi}, {Oppermann}, {Oram}, {O'Reilly}, {Ormiston}, {Ortega}, {O'Shaughnessy}, {Oshino}, {Ossokine}, {Ottaway}, {Overmier}, {Owen}, {Pace}, {Pagano}, {Page}, {Pai}, {Pai}, {Palamos}, {Palashov}, {Palomba}, {Pal-Singh}, {Pan}, {Pan}, {Pang}, {Pang}, {Pang}, {Pankow}, {Pannarale}, {Pant}, {Paoletti}, {Paoli}, {Papa}, {Parida}, {Park}, {Parker}, {Pascucci}, {Pasqualetti}, {Passaquieti}, {Passuello}, {Patil}, {Patricelli}, {Pearlstone}, {Pedersen}, {Pedraza}, {Pedurand}, {Pele}, {Arellano}, {Penn}, {Perez}, {Perreca}, {Pfeiffer}, {Phelps}, {Phukon}, {Piccinni}, {Pichot}, {Piergiovanni}, {Pillant}, {Pinard}, {Pinto}, {Pirello}, {Pitkin}, {Poggiani}, {Pong}, {Ponrathnam}, {Popolizio}, {Porter}, {Powell}, {Prajapati}, {Prasad}, {Prasai}, {Prasanna}, {Pratten}, {Prestegard}, {Privitera}, {Prodi}, {Prokhorov}, {Puncken}, {Punturo}, {Puppo}, {P{\"u}rrer}, {Qi}, {Quetschke}, {Quinonez}, {Quintero}, {Quitzow-James}, {Raab}, {Radkins}, {Radulescu}, {Raffai}, {Raja}, {Rajan}, {Rajbhandari},
  {Rakhmanov}, {Ramirez}, {Ramos-Buades}, {Rana}, {Rao}, {Rapagnani}, {Raymond}, {Razzano}, {Read}, {Regimbau}, {Rei}, {Reid}, {Reitze}, {Ren}, {Ricci}, {Richardson}, {Richardson}, {Ricker}, {Riles}, {Rizzo}, {Robertson}, {Robie}, {Robinet}, {Rocchi}, {Rolland}, {Rollins}, {Roma}, {Romanelli}, {Romano}, {Romel}, {Romie}, {Rose}, {Rosi{\'n}ska}, {Rosofsky}, {Ross}, {Rowan}, {R{\"u}diger}, {Ruggi}, {Rutins}, {Ryan}, {Sachdev}, {Sadecki}, {Sago}, {Saito}, {Saito}, {Sakai}, {Sakai}, {Sakamoto}, {Sakellariadou}, {Sakuno}, {Salconi}, {Saleem}, {Samajdar}, {Sammut}, {Sanchez}, {Sanchez}, {Sanchis-Gual}, {Sandberg}, {Sanders}, {Santiago}, {Sarin}, {Sassolas}, {Sathyaprakash}, {Sato}, {Sato}, {Sauter}, {Savage}, {Sawada}, {Schale}, {Scheel}, {Scheuer}, {Schmidt}, {Schnabel}, {Schofield}, {Sch{\"o}nbeck}, {Schreiber}, {Schulte}, {Schutz}, {Schwalbe}, {Scott}, {Scott}, {Seidel}, {Sekiguchi}, {Sekiguchi}, {Sellers}, {Sengupta}, {Sennett}, {Sentenac}, {Sequino}, {Sergeev}, {Setyawati}, {Shaddock}, {Shaffer}, {Shahriar},
  {Shaner}, {Shao}, {Sharma}, {Shawhan}, {Shen}, {Shibagaki}, {Shimizu}, {Shimoda}, {Shimode}, {Shink}, {Shinkai}, {Shishido}, {Shoda}, {Shoemaker}, {Shoemaker}, {Shyamsundar}, {Siellez}, {Sieniawska}, {Sigg}, {Silva}, {Singer}, {Singh}, {Singhal}, {Sintes}, {Sitmukhambetov}, {Skliris}, {Slagmolen}, {Slaven-Blair}, {Smith}, {Smith}, {Somala}, {Somiya}, {Son}, {Sorazu}, {Sorrentino}, {Sotani}, {Souradeep}, {Sowell}, {Spencer}, {Srivastava}, {Srivastava}, {Staats}, {Stachie}, {Standke}, {Steer}, {Steinke}, {Steinlechner}, {Steinlechner}, {Steinmeyer}, {Stevenson}, {Stocks}, {Stone}, {Stops}, {Strain}, {Stratta}, {Strigin}, {Strunk}, {Sturani}, {Stuver}, {Sudhir}, {Sugimoto}, {Summerscales}, {Sun}, {Sunil}, {Suresh}, {Sutton}, {Suzuki}, {Suzuki}, {Swinkels}, {Szczepa{\'n}czyk}, {Tacca}, {Tagoshi}, {Tait}, {Takahashi}, {Takahashi}, {Takamori}, {Takano}, {Takeda}, {Takeda}, {Talbot}, {Talukder}, {Tanaka}, {Tanaka}, {Tanaka}, {Tanaka}, {Tanaka}, {Tanioka}, {Tanner}, {T{\'a}pai}, {Tapia San Martin}, {Taracchini},
  {Tasson}, {Taylor}, {Telada}, {Thies}, {Thomas}, {Thomas}, {Thondapu}, {Thorne}, {Thrane}, {Tiwari}, {Tiwari}, {Tiwari}, {Toland}, {Tomaru}, {Tomigami}, {Tomura}, {Tonelli}, {Tornasi}, {Torres-Forn{\'e}}, {Torrie}, {T{\"o}yr{\"a}}, {Travasso}, {Traylor}, {Tringali}, {Trovato}, {Trozzo}, {Trudeau}, {Tsang}, {Tsang}, {Tse}, {Tso}, {Tsubono}, {Tsuchida}, {Tsukada}, {Tsuna}, {Tsuzuki}, {Tuyenbayev}, {Uchikata}, {Uchiyama}, {Ueda}, {Uehara}, {Ueno}, {Ueshima}, {Ugolini}, {Unnikrishnan}, {Uraguchi}, {Urban}, {Ushiba}, {Usman}, {Vahlbruch}, {Vajente}, {Valdes}, {van Bakel}, {van Beuzekom}, {van den Brand}, {van den Broeck}, {Vander-Hyde}, {van der Schaaf}, {van Heijningen}, {van Putten}, {van Veggel}, {Vardaro}, {Varma}, {Vass}, {Vas{\'u}th}, {Vecchio}, {Vedovato}, {Veitch}, {Veitch}, {Venkateswara}, {Venugopalan}, {Verkindt}, {Vetrano}, {Vicer{\'e}}, {Viets}, {Vine}, {Vinet}, {Vitale}, {Vivanco}, {Vo}, {Vocca}, {Vorvick}, {Vyatchanin}, {Wade}, {Wade}, {Wade}, {Walet}, {Walker}, {Wallace}, {Walsh}, {Wang}, {Wang},
  {Wang}, {Wang}, {Wang}, {Wang}, {Ward}, {Warden}, {Warner}, {Was}, {Watchi}, {Weaver}, {Wei}, {Weinert}, {Weinstein}, {Weiss}, {Wellmann}, {Wen}, {Wessel}, {We{\ss}els}, {Westhouse}, {Wette}, {Whelan}, {Whiting}, {Whittle}, {Wilken}, {Williams}, {Williamson}, {Willis}, {Willke}, {Wimmer}, {Winkler}, {Wipf}, {Wittel}, {Woan}, {Woehler}, {Wofford}, {Worden}, {Wright}, {Wu}, {Wu}, {Wu}, {Wu}, {Wysocki}, {Xiao}, {Xu}, {Yamada}, {Yamamoto}, {Yamamoto}, {Yamamoto}, {Yamamoto}, {Yancey}, {Yang}, {Yap}, {Yazback}, {Yeeles}, {Yokogawa}, {Yokoyama}, {Yokozawa}, {Yoshioka}, {Yu}, {Yu}, {Yuen}, {Yuzurihara}, {Yvert}, {Zadro{\.z}ny}, {Zanolin}, {Zeidler}, {Zelenova}, {Zendri}, {Zevin}, {Zhang}, {Zhang}, {Zhang}, {Zhao}, {Zhao}, {Zhou}, {Zhou}, {Zhu}, {Zhu}, {Zimmerman}, {Zucker}, {Zweizig}, {Kagra Collaboration}, \& {VIRGO Collaboration}}]{2020LRR....23....3A}
---. 2020, Living Reviews in Relativity, 23, 3, \dodoi{10.1007/s41114-020-00026-9}

\bibitem[{{Abbott} {et~al.}(2021){Abbott}, {Abbott}, {Abraham}, {Acernese}, {Ackley}, {Adams}, {Adams}, {Adhikari}, {Adya}, {Affeldt}, {Agathos}, {Agatsuma}, {Aggarwal}, {Aguiar}, {Aiello}, {Ain}, {Ajith}, {Allen}, {Allocca}, {Altin}, {Amato}, {Anand}, {Ananyeva}, {Anderson}, {Anderson}, {Angelova}, {Ansoldi}, {Antelis}, {Antier}, {Appert}, {Arai}, {Araya}, {Areeda}, {Ar{\`e}ne}, {Arnaud}, {Aronson}, {Arun}, {Asali}, {Ascenzi}, {Ashton}, {Aston}, {Astone}, {Aubin}, {Aufmuth}, {AultONeal}, {Austin}, {Avendano}, {Babak}, {Badaracco}, {Bader}, {Bae}, {Baer}, {Bagnasco}, {Baird}, {Ball}, {Ballardin}, {Ballmer}, {Bals}, {Balsamo}, {Baltus}, {Banagiri}, {Bankar}, {Bankar}, {Barayoga}, {Barbieri}, {Barish}, {Barker}, {Barneo}, {Barnum}, {Barone}, {Barr}, {Barsotti}, {Barsuglia}, {Barta}, {Bartlett}, {Bartos}, {Bassiri}, {Basti}, {Bawaj}, {Bayley}, {Bazzan}, {Becher}, {B{\'e}csy}, {Bedakihale}, {Bejger}, {Belahcene}, {Beniwal}, {Benjamin}, {Bennett}, {Bentley}, {Bergamin}, {Berger}, {Bergmann}, {Bernuzzi}, {Berry},
  {Bersanetti}, {Bertolini}, {Betzwieser}, {Bhandare}, {Bhandari}, {Bhattacharjee}, {Bidler}, {Bilenko}, {Billingsley}, {Birney}, {Birnholtz}, {Biscans}, {Bischi}, {Biscoveanu}, {Bisht}, {Bitossi}, {Bizouard}, {Blackburn}, {Blackman}, {Blair}, {Blair}, {Blair}, {Blanch}, {Bobba}, {Bode}, {Boer}, {Boetzel}, {Bogaert}, {Boldrini}, {Bondu}, {Bonilla}, {Bonnand}, {Booker}, {Boom}, {Bork}, {Boschi}, {Bose}, {Bossilkov}, {Boudart}, {Bouffanais}, {Bozzi}, {Bradaschia}, {Brady}, {Bramley}, {Branchesi}, {Brau}, {Breschi}, {Briant}, {Briggs}, {Brighenti}, {Brillet}, {Brinkmann}, {Brockill}, {Brooks}, {Brooks}, {Brown}, {Brunett}, {Bruno}, {Bruntz}, {Buikema}, {Bulik}, {Bulten}, {Buonanno}, {Buscicchio}, {Buskulic}, {Byer}, {Cabero}, {Cadonati}, {Caesar}, {Cagnoli}, {Cahillane}, {Calder{\'o}n Bustillo}, {Callaghan}, {Callister}, {Calloni}, {Camp}, {Canepa}, {Cannon}, {Cao}, {Cao}, {Carapella}, {Carbognani}, {Carney}, {Carpinelli}, {Carullo}, {Carver}, {Casanueva Diaz}, {Casentini}, {Caudill}, {Cavagli{\`a}}, {Cavalier},
  {Cavalieri}, {Cella}, {Cerd{\'a}-Dur{\'a}n}, {Cesarini}, {Chaibi}, {Chakravarti}, {Chan}, {Chan}, {Chandra}, {Chanial}, {Chao}, {Charlton}, {Chase}, {Chassande-Mottin}, {Chatterjee}, {Chattopadhyay}, {Chaturvedi}, {Chatziioannou}, {Chen}, {Chen}, {Chen}, {Chen}, {Cheng}, {Cheong}, {Chia}, {Chiadini}, {Chierici}, {Chincarini}, {Chiummo}, {Cho}, {Cho}, {Cho}, {Choate}, {Christensen}, {Chu}, {Chua}, {Chung}, {Chung}, {Ciani}, {Ciecielag}, {Cie{\'s}lar}, {Cifaldi}, {Ciobanu}, {Ciolfi}, {Cipriano}, {Cirone}, {Clara}, {Clark}, {Clark}, {Clarke}, {Clearwater}, {Clesse}, {Cleva}, {Coccia}, {Cohadon}, {Cohen}, {Colleoni}, {Collette}, {Collins}, {Colpi}, {Constancio}, {Conti}, {Cooper}, {Corban}, {Corbitt}, {Cordero-Carri{\'o}n}, {Corezzi}, {Corley}, {Cornish}, {Corre}, {Corsi}, {Cortese}, {Costa}, {Cotesta}, {Coughlin}, {Coughlin}, {Coulon}, {Countryman}, {Couvares}, {Covas}, {Coward}, {Cowart}, {Coyne}, {Coyne}, {Creighton}, {Creighton}, {Croquette}, {Crowder}, {Cudell}, {Cullen}, {Cumming}, {Cummings},
  {Cunningham}, {Cuoco}, {Curylo}, {Dal Canton}, {D{\'a}lya}, {Dana}, {DaneshgaranBajastani}, {D'Angelo}, {Danilishin}, {D'Antonio}, {Danzmann}, {Darsow-Fromm}, {Dasgupta}, {Datrier}, {Dattilo}, {Dave}, {Davier}, {Davies}, {Davis}, {Daw}, {Dean}, {DeBra}, {Deenadayalan}, {Degallaix}, {De Laurentis}, {Del{\'e}glise}, {Del Favero}, {De Lillo}, {De Lillo}, {Del Pozzo}, {DeMarchi}, {De Matteis}, {D'Emilio}, {Demos}, {Denker}, {Dent}, {Depasse}, {De Pietri}, {De Rosa}, {De Rossi}, {DeSalvo}, {de Varona}, {Dhurandhar}, {D{\'\i}az}, {Diaz-Ortiz}, {Didio}, {Dietrich}, {Di Fiore}, {DiFronzo}, {Di Giorgio}, {Di Giovanni}, {Di Giovanni}, {Di Girolamo}, {Di Lieto}, {Ding}, {Di Pace}, {Di Palma}, {Di Renzo}, {Divakarla}, {Dmitriev}, {Doctor}, {D'Onofrio}, {Donovan}, {Dooley}, {Doravari}, {Dorrington}, {Downes}, {Drago}, {Driggers}, {Du}, {Ducoin}, {Dupej}, {Durante}, {D'Urso}, {Duverne}, {Dwyer}, {Easter}, {Eddolls}, {Edelman}, {Edo}, {Edy}, {Effler}, {Eichholz}, {Eikenberry}, {Eisenmann}, {Eisenstein}, {Ejlli}, {Errico},
  {Essick}, {Estell{\'e}s}, {Estevez}, {Etienne}, {Etzel}, {Evans}, {Evans}, {Ewing}, {Fafone}, {Fair}, {Fairhurst}, {Fan}, {Farah}, {Farinon}, {Farr}, {Farr}, {Fauchon-Jones}, {Favata}, {Fays}, {Fazio}, {Feicht}, {Fejer}, {Feng}, {Fenyvesi}, {Ferguson}, {Fernandez-Galiana}, {Ferrante}, {Ferreira}, {Fidecaro}, {Figura}, {Fiori}, {Fiorucci}, {Fishbach}, {Fisher}, {Fishner}, {Fittipaldi}, {Fitz-Axen}, {Fiumara}, {Flaminio}, {Floden}, {Flynn}, {Fong}, {Font}, {Forsyth}, {Fournier}, {Frasca}, {Frasconi}, {Frei}, {Freise}, {Frey}, {Frey}, {Fritschel}, {Frolov}, {Fronz{\'e}}, {Fulda}, {Fyffe}, {Gabbard}, {Gadre}, {Gaebel}, {Gair}, {Gais}, {Galaudage}, {Gamba}, {Ganapathy}, {Ganguly}, {Gaonkar}, {Garaventa}, {Garc{\'\i}a-Quir{\'o}s}, {Garufi}, {Gateley}, {Gaudio}, {Gayathri}, {Gemme}, {Gennai}, {George}, {George}, {Gergely}, {Ghonge}, {Ghosh}, {Ghosh}, {Ghosh}, {Giacomazzo}, {Giacoppo}, {Giaime}, {Giardina}, {Gibson}, {Gier}, {Gill}, {Giri}, {Glanzer}, {Gleckl}, {Godwin}, {Goetz}, {Goetz}, {Gohlke}, {Goncharov},
  {Gonz{\'a}lez}, {Gopakumar}, {Gossan}, {Gosselin}, {Gouaty}, {Grace}, {Grado}, {Granata}, {Granata}, {Grant}, {Gras}, {Grassia}, {Gray}, {Gray}, {Greco}, {Green}, {Green}, {Gretarsson}, {Griggs}, {Grignani}, {Grimaldi}, {Grimes}, {Grimm}, {Grote}, {Grunewald}, {Gruning}, {Guerrero}, {Guidi}, {Guimaraes}, {Guix{\'e}}, {Gulati}, {Guo}, {Gupta}, {Gupta}, {Gupta}, {Gustafson}, {Gustafson}, {Guzman}, {Haegel}, {Halim}, {Hall}, {Hamilton}, {Hammond}, {Haney}, {Hanke}, {Hanks}, {Hanna}, {Hannuksela}, {Hannuksela}, {Hansen}, {Hansen}, {Hanson}, {Harder}, {Hardwick}, {Haris}, {Harms}, {Harry}, {Harry}, {Hartwig}, {Hasskew}, {Haster}, {Haughian}, {Hayes}, {Healy}, {Heidmann}, {Heintze}, {Heinze}, {Heinzel}, {Heitmann}, {Hellman}, {Hello}, {Helmling-Cornell}, {Hemming}, {Hendry}, {Heng}, {Hennes}, {Hennig}, {Hennig}, {Hernandez Vivanco}, {Heurs}, {Hild}, {Hill}, {Hines}, {Hochheim}, {Hofgard}, {Hofman}, {Hohmann}, {Holgado}, {Holland}, {Hollows}, {Holmes}, {Holt}, {Holz}, {Hopkins}, {Horst}, {Hough}, {Howell}, {Hoy},
  {Hoyland}, {Huang}, {H{\"u}bner}, {Huddart}, {Huerta}, {Hughey}, {Hui}, {Husa}, {Huttner}, {Hutzler}, {Huxford}, {Huynh-Dinh}, {Idzkowski}, {Iess}, {Imperato}, {Inchauspe}, {Ingram}, {Intini}, {Isi}, {Iyer}, {JaberianHamedan}, {Jacqmin}, {Jadhav}, {Jadhav}, {James}, {Jani}, {Janssens}, {Janthalur}, {Jaranowski}, {Jariwala}, {Jaume}, {Jenkins}, {Jeunon}, {Jiang}, {Johns}, {Jones}, {Jones}, {Jones}, {Jones}, {Jones}, {Jonker}, {Ju}, {Junker}, {Kalaghatgi}, {Kalogera}, {Kamai}, {Kandhasamy}, {Kang}, {Kanner}, {Kapadia}, {Kapasi}, {Karathanasis}, {Karki}, {Kashyap}, {Kasprzack}, {Kastaun}, {Katsanevas}, {Katsavounidis}, {Katzman}, {Kawabe}, {K{\'e}f{\'e}lian}, {Keitel}, {Key}, {Khadka}, {Khalili}, {Khan}, {Khan}, {Khazanov}, {Khetan}, {Khursheed}, {Kijbunchoo}, {Kim}, {Kim}, {Kim}, {Kim}, {Kim}, {Kim}, {Kimball}, {King}, {Kinley-Hanlon}, {Kirchhoff}, {Kissel}, {Kleybolte}, {Klimenko}, {Knowles}, {Knyazev}, {Koch}, {Koehlenbeck}, {Koekoek}, {Koley}, {Kolstein}, {Komori}, {Kondrashov}, {Kontos}, {Koper},
  {Korobko}, {Korth}, {Kovalam}, {Kozak}, {Kr{\"a}mer}, {Kringel}, {Krishnendu}, {Kr{\'o}lak}, {Kuehn}, {Kumar}, {Kumar}, {Kumar}, {Kumar}, {Kuns}, {Kwang}, {Lackey}, {Laghi}, {Lalande}, {Lam}, {Lamberts}, {Landry}, {Lane}, {Lang}, {Lange}, {Lantz}, {Lanza}, {La Rosa}, {Lartaux-Vollard}, {Lasky}, {Laxen}, {Lazzarini}, {Lazzaro}, {Leaci}, {Leavey}, {Lecoeuche}, {Lee}, {Lee}, {Lee}, {Lee}, {Lehmann}, {Leon}, {Leroy}, {Letendre}, {Levin}, {Li}, {Li}, {Li}, {Li}, {Li}, {Linde}, {Linker}, {Linley}, {Littenberg}, {Liu}, {Liu}, {Llorens-Monteagudo}, {Lo}, {Lockwood}, {London}, {Longo}, {Lorenzini}, {Loriette}, {Lormand}, {Losurdo}, {Lough}, {Lousto}, {Lovelace}, {L{\"u}ck}, {Lumaca}, {Lundgren}, {Ma}, {Macas}, {MacInnis}, {Macleod}, {MacMillan}, {Macquet}, {Maga{\~n}a Hernandez}, {Maga{\~n}a-Sandoval}, {Magazz{\`u}}, {Magee}, {Majorana}, {Maksimovic}, {Maliakal}, {Malik}, {Man}, {Mandic}, {Mangano}, {Mansell}, {Manske}, {Mantovani}, {Mapelli}, {Marchesoni}, {Marion}, {M{\'a}rka}, {M{\'a}rka}, {Markakis},
  {Markosyan}, {Markowitz}, {Maros}, {Marquina}, {Marsat}, {Martelli}, {Martin}, {Martin}, {Martinez}, {Martinez}, {Martynov}, {Masalehdan}, {Mason}, {Massera}, {Masserot}, {Massinger}, {Masso-Reid}, {Mastrogiovanni}, {Matas}, {Mateu-Lucena}, {Matichard}, {Matiushechkina}, {Mavalvala}, {Maynard}, {McCann}, {McCarthy}, {McClelland}, {McCormick}, {McCuller}, {McGuire}, {McIsaac}, {McIver}, {McManus}, {McRae}, {McWilliams}, {Meacher}, {Meadors}, {Mehmet}, {Mehta}, {Melatos}, {Melchor}, {Mendell}, {Menendez-Vazquez}, {Mercer}, {Mereni}, {Merfeld}, {Merilh}, {Merritt}, {Merzougui}, {Meshkov}, {Messenger}, {Messick}, {Metzdorff}, {Meyers}, {Meylahn}, {Mhaske}, {Miani}, {Miao}, {Michaloliakos}, {Michel}, {Middleton}, {Milano}, {Miller}, {Miller}, {Millhouse}, {Mills}, {Milotti}, {Milovich-Goff}, {Minazzoli}, {Minenkov}, {Mir}, {Mishkin}, {Mishra}, {Mistry}, {Mitra}, {Mitrofanov}, {Mitselmakher}, {Mittleman}, {Mo}, {Mogushi}, {Mohapatra}, {Mohite}, {Molina}, {Molina-Ruiz}, {Mondin}, {Montani}, {Moore}, {Moraru},
  {Morawski}, {Moreno}, {Morisaki}, {Mours}, {Mow-Lowry}, {Mozzon}, {Muciaccia}, {Mukherjee}, {Mukherjee}, {Mukherjee}, {Mukherjee}, {Mukund}, {Mullavey}, {Munch}, {Mu{\~n}iz}, {Murray}, {Nadji}, {Nagar}, {Nardecchia}, {Naticchioni}, {Nayak}, {Neil}, {Neilson}, {Nelemans}, {Nelson}, {Nery}, {Neunzert}, {Ng}, {Ng}, {Nguyen}, {Nguyen}, {Nguyen}, {Nichols}, {Nissanke}, {Nocera}, {Noh}, {North}, {Nothard}, {Nuttall}, {Oberling}, {O'Brien}, {O'Dell}, {Oganesyan}, {Ogin}, {Oh}, {Oh}, {Ohme}, {Ohta}, {Okada}, {Olivetto}, {Oppermann}, {Oram}, {O'Reilly}, {Ormiston}, {Ormsby}, {Ortega}, {O'Shaughnessy}, {Ossokine}, {Osthelder}, {Ottaway}, {Overmier}, {Owen}, {Pace}, {Pagano}, {Page}, {Pagliaroli}, {Pai}, {Pai}, {Palamos}, {Palashov}, {Palomba}, {Pan}, {Panda}, {Pang}, {Pankow}, {Pannarale}, {Pant}, {Paoletti}, {Paoli}, {Paolone}, {Parker}, {Pascucci}, {Pasqualetti}, {Passaquieti}, {Passuello}, {Patel}, {Patricelli}, {Payne}, {Pechsiri}, {Pedraza}, {Pegoraro}, {Pele}, {Penn}, {Perego}, {Perez}, {P{\'e}rigois},
  {Perreca}, {Perri{\`e}s}, {Petermann}, {Petterson}, {Pfeiffer}, {Pham}, {Phukon}, {Piccinni}, {Pichot}, {Piendibene}, {Piergiovanni}, {Pierini}, {Pierro}, {Pillant}, {Pilo}, {Pinard}, {Pinto}, {Piotrzkowski}, {Pirello}, {Pitkin}, {Placidi}, {Plastino}, {Pluchar}, {Poggiani}, {Polini}, {Pong}, {Ponrathnam}, {Popolizio}, {Porter}, {Poverman}, {Powell}, {Pracchia}, {Prajapati}, {Prasai}, {Prasanna}, {Pratten}, {Prestegard}, {Principe}, {Prodi}, {Prokhorov}, {Prosposito}, {Puecher}, {Punturo}, {Puosi}, {Puppo}, {P{\"u}rrer}, {Qi}, {Quetschke}, {Quinonez}, {Quitzow-James}, {Raab}, {Raaijmakers}, {Radkins}, {Radulesco}, {Raffai}, {Rafferty}, {Rail}, {Raja}, {Rajan}, {Rajbhandari}, {Rakhmanov}, {Ramirez}, {Ramirez}, {Ramos-Buades}, {Rana}, {Rao}, {Rapagnani}, {Rapol}, {Ratto}, {Raymond}, {Razzano}, {Read}, {Regimbau}, {Rei}, {Reid}, {Reitze}, {Rettegno}, {Ricci}, {Richardson}, {Richardson}, {Richardson}, {Ricker}, {Riemenschneider}, {Riles}, {Rizzo}, {Robertson}, {Robinet}, {Rocchi}, {Rocha}, {Rodriguez},
  {Rodriguez-Soto}, {Rolland}, {Rollins}, {Roma}, {Romanelli}, {Romano}, {Romel}, {Romero}, {Romero-Shaw}, {Romie}, {Ronchini}, {Rose}, {Rose}, {Rose}, {Rosell}, {Rosi{\'n}ska}, {Rosofsky}, {Ross}, {Rowan}, {Rowlinson}, {Roy}, {Roy}, {Ruggi}, {Ryan}, {Sachdev}, {Sadecki}, {Sakellariadou}, {Salafia}, {Salconi}, {Saleem}, {Samajdar}, {Sanchez}, {Sanchez}, {Sanchez}, {Sanchis-Gual}, {Sanders}, {Santiago}, {Santos}, {Saravanan}, {Sarin}, {Sassolas}, {Sathyaprakash}, {Sauter}, {Savage}, {Savant}, {Sawant}, {Sayah}, {Schaetzl}, {Schale}, {Scheel}, {Scheuer}, {Schindler-Tyka}, {Schmidt}, {Schnabel}, {Schofield}, {Sch{\"o}nbeck}, {Schreiber}, {Schulte}, {Schutz}, {Schwarm}, {Schwartz}, {Scott}, {Scott}, {Seglar-Arroyo}, {Seidel}, {Sellers}, {Sengupta}, {Sennett}, {Sentenac}, {Sequino}, {Sergeev}, {Setyawati}, {Shaffer}, {Shahriar}, {Sharifi}, {Sharma}, {Sharma}, {Shawhan}, {Shen}, {Shikauchi}, {Shink}, {Shoemaker}, {Shoemaker}, {Shukla}, {ShyamSundar}, {Sieniawska}, {Sigg}, {Singer}, {Singh}, {Singh}, {Singha},
  {Singhal}, {Sintes}, {Sipala}, {Skliris}, {Slagmolen}, {Slaven-Blair}, {Smetana}, {Smith}, {Smith}, {Somala}, {Son}, {Soni}, {Sorazu}, {Sordini}, {Sorrentino}, {Sorrentino}, {Soulard}, {Souradeep}, {Sowell}, {Spencer}, {Spera}, {Srivastava}, {Srivastava}, {Staats}, {Stachie}, {Steer}, {Steinke}, {Steinlechner}, {Steinlechner}, {Steinmeyer}, {Stevenson}, {Stolle-McAllister}, {Stops}, {Stover}, {Strain}, {Stratta}, {Strunk}, {Sturani}, {Stuver}, {S{\"u}dbeck}, {Sudhagar}, {Sudhir}, {Suh}, {Summerscales}, {Sun}, {Sun}, {Sunil}, {Sur}, {Suresh}, {Sutton}, {Swinkels}, {Szczepa{\'n}czyk}, {Tacca}, {Tait}, {Talbot}, {Tanasijczuk}, {Tanner}, {Tao}, {Tapia}, {Tapia San Martin}, {Tasson}, {Taylor}, {Tenorio}, {Terkowski}, {Thirugnanasambandam}, {Thomas}, {Thomas}, {Thomas}, {Thompson}, {Thondapu}, {Thorne}, {Thrane}, {Tiwari}, {Tiwari}, {Tiwari}, {Toland}, {Tolley}, {Tonelli}, {Tornasi}, {Torres-Forn{\'e}}, {Torrie}, {Tosta e Melo}, {T{\"o}yr{\"a}}, {Tran}, {Trapananti}, {Travasso}, {Traylor}, {Tringali},
  {Tripathee}, {Trovato}, {Trudeau}, {Tsai}, {Tsang}, {Tse}, {Tso}, {Tsukada}, {Tsuna}, {Tsutsui}, {Turconi}, {Ubhi}, {Udall}, {Ueno}, {Ugolini}, {Unnikrishnan}, {Urban}, {Usman}, {Utina}, {Vahlbruch}, {Vajente}, {Vajpeyi}, {Valdes}, {Valentini}, {Valsan}, {van Bakel}, {Beuzekom}, {van den Brand}, {Van Den Broeck}, {Vander-Hyde}, {van der Schaaf}, {van Heijningen}, {Vardaro}, {Vargas}, {Varma}, {Vass}, {Vas{\'u}th}, {Vecchio}, {Vedovato}, {Veitch}, {Veitch}, {Venkateswara}, {Venneberg}, {Venugopalan}, {Verkindt}, {Verma}, {Veske}, {Vetrano}, {Vicer{\'e}}, {Viets}, {Villa-Ortega}, {Vinet}, {Vitale}, {Vo}, {Vocca}, {Vorvick}, {Vyatchanin}, {Wade}, {Wade}, {Wade}, {Walet}, {Walker}, {Wallace}, {Wallace}, {Walsh}, {Wang}, {Wang}, {Wang}, {Wang}, {Ward}, {Warner}, {Was}, {Washington}, {Watchi}, {Weaver}, {Wei}, {Weinert}, {Weinstein}, {Weiss}, {Wellmann}, {Wen}, {We{\ss}els}, {Westhouse}, {Wette}, {Whelan}, {White}, {White}, {Whiting}, {Whittle}, {Wilken}, {Williams}, {Williams}, {Williamson}, {Willis}, {Willke},
  {Wilson}, {Wimmer}, {Winkler}, {Wipf}, {Woan}, {Woehler}, {Wofford}, {Wong}, {Wrangel}, {Wright}, {Wu}, {Wysocki}, {Xiao}, {Yamamoto}, {Yang}, {Yang}, {Yang}, {Yap}, {Yeeles}, {Yoon}, {Yu}, {Yu}, {Yuen}, {Zadro{\.z}ny}, {Zanolin}, {Zelenova}, {Zendri}, {Zevin}, {Zhang}, {Zhang}, {Zhang}, {Zhang}, {Zhao}, {Zhao}, {Zhou}, {Zhou}, {Zhu}, {Zimmerman}, {Zucker}, {Zweizig}, {LIGO Scientific Collaboration}, \& {Virgo Collaboration}}]{2021ApJ...913L...7A}
{Abbott}, R., {Abbott}, T.~D., {Abraham}, S., {et~al.} 2021, \apjl, 913, L7, \dodoi{10.3847/2041-8213/abe949}

\bibitem[{{Abbott} {et~al.}(2023){Abbott}, {Abbott}, {Acernese}, {Ackley}, {Adams}, {Adhikari}, {Adhikari}, {Adya}, {Affeldt}, {Agarwal}, {Agathos}, {Agatsuma}, {Aggarwal}, {Aguiar}, {Aiello}, {Ain}, {Ajith}, {Akutsu}, {de Alarc{\'o}n}, {Akcay}, {Albanesi}, {Allocca}, {Altin}, {Amato}, {Anand}, {Anand}, {Ananyeva}, {Anderson}, {Anderson}, {Ando}, {Andrade}, {Andres}, {Andri{\'c}}, {Angelova}, {Ansoldi}, {Antelis}, {Antier}, {Antonini}, {Appert}, {Arai}, {Arai}, {Arai}, {Araki}, {Araya}, {Araya}, {Areeda}, {Ar{\`e}ne}, {Aritomi}, {Arnaud}, {Arogeti}, {Aronson}, {Arun}, {Asada}, {Asali}, {Ashton}, {Aso}, {Assiduo}, {Aston}, {Astone}, {Aubin}, {Austin}, {Babak}, {Badaracco}, {Bader}, {Badger}, {Bae}, {Bae}, {Baer}, {Bagnasco}, {Bai}, {Baiotti}, {Baird}, {Bajpai}, {Ball}, {Ballardin}, {Ballmer}, {Balsamo}, {Baltus}, {Banagiri}, {Bankar}, {Barayoga}, {Barbieri}, {Barish}, {Barker}, {Barneo}, {Barone}, {Barr}, {Barsotti}, {Barsuglia}, {Barta}, {Bartlett}, {Barton}, {Bartos}, {Bassiri}, {Basti}, {Bawaj}, {Bayley},
  {Baylor}, {Bazzan}, {B{\'e}csy}, {Bedakihale}, {Bejger}, {Belahcene}, {Benedetto}, {Beniwal}, {Bennett}, {Bentley}, {Benyaala}, {Bergamin}, {Berger}, {Bernuzzi}, {Berry}, {Bersanetti}, {Bertolini}, {Betzwieser}, {Beveridge}, {Bhandare}, {Bhardwaj}, {Bhattacharjee}, {Bhaumik}, {Bilenko}, {Billingsley}, {Bini}, {Birney}, {Birnholtz}, {Biscans}, {Bischi}, {Biscoveanu}, {Bisht}, {Biswas}, {Bitossi}, {Bizouard}, {Blackburn}, {Blair}, {Blair}, {Blair}, {Bobba}, {Bode}, {Boer}, {Bogaert}, {Boldrini}, {Bonavena}, {Bondu}, {Bonilla}, {Bonnand}, {Booker}, {Boom}, {Bork}, {Boschi}, {Bose}, {Bose}, {Bossilkov}, {Boudart}, {Bouffanais}, {Bozzi}, {Bradaschia}, {Brady}, {Bramley}, {Branch}, {Branchesi}, {Brandt}, {Brau}, {Breschi}, {Briant}, {Briggs}, {Brillet}, {Brinkmann}, {Brockill}, {Brooks}, {Brooks}, {Brown}, {Brunett}, {Bruno}, {Bruntz}, {Bryant}, {Bulik}, {Bulten}, {Buonanno}, {Buscicchio}, {Buskulic}, {Buy}, {Byer}, {Cadonati}, {Cagnoli}, {Cahillane}, {Bustillo}, {Callaghan}, {Callister}, {Calloni}, {Cameron},
  {Camp}, {Canepa}, {Canevarolo}, {Cannavacciuolo}, {Cannon}, {Cao}, {Cao}, {Capocasa}, {Capote}, {Carapella}, {Carbognani}, {Carlin}, {Carney}, {Carpinelli}, {Carrillo}, {Carullo}, {Carver}, {Diaz}, {Casentini}, {Castaldi}, {Caudill}, {Cavagli{\`a}}, {Cavalier}, {Cavalieri}, {Ceasar}, {Cella}, {Cerd{\'a}-Dur{\'a}n}, {Cesarini}, {Chaibi}, {Chakravarti}, {Subrahmanya}, {Champion}, {Chan}, {Chan}, {Chan}, {Chan}, {Chan}, {Chandra}, {Chanial}, {Chao}, {Chapman-Bird}, {Charlton}, {Chase}, {Chassande-Mottin}, {Chatterjee}, {Chatterjee}, {Chatterjee}, {Chaturvedi}, {Chaty}, {Chatziioannou}, {Chen}, {Chen}, {Chen}, {Chen}, {Chen}, {Chen}, {Chen}, {Chen}, {Cheng}, {Cheong}, {Cheung}, {Chia}, {Chiadini}, {Chiang}, {Chiarini}, {Chierici}, {Chincarini}, {Chiofalo}, {Chiummo}, {Cho}, {Cho}, {Choudhary}, {Choudhary}, {Christensen}, {Chu}, {Chu}, {Chu}, {Chua}, {Chung}, {Ciani}, {Ciecielag}, {Cie{\'s}lar}, {Cifaldi}, {Ciobanu}, {Ciolfi}, {Cipriano}, {Cirone}, {Clara}, {Clark}, {Clark}, {Clarke}, {Clearwater}, {Clesse},
  {Cleva}, {Coccia}, {Codazzo}, {Cohadon}, {Cohen}, {Cohen}, {Colleoni}, {Collette}, {Colombo}, {Colpi}, {Compton}, {Constancio}, {Conti}, {Cooper}, {Corban}, {Corbitt}, {Cordero-Carri{\'o}n}, {Corezzi}, {Corley}, {Cornish}, {Corre}, {Corsi}, {Cortese}, {Costa}, {Cotesta}, {Coughlin}, {Coulon}, {Countryman}, {Cousins}, {Couvares}, {Coward}, {Cowart}, {Coyne}, {Coyne}, {Creighton}, {Creighton}, {Criswell}, {Croquette}, {Crowder}, {Cudell}, {Cullen}, {Cumming}, {Cummings}, {Cunningham}, {Cuoco}, {Cury{\l}o}, {Dabadie}, {Canton}, {Dall'Osso}, {D{\'a}lya}, {Dana}, {Daneshgaranbajastani}, {D'Angelo}, {Danila}, {Danilishin}, {D'Antonio}, {Danzmann}, {Darsow-Fromm}, {Dasgupta}, {Datrier}, {Datta}, {Dattilo}, {Dave}, {Davier}, {Davies}, {Davis}, {Davis}, {Daw}, {Dean}, {Debra}, {Deenadayalan}, {Degallaix}, {de Laurentis}, {Del{\'e}glise}, {Del Favero}, {de Lillo}, {de Lillo}, {Del Pozzo}, {Demarchi}, {de Matteis}, {D'Emilio}, {Demos}, {Dent}, {Depasse}, {de Pietri}, {De Rosa}, {de Rossi}, {Desalvo}, {de Simone},
  {Dhurandhar}, {D{\'\i}az}, {Diaz-Ortiz}, {Didio}, {Dietrich}, {di Fiore}, {di Fronzo}, {di Giorgio}, {di Giovanni}, {di Giovanni}, {di Girolamo}, {di Lieto}, {Ding}, {di Pace}, {di Palma}, {di Renzo}, {Divakarla}, {Dmitriev}, {Doctor}, {D'Onofrio}, {Donovan}, {Dooley}, {Doravari}, {Dorrington}, {Drago}, {Driggers}, {Drori}, {Ducoin}, {Dupej}, {Durante}, {D'Urso}, {Duverne}, {Dwyer}, {Eassa}, {Easter}, {Ebersold}, {Eckhardt}, {Eddolls}, {Edelman}, {Edo}, {Edy}, {Effler}, {Eguchi}, {Eichholz}, {Eikenberry}, {Eisenmann}, {Eisenstein}, {Ejlli}, {Engelby}, {Enomoto}, {Errico}, {Essick}, {Estell{\'e}s}, {Estevez}, {Etienne}, {Etzel}, {Evans}, {Evans}, {Ewing}, {Fafone}, {Fair}, {Fairhurst}, {Farah}, {Farinon}, {Farr}, {Farr}, {Farrow}, {Fauchon-Jones}, {Favaro}, {Favata}, {Fays}, {Fazio}, {Feicht}, {Fejer}, {Fenyvesi}, {Ferguson}, {Fernandez-Galiana}, {Ferrante}, {Ferreira}, {Fidecaro}, {Figura}, {Fiori}, {Fishbach}, {Fisher}, {Fittipaldi}, {Fiumara}, {Flaminio}, {Floden}, {Fong}, {Font}, {Fornal}, {Forsyth},
  {Franke}, {Frasca}, {Frasconi}, {Frederick}, {Freed}, {Frei}, {Freise}, {Frey}, {Fritschel}, {Frolov}, {Fronz{\'e}}, {Fujii}, {Fujikawa}, {Fukunaga}, {Fukushima}, {Fulda}, {Fyffe}, {Gabbard}, {Gadre}, {Gair}, {Gais}, {Galaudage}, {Gamba}, {Ganapathy}, {Ganguly}, {Gao}, {Gaonkar}, {Garaventa}, {Garc{\'\i}a}, {Garc{\'\i}a-N{\'u}{\~n}ez}, {Garc{\'\i}a-Quir{\'o}s}, {Garufi}, {Gateley}, {Gaudio}, {Gayathri}, {Ge}, {Gemme}, {Gennai}, {George}, {George}, {Gerberding}, {Gergely}, {Gewecke}, {Ghonge}, {Ghosh}, {Ghosh}, {Ghosh}, {Ghosh}, {Giacomazzo}, {Giacoppo}, {Giaime}, {Giardina}, {Gibson}, {Gier}, {Giesler}, {Giri}, {Gissi}, {Glanzer}, {Gleckl}, {Godwin}, {Golomb}, {Goetz}, {Goetz}, {Gohlke}, {Goncharov}, {Gonz{\'a}lez}, {Gopakumar}, {Gosselin}, {Gouaty}, {Gould}, {Grace}, {Grado}, {Granata}, {Granata}, {Grant}, {Gras}, {Grassia}, {Gray}, {Gray}, {Greco}, {Green}, {Green}, {Gretarsson}, {Gretarsson}, {Griffith}, {Griffiths}, {Griggs}, {Grignani}, {Grimaldi}, {Grimm}, {Grote}, {Grunewald}, {Gruning}, {Guerra},
  {Guidi}, {Guimaraes}, {Guix{\'e}}, {Gulati}, {Guo}, {Guo}, {Gupta}, {Gupta}, {Gupta}, {Gustafson}, {Gustafson}, {Guzman}, {Ha}, {Haegel}, {Hagiwara}, {Haino}, {Halim}, {Hall}, {Hamilton}, {Hammond}, {Han}, {Haney}, {Hanks}, {Hanna}, {Hannam}, {Hannuksela}, {Hansen}, {Hansen}, {Hanson}, {Harder}, {Hardwick}, {Haris}, {Harms}, {Harry}, {Harry}, {Hartwig}, {Hasegawa}, {Haskell}, {Hasskew}, {Haster}, {Hattori}, {Haughian}, {Hayakawa}, {Hayama}, {Hayes}, {Healy}, {Heidmann}, {Heidt}, {Heintze}, {Heinze}, {Heinzel}, {Heitmann}, {Hellman}, {Hello}, {Helmling-Cornell}, {Hemming}, {Hendry}, {Heng}, {Hennes}, {Hennig}, {Hennig}, {Hernandez}, {Vivanco}, {Heurs}, {Hild}, {Hill}, {Himemoto}, {Hines}, {Hiranuma}, {Hirata}, {Hirose}, {Hochheim}, {Hofman}, {Hohmann}, {Holcomb}, {Holland}, {Hollows}, {Holmes}, {Holt}, {Holz}, {Hong}, {Hopkins}, {Hough}, {Hourihane}, {Howell}, {Hoy}, {Hoyland}, {Hreibi}, {Hsieh}, {Hsu}, {Huang}, {Huang}, {Huang}, {Huang}, {Huang}, {Huang}, {H{\"u}bner}, {Huddart}, {Hughey}, {Hui}, {Hui},
  {Husa}, {Huttner}, {Huxford}, {Huynh-Dinh}, {Ide}, {Idzkowski}, {Iess}, {Ikenoue}, {Imam}, {Inayoshi}, {Ingram}, {Inoue}, {Ioka}, {Isi}, {Isleif}, {Ito}, {Itoh}, {Iyer}, {Izumi}, {Jaberianhamedan}, {Jacqmin}, {Jadhav}, {Jadhav}, {James}, {Jan}, {Jani}, {Janquart}, {Janssens}, {Janthalur}, {Jaranowski}, {Jariwala}, {Jaume}, {Jenkins}, {Jenner}, {Jeon}, {Jeunon}, {Jia}, {Jin}, {Johns}, {Jones}, {Jones}, {Jones}, {Jones}, {Jones}, {Jonker}, {Ju}, {Jung}, {Jung}, {Junker}, {Juste}, {Kaihotsu}, {Kajita}, {Kakizaki}, {Kalaghatgi}, {Kalogera}, {Kamai}, {Kamiizumi}, {Kanda}, {Kandhasamy}, {Kang}, {Kanner}, {Kao}, {Kapadia}, {Kapasi}, {Karat}, {Karathanasis}, {Karki}, {Kashyap}, {Kasprzack}, {Kastaun}, {Katsanevas}, {Katsavounidis}, {Katzman}, {Kaur}, {Kawabe}, {Kawaguchi}, {Kawai}, {Kawasaki}, {K{\'e}f{\'e}lian}, {Keitel}, {Key}, {Khadka}, {Khalili}, {Khan}, {Khazanov}, {Khetan}, {Khursheed}, {Kijbunchoo}, {Kim}, {Kim}, {Kim}, {Kim}, {Kim}, {Kim}, {Kimball}, {Kimura}, {Kinley-Hanlon}, {Kirchhoff}, {Kissel}, {Kita},
  {Kitazawa}, {Kleybolte}, {Klimenko}, {Knee}, {Knowles}, {Knyazev}, {Koch}, {Koekoek}, {Kojima}, {Kokeyama}, {Koley}, {Kolitsidou}, {Kolstein}, {Komori}, {Kondrashov}, {Kong}, {Kontos}, {Koper}, {Korobko}, {Kotake}, {Kovalam}, {Kozak}, {Kozakai}, {Kozu}, {Kringel}, {Krishnendu}, {Kr{\'o}lak}, {Kuehn}, {Kuei}, {Kuijer}, {Kulkarni}, {Kumar}, {Kumar}, {Kumar}, {Kumar}, {Kume}, {Kuns}, {Kuo}, {Kuo}, {Kuromiya}, {Kuroyanagi}, {Kusayanagi}, {Kuwahara}, {Kwak}, {Lagabbe}, {Laghi}, {Lalande}, {Lam}, {Lamberts}, {Landry}, {Landry}, {Lane}, {Lang}, {Lange}, {Lantz}, {La Rosa}, {Lartaux-Vollard}, {Lasky}, {Laxen}, {Lazzarini}, {Lazzaro}, {Leaci}, {Leavey}, {Lecoeuche}, {Lee}, {Lee}, {Lee}, {Lee}, {Lee}, {Lee}, {Lehmann}, {Lema{\^\i}tre}, {Leonardi}, {Leroy}, {Letendre}, {Levesque}, {Levin}, {Leviton}, {Leyde}, {Li}, {Li}, {Li}, {Li}, {Li}, {Li}, {Lin}, {Lin}, {Lin}, {Lin}, {Lin}, {Linde}, {Linker}, {Linley}, {Littenberg}, {Liu}, {Liu}, {Liu}, {Liu}, {Llamas}, {Llorens-Monteagudo}, {Lo}, {Lockwood}, {Loh}, {London},
  {Longo}, {Lopez}, {Portilla}, {Lorenzini}, {Loriette}, {Lormand}, {Losurdo}, {Lott}, {Lough}, {Lousto}, {Lovelace}, {Lucaccioni}, {L{\"u}ck}, {Lumaca}, {Lundgren}, {Luo}, {Lynam}, {Macas}, {Macinnis}, {MacLeod}, {MacMillan}, {Macquet}, {Hernandez}, {Magazz{\`u}}, {Magee}, {Maggiore}, {Magnozzi}, {Mahesh}, {Majorana}, {Makarem}, {Maksimovic}, {Maliakal}, {Malik}, {Man}, {Mandic}, {Mangano}, {Mango}, {Mansell}, {Manske}, {Mantovani}, {Mapelli}, {Marchesoni}, {Marchio}, {Marion}, {Mark}, {M{\'a}rka}, {M{\'a}rka}, {Markakis}, {Markosyan}, {Markowitz}, {Maros}, {Marquina}, {Marsat}, {Martelli}, {Martin}, {Martin}, {Martinez}, {Martinez}, {Martinez}, {Martinovic}, {Martynov}, {Marx}, {Masalehdan}, {Mason}, {Massera}, {Masserot}, {Massinger}, {Masso-Reid}, {Mastrogiovanni}, {Matas}, {Mateu-Lucena}, {Matichard}, {Matiushechkina}, {Mavalvala}, {McCann}, {McCarthy}, {McClelland}, {McClincy}, {McCormick}, {McCuller}, {McGhee}, {McGuire}, {McIsaac}, {McIver}, {McRae}, {McWilliams}, {Meacher}, {Mehmet}, {Mehta},
  {Meijer}, {Melatos}, {Melchor}, {Mendell}, {Menendez-Vazquez}, {Menoni}, {Mercer}, {Mereni}, {Merfeld}, {Merilh}, {Merritt}, {Merzougui}, {Meshkov}, {Messenger}, {Messick}, {Meyers}, {Meylahn}, {Mhaske}, {Miani}, {Miao}, {Michaloliakos}, {Michel}, {Michimura}, {Middleton}, {Milano}, {Miller}, {Miller}, {Miller}, {Miller}, {Millhouse}, {Mills}, {Milotti}, {Minazzoli}, {Minenkov}, {Mio}, {Mir}, {Miravet-Ten{\'e}s}, {Mishra}, {Mishra}, {Mistry}, {Mitra}, {Mitrofanov}, {Mitselmakher}, {Mittleman}, {Miyakawa}, {Miyamoto}, {Miyazaki}, {Miyo}, {Miyoki}, {Mo}, {Modafferi}, {Moguel}, {Mogushi}, {Mohapatra}, {Mohite}, {Molina}, {Molina-Ruiz}, {Mondin}, {Montani}, {Moore}, {Moraru}, {Morawski}, {More}, {Moreno}, {Moreno}, {Mori}, {Morisaki}, {Moriwaki}, {Morr{\'a}s}, {Mours}, {Mow-Lowry}, {Mozzon}, {Muciaccia}, {Mukherjee}, {Mukherjee}, {Mukherjee}, {Mukherjee}, {Mukherjee}, {Mukund}, {Mullavey}, {Munch}, {Mu{\~n}iz}, {Murray}, {Musenich}, {Muusse}, {Nadji}, {Nagano}, {Nagano}, {Nagar}, {Nakamura}, {Nakano}, {Nakano},
  {Nakashima}, {Nakayama}, {Napolano}, {Nardecchia}, {Narikawa}, {Naticchioni}, {Nayak}, {Nayak}, {Negishi}, {Neil}, {Neilson}, {Nelemans}, {Nelson}, {Nery}, {Neubauer}, {Neunzert}, {Ng}, {Ng}, {Nguyen}, {Nguyen}, {Nguyen}, {Quynh}, {Ni}, {Nichols}, {Nishizawa}, {Nissanke}, {Nitoglia}, {Nocera}, {Norman}, {North}, {Nozaki}, {Siles}, {Nuttall}, {Oberling}, {O'Brien}, {Obuchi}, {O'Dell}, {Oelker}, {Ogaki}, {Oganesyan}, {Oh}, {Oh}, {Oh}, {Ohashi}, {Ohishi}, {Ohkawa}, {Ohme}, {Ohta}, {Okada}, {Okutani}, {Okutomi}, {Olivetto}, {Oohara}, {Ooi}, {Oram}, {O'Reilly}, {Ormiston}, {Ormsby}, {Ortega}, {O'Shaughnessy}, {O'Shea}, {Oshino}, {Ossokine}, {Osthelder}, {Otabe}, {Ottaway}, {Overmier}, {Pace}, {Pagano}, {Page}, {Pagliaroli}, {Pai}, {Pai}, {Palamos}, {Palashov}, {Palomba}, {Pan}, {Pan}, {Panda}, {Pang}, {Pang}, {Pankow}, {Pannarale}, {Pant}, {Panther}, {Paoletti}, {Paoli}, {Paolone}, {Parisi}, {Park}, {Park}, {Parker}, {Pascucci}, {Pasqualetti}, {Passaquieti}, {Passuello}, {Patel}, {Pathak}, {Patricelli},
  {Patron}, {Paul}, {Payne}, {Pedraza}, {Pegoraro}, {Pele}, {Arellano}, {Penn}, {Perego}, {Pereira}, {Pereira}, {Perez}, {P{\'e}rigois}, {Perkins}, {Perreca}, {Perri{\`e}s}, {Petermann}, {Petterson}, {Pfeiffer}, {Pham}, {Phukon}, {Piccinni}, {Pichot}, {Piendibene}, {Piergiovanni}, {Pierini}, {Pierro}, {Pillant}, {Pillas}, {Pilo}, {Pinard}, {Pinto}, {Pinto}, {Piotrzkowski}, {Piotrzkowski}, {Pirello}, {Pitkin}, {Placidi}, {Planas}, {Plastino}, {Pluchar}, {Poggiani}, {Polini}, {Pong}, {Ponrathnam}, {Popolizio}, {Porter}, {Poulton}, {Powell}, {Pracchia}, {Pradier}, {Prajapati}, {Prasai}, {Prasanna}, {Pratten}, {Principe}, {Prodi}, {Prokhorov}, {Prosposito}, {Prudenzi}, {Puecher}, {Punturo}, {Puosi}, {Puppo}, {P{\"u}rrer}, {Qi}, {Quetschke}, {Quitzow-James}, {Raab}, {Raaijmakers}, {Radkins}, {Radulesco}, {Raffai}, {Rail}, {Raja}, {Rajan}, {Ramirez}, {Ramirez}, {Ramos-Buades}, {Rana}, {Rapagnani}, {Rapol}, {Ray}, {Raymond}, {Raza}, {Razzano}, {Read}, {Rees}, {Regimbau}, {Rei}, {Reid}, {Reid}, {Reitze}, {Relton},
  {Renzini}, {Rettegno}, {Reza}, {Rezac}, {Ricci}, {Richards}, {Richardson}, {Richardson}, {Riemenschneider}, {Riles}, {Rinaldi}, {Rink}, {Rizzo}, {Robertson}, {Robie}, {Robinet}, {Rocchi}, {Rodriguez}, {Rolland}, {Rollins}, {Romanelli}, {Romano}, {Romel}, {Romero-Rodr{\'\i}guez}, {Romero-Shaw}, {Romie}, {Ronchini}, {Rosa}, {Rose}, {Rosi{\'n}ska}, {Ross}, {Rowan}, {Rowlinson}, {Roy}, {Roy}, {Roy}, {Rozza}, {Ruggi}, {Ryan}, {Sachdev}, {Sadecki}, {Sadiq}, {Sago}, {Saito}, {Saito}, {Sakai}, {Sakai}, {Sakellariadou}, {Sakuno}, {Salafia}, {Salconi}, {Saleem}, {Salemi}, {Samajdar}, {Sanchez}, {Sanchez}, {Sanchez}, {Sanchis-Gual}, {Sanders}, {Sanuy}, {Saravanan}, {Sarin}, {Sassolas}, {Satari}, {Sathyaprakash}, {Sato}, {Sato}, {Sauter}, {Savage}, {Sawada}, {Sawant}, {Sawant}, {Sayah}, {Schaetzl}, {Scheel}, {Scheuer}, {Schiworski}, {Schmidt}, {Schmidt}, {Schnabel}, {Schneewind}, {Schofield}, {Sch{\"o}nbeck}, {Schulte}, {Schutz}, {Schwartz}, {Scott}, {Scott}, {Seglar-Arroyo}, {Sekiguchi}, {Sekiguchi}, {Sellers},
  {Sengupta}, {Sentenac}, {Seo}, {Sequino}, {Sergeev}, {Setyawati}, {Shaffer}, {Shahriar}, {Shams}, {Shao}, {Sharma}, {Sharma}, {Shawhan}, {Shcheblanov}, {Shibagaki}, {Shikauchi}, {Shimizu}, {Shimoda}, {Shimode}, {Shinkai}, {Shishido}, {Shoda}, {Shoemaker}, {Shoemaker}, {Shyamsundar}, {Sieniawska}, {Sigg}, {Singer}, {Singh}, {Singh}, {Singha}, {Sintes}, {Sipala}, {Skliris}, {Slagmolen}, {Slaven-Blair}, {Smetana}, {Smith}, {Smith}, {Soldateschi}, {Somala}, {Somiya}, {Son}, {Soni}, {Soni}, {Sordini}, {Sorrentino}, {Sorrentino}, {Sotani}, {Soulard}, {Souradeep}, {Sowell}, {Spagnuolo}, {Spencer}, {Spera}, {Srinivasan}, {Srivastava}, {Srivastava}, {Staats}, {Stachie}, {Steer}, {Steinhoff}, {Steinlechner}, {Steinlechner}, {Stevenson}, {Stops}, {Stover}, {Strain}, {Strang}, {Stratta}, {Strunk}, {Sturani}, {Stuver}, {Sudhagar}, {Sudhir}, {Sugimoto}, {Suh}, {Sullivan}, {Summerscales}, {Sun}, {Sun}, {Sunil}, {Sur}, {Suresh}, {Sutton}, {Suzuki}, {Suzuki}, {Swinkels}, {Szczepa{\'n}czyk}, {Szewczyk}, {Tacca}, {Tagoshi},
  {Tait}, {Takahashi}, {Takahashi}, {Takamori}, {Takano}, {Takeda}, {Takeda}, {Talbot}, {Talbot}, {Tanaka}, {Tanaka}, {Tanaka}, {Tanaka}, {Tanaka}, {Tanasijczuk}, {Tanioka}, {Tanner}, {Tao}, {Tao}, {Mart{\'\i}n}, {Taranto}, {Tasson}, {Telada}, {Tenorio}, {Terhune}, {Terkowski}, {Thirugnanasambandam}, {Thomas}, {Thomas}, {Thomas}, {Thompson}, {Thondapu}, {Thorne}, {Thrane}, {Tiwari}, {Tiwari}, {Tiwari}, {Toivonen}, {Toland}, {Tolley}, {Tomaru}, {Tomigami}, {Tomura}, {Tonelli}, {Torres-Forn{\'e}}, {Torrie}, {E Melo}, {T{\"o}yr{\"a}}, {Trapananti}, {Travasso}, {Traylor}, {Trevor}, {Tringali}, {Tripathee}, {Troiano}, {Trovato}, {Trozzo}, {Trudeau}, {Tsai}, {Tsai}, {Tsang}, {Tsang}, {Tsao}, {Tse}, {Tso}, {Tsubono}, {Tsuchida}, {Tsukada}, {Tsuna}, {Tsutsui}, {Tsuzuki}, {Turbang}, {Turconi}, {Tuyenbayev}, {Ubhi}, {Uchikata}, {Uchiyama}, {Udall}, {Ueda}, {Uehara}, {Ueno}, {Ueshima}, {Unnikrishnan}, {Uraguchi}, {Urban}, {Ushiba}, {Utina}, {Vahlbruch}, {Vajente}, {Vajpeyi}, {Valdes}, {Valentini}, {Valsan}, {van Bakel},
  {van Beuzekom}, {van den Brand}, {van den Broeck}, {Vander-Hyde}, {van der Schaaf}, {van Heijningen}, {Vanosky}, {van Putten}, {van Remortel}, {Vardaro}, {Vargas}, {Varma}, {Vas{\'u}th}, {Vecchio}, {Vedovato}, {Veitch}, {Veitch}, {Venneberg}, {Venugopalan}, {Verkindt}, {Verma}, {Verma}, {Veske}, {Vetrano}, {Vicer{\'e}}, {Vidyant}, {Viets}, {Vijaykumar}, {Villa-Ortega}, {Vinet}, {Virtuoso}, {Vitale}, {Vo}, {Vocca}, {von Reis}, {von Wrangel}, {Vorvick}, {Vyatchanin}, {Wade}, {Wade}, {Wagner}, {Walet}, {Walker}, {Wallace}, {Wallace}, {Walsh}, {Wang}, {Wang}, {Wang}, {Ward}, {Warner}, {Was}, {Washimi}, {Washington}, {Watchi}, {Weaver}, {Webster}, {Weinert}, {Weinstein}, {Weiss}, {Weller}, {Wellmann}, {Wen}, {We{\ss}els}, {Wette}, {Whelan}, {White}, {Whiting}, {Whittle}, {Wilken}, {Williams}, {Williams}, {Williamson}, {Willis}, {Willke}, {Wilson}, {Winkler}, {Wipf}, {Wlodarczyk}, {Woan}, {Woehler}, {Wofford}, {Wong}, {Wu}, {Wu}, {Wu}, {Wu}, {Wysocki}, {Xiao}, {Xu}, {Yamada}, {Yamamoto}, {Yamamoto}, {Yamamoto},
  {Yamamoto}, {Yamashita}, {Yamazaki}, {Yang}, {Yang}, {Yang}, {Yang}, {Yang}, {Yap}, {Yeeles}, {Yelikar}, {Ying}, {Yokogawa}, {Yokoyama}, {Yokozawa}, {Yoo}, {Yoshioka}, {Yu}, {Yu}, {Yuzurihara}, {Zadro{\.z}ny}, {Zanolin}, {Zeidler}, {Zelenova}, {Zendri}, {Zevin}, {Zhan}, {Zhang}, {Zhang}, {Zhang}, {Zhang}, {Zhang}, {Zhao}, {Zhao}, {Zhao}, {Zhao}, {Zheng}, {Zhou}, {Zhou}, {Zhu}, {Zhu}, {Zimmerman}, {Zlochower}, {Zucker}, {Zweizig}, {LIGO Scientific Collaboration}, {VIRGO Collaboration}, \& {KAGRA Collaboration}}]{2023PhRvX..13a1048A}
{Abbott}, R., {Abbott}, T.~D., {Acernese}, F., {et~al.} 2023, Physical Review X, 13, 011048, \dodoi{10.1103/PhysRevX.13.011048}

\bibitem[{{Agudo} {et~al.}(2023){Agudo}, {Amati}, {An}, {Bauer}, {Benetti}, {Bernardini}, {Beswick}, {Bhirombhakdi}, {de Boer}, {Branchesi}, {Brennan}, {Brocato}, {Caballero-Garc{\'\i}a}, {Cappellaro}, {Castro Rodr{\'\i}guez}, {Castro-Tirado}, {Chambers}, {Chassande-Mottin}, {Chaty}, {Chen}, {Coleiro}, {Covino}, {D'Ammando}, {D'Avanzo}, {D'Elia}, {Fiore}, {Fl{\"o}rs}, {Fraser}, {Frey}, {Frohmaier}, {Fulton}, {Galbany}, {Gall}, {Gao}, {Garc{\'\i}a-Rojas}, {Ghirlanda}, {Giarratana}, {Gillanders}, {Giroletti}, {Gompertz}, {Gromadzki}, {Heintz}, {Hjorth}, {Hu}, {Huber}, {Inkenhaag}, {Izzo}, {Jin}, {Jonker}, {Kann}, {Kool}, {Kotak}, {Leloudas}, {Levan}, {Lin}, {Lyman}, {Magnier}, {Maguire}, {Mandel}, {Marcote}, {Mata S{\'a}nchez}, {Mattila}, {Melandri}, {Micha{\l}owski}, {Moldon}, {Nicholl}, {Nicuesa Guelbenzu}, {Oates}, {Onori}, {Orienti}, {Paladino}, {Paragi}, {Perez-Torres}, {Pian}, {Pignata}, {Piranomonte}, {Quirola-V{\'a}squez}, {Ragosta}, {Rau}, {Ronchini}, {Rossi}, {S{\'a}nchez-Ram{\'\i}rez}, {Salafia},
  {Schulze}, {Smartt}, {Smith}, {Sollerman}, {Srivastav}, {Starling}, {Steeghs}, {Stevance}, {Tanvir}, {Testa}, {Torres}, {Valeev}, {Vergani}, {Vescovi}, {Wainscost}, {Watson}, {Wiersema}, {Wyrzykowski}, {Yang}, {Yang}, \& {Young}}]{2023A&A...675A.201A}
{Agudo}, I., {Amati}, L., {An}, T., {et~al.} 2023, \aap, 675, A201, \dodoi{10.1051/0004-6361/202244751}

\bibitem[{{Arcavi} {et~al.}(2017){Arcavi}, {McCully}, {Hosseinzadeh}, {Howell}, {Vasylyev}, {Poznanski}, {Zaltzman}, {Maoz}, {Singer}, {Valenti}, {Kasen}, {Barnes}, {Piran}, \& {Fong}}]{2017ApJ...848L..33A}
{Arcavi}, I., {McCully}, C., {Hosseinzadeh}, G., {et~al.} 2017, \apjl, 848, L33, \dodoi{10.3847/2041-8213/aa910f}

\bibitem[{{Atteia} {et~al.}(2022){Atteia}, {Cordier}, \& {Wei}}]{2022IJMPD..3130008A}
{Atteia}, J.~L., {Cordier}, B., \& {Wei}, J. 2022, International Journal of Modern Physics D, 31, 2230008, \dodoi{10.1142/S0218271822300087}

\bibitem[{{Banerjee} {et~al.}(2023){Banerjee}, {Tanaka}, {Kato}, \& {Gaigalas}}]{2023arXiv230405810B}
{Banerjee}, S., {Tanaka}, M., {Kato}, D., \& {Gaigalas}, G. 2023, arXiv e-prints, arXiv:2304.05810, \dodoi{10.48550/arXiv.2304.05810}

\bibitem[{{Banerjee} {et~al.}(2020){Banerjee}, {Tanaka}, {Kawaguchi}, {Kato}, \& {Gaigalas}}]{2020ApJ...901...29B}
{Banerjee}, S., {Tanaka}, M., {Kawaguchi}, K., {Kato}, D., \& {Gaigalas}, G. 2020, \apj, 901, 29, \dodoi{10.3847/1538-4357/abae61}

\bibitem[{{Barnes} \& {Kasen}(2013)}]{2013ApJ...775...18B}
{Barnes}, J., \& {Kasen}, D. 2013, \apj, 775, 18, \dodoi{10.1088/0004-637X/775/1/18}

\bibitem[{{Barnes} {et~al.}(2016){Barnes}, {Kasen}, {Wu}, \& {Mart{\'\i}nez-Pinedo}}]{2016ApJ...829..110B}
{Barnes}, J., {Kasen}, D., {Wu}, M.-R., \& {Mart{\'\i}nez-Pinedo}, G. 2016, \apj, 829, 110, \dodoi{10.3847/0004-637X/829/2/110}

\bibitem[{{Barthelmy} {et~al.}(2005){Barthelmy}, {Cannizzo}, {Gehrels}, {Cusumano}, {Mangano}, {O'Brien}, {Vaughan}, {Zhang}, {Burrows}, {Campana}, {Chincarini}, {Goad}, {Kouveliotou}, {Kumar}, {M{\'e}sz{\'a}ros}, {Nousek}, {Osborne}, {Panaitescu}, {Reeves}, {Sakamoto}, {Tagliaferri}, \& {Wijers}}]{2005ApJ...635L.133B}
{Barthelmy}, S.~D., {Cannizzo}, J.~K., {Gehrels}, N., {et~al.} 2005, \apjl, 635, L133, \dodoi{10.1086/499432}

\bibitem[{{Bauer} {et~al.}(2017){Bauer}, {Treister}, {Schawinski}, {Schulze}, {Luo}, {Alexander}, {Brandt}, {Comastri}, {Forster}, {Gilli}, {Kann}, {Maeda}, {Nomoto}, {Paolillo}, {Ranalli}, {Schneider}, {Shemmer}, {Tanaka}, {Tolstov}, {Tominaga}, {Tozzi}, {Vignali}, {Wang}, {Xue}, \& {Yang}}]{2017MNRAS.467.4841B}
{Bauer}, F.~E., {Treister}, E., {Schawinski}, K., {et~al.} 2017, \mnras, 467, 4841, \dodoi{10.1093/mnras/stx417}

\bibitem[{{Bauswein} {et~al.}(2013){Bauswein}, {Goriely}, \& {Janka}}]{2013ApJ...773...78B}
{Bauswein}, A., {Goriely}, S., \& {Janka}, H.~T. 2013, \apj, 773, 78, \dodoi{10.1088/0004-637X/773/1/78}

\bibitem[{{Becerra} {et~al.}(2019){Becerra}, {Watson}, {Fraija}, {Butler}, {Lee}, {Troja}, {Rom{\'a}n-Z{\'u}{\~n}iga}, {Kutyrev}, {{\'A}lvarez Nu{\~n}ez}, {{\'A}ngeles}, {Chapa}, {Cuevas}, {Farah}, {Fuentes-Fern{\'a}ndez}, {Figueroa}, {Langarica}, {Quir{\'o}s}, {Ru{\'\i}z-D{\'\i}az-Soto}, {Tejada}, \& {Tinoco}}]{2019ApJ...872..118B}
{Becerra}, R.~L., {Watson}, A.~M., {Fraija}, N., {et~al.} 2019, \apj, 872, 118, \dodoi{10.3847/1538-4357/ab0026}

\bibitem[{{Begelman} \& {Cioffi}(1989)}]{1989ApJ...345L..21B}
{Begelman}, M.~C., \& {Cioffi}, D.~F. 1989, \apjl, 345, L21, \dodoi{10.1086/185542}

\bibitem[{{Beniamini} {et~al.}(2020){Beniamini}, {Duque}, {Daigne}, \& {Mochkovitch}}]{2020MNRAS.492.2847B}
{Beniamini}, P., {Duque}, R., {Daigne}, F., \& {Mochkovitch}, R. 2020, \mnras, 492, 2847, \dodoi{10.1093/mnras/staa070}

\bibitem[{{Beniamini} {et~al.}(2016){Beniamini}, {Nava}, \& {Piran}}]{2016MNRAS.461...51B}
{Beniamini}, P., {Nava}, L., \& {Piran}, T. 2016, \mnras, 461, 51, \dodoi{10.1093/mnras/stw1331}

\bibitem[{{Berger} {et~al.}(2013){Berger}, {Fong}, \& {Chornock}}]{2013ApJ...774L..23B}
{Berger}, E., {Fong}, W., \& {Chornock}, R. 2013, \apjl, 774, L23, \dodoi{10.1088/2041-8205/774/2/L23}

\bibitem[{{Bromberg} {et~al.}(2011){Bromberg}, {Nakar}, {Piran}, \& {Sari}}]{2011ApJ...740..100B}
{Bromberg}, O., {Nakar}, E., {Piran}, T., \& {Sari}, R. 2011, \apj, 740, 100, \dodoi{10.1088/0004-637X/740/2/100}

\bibitem[{{Bromberg} {et~al.}(2013){Bromberg}, {Nakar}, {Piran}, \& {Sari}}]{2013ApJ...764..179B}
---. 2013, \apj, 764, 179, \dodoi{10.1088/0004-637X/764/2/179}

\bibitem[{{Brown} {et~al.}(2010){Brown}, {Roming}, {Milne}, {Bufano}, {Ciardullo}, {Elias-Rosa}, {Filippenko}, {Foley}, {Gehrels}, {Gronwall}, {Hicken}, {Holland}, {Hoversten}, {Immler}, {Kirshner}, {Li}, {Mazzali}, {Phillips}, {Pritchard}, {Still}, {Turatto}, \& {Vanden Berk}}]{2010ApJ...721.1608B}
{Brown}, P.~J., {Roming}, P. W.~A., {Milne}, P., {et~al.} 2010, \apj, 721, 1608, \dodoi{10.1088/0004-637X/721/2/1608}

\bibitem[{{Burdge} {et~al.}(2020){Burdge}, {Prince}, {Fuller}, {Kaplan}, {Marsh}, {Tremblay}, {Zhuang}, {Bellm}, {Caiazzo}, {Coughlin}, {Dhillon}, {Gaensicke}, {Rodr{\'\i}guez-Gil}, {Graham}, {Hermes}, {Kupfer}, {Littlefair}, {Mr{\'o}z}, {Phinney}, {van Roestel}, {Yao}, {Dekany}, {Drake}, {Duev}, {Hale}, {Feeney}, {Helou}, {Kaye}, {Mahabal}, {Masci}, {Riddle}, {Smith}, {Soumagnac}, \& {Kulkarni}}]{2020ApJ...905...32B}
{Burdge}, K.~B., {Prince}, T.~A., {Fuller}, J., {et~al.} 2020, \apj, 905, 32, \dodoi{10.3847/1538-4357/abc261}

\bibitem[{Chornock {et~al.}(2017)}]{Chornock:2017sdf}
Chornock, R., {et~al.} 2017, Astrophys. J., 848, L19, \dodoi{10.3847/2041-8213/aa905c}

\bibitem[{{Combi} \& {Siegel}(2023)}]{2023arXiv230312284C}
{Combi}, L., \& {Siegel}, D.~M. 2023, arXiv e-prints, arXiv:2303.12284, \dodoi{10.48550/arXiv.2303.12284}

\bibitem[{Coulter {et~al.}(2017)}]{Coulter:2017wya}
Coulter, D.~A., {et~al.} 2017, Science, \dodoi{10.1126/science.aap9811}

\bibitem[{{D'Ai} {et~al.}(2016){D'Ai}, {Melandri}, {McCauley}, {Kennea}, {Sbarufatti}, {Beardmore}, {Gibson}, {D'Avanzo}, \& {Evans}}]{2016GCN.20206....1D}
{D'Ai}, A., {Melandri}, A., {McCauley}, L.~M., {et~al.} 2016, GRB Coordinates Network, 20206, 1

\bibitem[{{Davis} {et~al.}(2021){Davis}, {Areeda}, {Berger}, {Bruntz}, {Effler}, {Essick}, {Fisher}, {Godwin}, {Goetz}, {Helmling-Cornell}, {Hughey}, {Katsavounidis}, {Lundgren}, {Macleod}, {M{\'a}rka}, {Massinger}, {Matas}, {McIver}, {Mo}, {Mogushi}, {Nguyen}, {Nuttall}, {Schofield}, {Shoemaker}, {Soni}, {Stuver}, {Urban}, {Valdes}, {Walker}, {Abbott}, {Adams}, {Adhikari}, {Ananyeva}, {Appert}, {Arai}, {Asali}, {Aston}, {Austin}, {Baer}, {Ball}, {Ballmer}, {Banagiri}, {Barker}, {Barschaw}, {Barsotti}, {Bartlett}, {Betzwieser}, {Beda}, {Bhattacharjee}, {Bidler}, {Billingsley}, {Biscans}, {Blair}, {Blair}, {Bode}, {Booker}, {Bork}, {Bramley}, {Brooks}, {Brown}, {Buikema}, {Cahillane}, {Callister}, {Caneva Santoro}, {Cannon}, {Carlin}, {Chandra}, {Chen}, {Christensen}, {Ciobanu}, {Clara}, {Compton}, {Cooper}, {Corley}, {Coughlin}, {Countryman}, {Covas}, {Coyne}, {Crowder}, {Dal Canton}, {Danila}, {Datrier}, {Davies}, {Dent}, {Didio}, {Di Fronzo}, {Dooley}, {Driggers}, {Dupej}, {Dwyer}, {Etzel}, {Evans},
  {Evans}, {Fairhurst}, {Feicht}, {Fernandez-Galiana}, {Frey}, {Fritschel}, {Frolov}, {Fulda}, {Fyffe}, {Gadre}, {Giaime}, {Giardina}, {Gonz{\'a}lez}, {Gras}, {Gray}, {Gray}, {Green}, {Gupta}, {Gustafson}, {Gustafson}, {Hanks}, {Hanson}, {Hardwick}, {Harry}, {Hasskew}, {Heintze}, {Heinzel}, {Holland}, {Hollows}, {Hoy}, {Hughey}, {Jadhav}, {Janssens}, {Johns}, {Jones}, {Kandhasamy}, {Karki}, {Kasprzack}, {Kawabe}, {Keitel}, {Kijbunchoo}, {Kim}, {King}, {Kissel}, {Kulkarni}, {Kumar}, {Landry}, {Lane}, {Lantz}, {Laxen}, {Lecoeuche}, {Leviton}, {Liu}, {Lormand}, {Macas}, {Macedo}, {MacInnis}, {Mandic}, {Mansell}, {M{\'a}rka}, {Martinez}, {Martinovic}, {Martynov}, {Mason}, {Matichard}, {Mavalvala}, {McCarthy}, {McClelland}, {McCormick}, {McCuller}, {McIsaac}, {McRae}, {Mendell}, {Merfeld}, {Merilh}, {Meyers}, {Meylahn}, {Michaloliakos}, {Middleton}, {Mills}, {Mistry}, {Mittleman}, {Moreno}, {Mow-Lowry}, {Mozzon}, {Mueller}, {Mukund}, {Mullavey}, {Muth}, {Nelson}, {Neunzert}, {Nichols}, {Nitoglia}, {Oberling},
  {Oh}, {Oh}, {Oram}, {Ormiston}, {Ormsby}, {Osthelder}, {Ottaway}, {Overmier}, {Pai}, {Palamos}, {Pannarale}, {Parker}, {Patane}, {Patel}, {Payne}, {Pele}, {Penhorwood}, {Perez}, {Phukon}, {Pillas}, {Pirello}, {Radkins}, {Ramirez}, {Richardson}, {Riles}, {Rink}, {Robertson}, {Rollins}, {Romel}, {Romie}, {Ross}, {Ryan}, {Sadecki}, {Sakellariadou}, {Sanchez}, {Sanchez}, {Sandles}, {Saravanan}, {Savage}, {Schaetzl}, {Schnabel}, {Schwartz}, {Sellers}, {Shaffer}, {Sigg}, {Sintes}, {Slagmolen}, {Smith}, {Soni}, {Sorazu}, {Spencer}, {Strain}, {Strom}, {Sun}, {Szczepa{\'n}czyk}, {Tasson}, {Tenorio}, {Thomas}, {Thomas}, {Thorne}, {Toland}, {Torrie}, {Tran}, {Traylor}, {Trevor}, {Tse}, {Vajente}, {van Remortel}, {Vander-Hyde}, {Vargas}, {Veitch}, {Veitch}, {Venkateswara}, {Venugopalan}, {Viets}, {Villa-Ortega}, {Vo}, {Vorvick}, {Wade}, {Wallace}, {Ward}, {Warner}, {Weaver}, {Weinstein}, {Weiss}, {Wette}, {White}, {White}, {Whittle}, {Williamson}, {Willke}, {Wipf}, {Xiao}, {Xu}, {Yamamoto}, {Yu}, {Yu}, {Zhang},
  {Zheng}, {Zucker}, \& {Zweizig}}]{2021CQGra..38m5014D}
{Davis}, D., {Areeda}, J.~S., {Berger}, B.~K., {et~al.} 2021, Classical and Quantum Gravity, 38, 135014, \dodoi{10.1088/1361-6382/abfd85}

\bibitem[{{De Colle} {et~al.}(2018){De Colle}, {Lu}, {Kumar}, {Ramirez-Ruiz}, \& {Smoot}}]{2018MNRAS.478.4553D}
{De Colle}, F., {Lu}, W., {Kumar}, P., {Ramirez-Ruiz}, E., \& {Smoot}, G. 2018, \mnras, 478, 4553, \dodoi{10.1093/mnras/sty1282}

\bibitem[{{D{\'{\i}}az} {et~al.}(2017){D{\'{\i}}az}, {Macri}, {Garcia Lambas}, {Mendes de Oliveira}, {Nilo Castell{\'o}n}, {Ribeiro}, {S{\'a}nchez}, {Schoenell}, {Abramo}, {Akras}, {Alcaniz}, {Artola}, {Beroiz}, {Bonoli}, {Cabral}, {Camuccio}, {Castillo}, {Chavushyan}, {Coelho}, {Colazo}, {Costa-Duarte}, {Cuevas Larenas}, {DePoy}, {Dom{\'{\i}}nguez Romero}, {Dultzin}, {Fern{\'a}ndez}, {Garc{\'{\i}}a}, {Girardini}, {Gon{\c c}alves}, {Gon{\c c}alves}, {Gurovich}, {Jim{\'e}nez-Teja}, {Kanaan}, {Lares}, {Lopes de Oliveira}, {L{\'o}pez-Cruz}, {Marshall}, {Melia}, {Molino}, {Padilla}, {Pe{\~n}uela}, {Placco}, {Qui{\~n}ones}, {Ram{\'{\i}}rez Rivera}, {Renzi}, {Riguccini}, {R{\'{\i}}os-L{\'o}pez}, {Rodriguez}, {Sampedro}, {Schneiter}, {Sodr{\'e}}, {Starck}, {Torres-Flores}, {Tornatore}, \& {Zadro{\.z}ny}}]{2017ApJ...848L..29D}
{D{\'{\i}}az}, M.~C., {Macri}, L.~M., {Garcia Lambas}, D., {et~al.} 2017, \apjl, 848, L29, \dodoi{10.3847/2041-8213/aa9060}

\bibitem[{{Drout} {et~al.}(2017){Drout}, {Piro}, {Shappee}, {Kilpatrick}, {Simon}, {Contreras}, {Coulter}, {Foley}, {Siebert}, {Morrell}, {Boutsia}, {Di Mille}, {Holoien}, {Kasen}, {Kollmeier}, {Madore}, {Monson}, {Murguia-Berthier}, {Pan}, {Prochaska}, {Ramirez-Ruiz}, {Rest}, {Adams}, {Alatalo}, {Ba{\~n}ados}, {Baughman}, {Beers}, {Bernstein}, {Bitsakis}, {Campillay}, {Hansen}, {Higgs}, {Ji}, {Maravelias}, {Marshall}, {Moni Bidin}, {Prieto}, {Rasmussen}, {Rojas-Bravo}, {Strom}, {Ulloa}, {Vargas-Gonz{\'a}lez}, {Wan}, \& {Whitten}}]{2017Sci...358.1570D}
{Drout}, M.~R., {Piro}, A.~L., {Shappee}, B.~J., {et~al.} 2017, Science, 358, 1570, \dodoi{10.1126/science.aaq0049}

\bibitem[{{Duffell} {et~al.}(2015){Duffell}, {Quataert}, \& {MacFadyen}}]{2015ApJ...813...64D}
{Duffell}, P.~C., {Quataert}, E., \& {MacFadyen}, A.~I. 2015, \apj, 813, 64, \dodoi{10.1088/0004-637X/813/1/64}

\bibitem[{{Eichler} {et~al.}(1989){Eichler}, {Livio}, {Piran}, \& {Schramm}}]{1989Natur.340..126E}
{Eichler}, D., {Livio}, M., {Piran}, T., \& {Schramm}, D.~N. 1989, \nat, 340, 126, \dodoi{10.1038/340126a0}

\bibitem[{{Fong} {et~al.}(2015){Fong}, {Berger}, {Margutti}, \& {Zauderer}}]{2015ApJ...815..102F}
{Fong}, W., {Berger}, E., {Margutti}, R., \& {Zauderer}, B.~A. 2015, \apj, 815, 102, \dodoi{10.1088/0004-637X/815/2/102}

\bibitem[{{Fujibayashi} {et~al.}(2018){Fujibayashi}, {Kiuchi}, {Nishimura}, {Sekiguchi}, \& {Shibata}}]{2018ApJ...860...64F}
{Fujibayashi}, S., {Kiuchi}, K., {Nishimura}, N., {Sekiguchi}, Y., \& {Shibata}, M. 2018, \apj, 860, 64, \dodoi{10.3847/1538-4357/aabafd}

\bibitem[{{Fujibayashi} {et~al.}(2020){Fujibayashi}, {Wanajo}, {Kiuchi}, {Kyutoku}, {Sekiguchi}, \& {Shibata}}]{2020ApJ...901..122F}
{Fujibayashi}, S., {Wanajo}, S., {Kiuchi}, K., {et~al.} 2020, \apj, 901, 122, \dodoi{10.3847/1538-4357/abafc2}

\bibitem[{{Gao} {et~al.}(2017){Gao}, {Zhang}, {L{\"u}}, \& {Li}}]{2017ApJ...837...50G}
{Gao}, H., {Zhang}, B., {L{\"u}}, H.-J., \& {Li}, Y. 2017, \apj, 837, 50, \dodoi{10.3847/1538-4357/aa5be3}

\bibitem[{{Gillanders} {et~al.}(2023){Gillanders}, {Troja}, {Fryer}, {Ristic}, {O'Connor}, {Fontes}, {Yang}, {Domoto}, {Rahmouni}, {Tanaka}, {Fox}, \& {Dichiara}}]{2023arXiv230800633G}
{Gillanders}, J.~H., {Troja}, E., {Fryer}, C.~L., {et~al.} 2023, arXiv e-prints, arXiv:2308.00633, \dodoi{10.48550/arXiv.2308.00633}

\bibitem[{{Gompertz} {et~al.}(2013){Gompertz}, {O'Brien}, {Wynn}, \& {Rowlinson}}]{2013MNRAS.431.1745G}
{Gompertz}, B.~P., {O'Brien}, P.~T., {Wynn}, G.~A., \& {Rowlinson}, A. 2013, \mnras, 431, 1745, \dodoi{10.1093/mnras/stt293}

\bibitem[{{Gompertz} {et~al.}(2018){Gompertz}, {Levan}, {Tanvir}, {Hjorth}, {Covino}, {Evans}, {Fruchter}, {Gonz{\'a}lez-Fern{\'a}ndez}, {Jin}, {Lyman}, {Oates}, {O'Brien}, \& {Wiersema}}]{2018ApJ...860...62G}
{Gompertz}, B.~P., {Levan}, A.~J., {Tanvir}, N.~R., {et~al.} 2018, \apj, 860, 62, \dodoi{10.3847/1538-4357/aac206}

\bibitem[{{Goodman}(1986)}]{1986ApJ...308L..47G}
{Goodman}, J. 1986, \apjl, 308, L47, \dodoi{10.1086/184741}

\bibitem[{{Gottlieb} \& {Loeb}(2020)}]{2020MNRAS.493.1753G}
{Gottlieb}, O., \& {Loeb}, A. 2020, \mnras, 493, 1753, \dodoi{10.1093/mnras/staa363}

\bibitem[{{Gottlieb} \& {Nakar}(2022)}]{2022MNRAS.517.1640G}
{Gottlieb}, O., \& {Nakar}, E. 2022, \mnras, 517, 1640, \dodoi{10.1093/mnras/stac2699}

\bibitem[{{Gottlieb} {et~al.}(2021){Gottlieb}, {Nakar}, \& {Bromberg}}]{2021MNRAS.500.3511G}
{Gottlieb}, O., {Nakar}, E., \& {Bromberg}, O. 2021, \mnras, 500, 3511, \dodoi{10.1093/mnras/staa3501}

\bibitem[{{Gottlieb} {et~al.}(2018){Gottlieb}, {Nakar}, \& {Piran}}]{2018MNRAS.473..576G}
{Gottlieb}, O., {Nakar}, E., \& {Piran}, T. 2018, \mnras, 473, 576, \dodoi{10.1093/mnras/stx2357}

\bibitem[{{Gottlieb} {et~al.}(2023){Gottlieb}, {Metzger}, {Quataert}, {Issa}, {Martineau}, {Foucart}, {Duez}, {Kidder}, {Pfeiffer}, \& {Scheel}}]{2023ApJ...958L..33G}
{Gottlieb}, O., {Metzger}, B.~D., {Quataert}, E., {et~al.} 2023, \apjl, 958, L33, \dodoi{10.3847/2041-8213/ad096e}

\bibitem[{{Granot} {et~al.}(2023){Granot}, {Levinson}, \& {Nakar}}]{2023arXiv230508575G}
{Granot}, A., {Levinson}, A., \& {Nakar}, E. 2023, arXiv e-prints, arXiv:2305.08575, \dodoi{10.48550/arXiv.2305.08575}

\bibitem[{{Granot} {et~al.}(2006){Granot}, {K{\"o}nigl}, \& {Piran}}]{2006MNRAS.370.1946G}
{Granot}, J., {K{\"o}nigl}, A., \& {Piran}, T. 2006, \mnras, 370, 1946, \dodoi{10.1111/j.1365-2966.2006.10621.x}

\bibitem[{{Hamidani} \& {Ioka}(2021)}]{2021MNRAS.500..627H}
{Hamidani}, H., \& {Ioka}, K. 2021, \mnras, 500, 627, \dodoi{10.1093/mnras/staa3276}

\bibitem[{{Hamidani} \& {Ioka}(2023{\natexlab{a}})}]{2023MNRAS.520.1111H}
---. 2023{\natexlab{a}}, \mnras, 520, 1111, \dodoi{10.1093/mnras/stad041}

\bibitem[{{Hamidani} \& {Ioka}(2023{\natexlab{b}})}]{2023MNRAS.524.4841H}
---. 2023{\natexlab{b}}, \mnras, 524, 4841, \dodoi{10.1093/mnras/stad1933}

\bibitem[{{Hamidani} {et~al.}(2020){Hamidani}, {Kiuchi}, \& {Ioka}}]{2020MNRAS.491.3192H}
{Hamidani}, H., {Kiuchi}, K., \& {Ioka}, K. 2020, \mnras, 491, 3192, \dodoi{10.1093/mnras/stz3231}

\bibitem[{{Hayes} {et~al.}(2023){Hayes}, {Heng}, {Lamb}, {Lin}, {Veitch}, \& {Williams}}]{2023ApJ...954...92H}
{Hayes}, F., {Heng}, I.~S., {Lamb}, G., {et~al.} 2023, \apj, 954, 92, \dodoi{10.3847/1538-4357/ace899}

\bibitem[{{Ho} {et~al.}(2023){Ho}, {Perley}, {Gal-Yam}, {Lunnan}, {Sollerman}, {Schulze}, {Das}, {Dobie}, {Yao}, {Fremling}, {Adams}, {Anand}, {Andreoni}, {Bellm}, {Bruch}, {Burdge}, {Castro-Tirado}, {Dahiwale}, {De}, {Dekany}, {Drake}, {Duev}, {Graham}, {Helou}, {Kaplan}, {Karambelkar}, {Kasliwal}, {Kool}, {Kulkarni}, {Mahabal}, {Medford}, {Miller}, {Nordin}, {Ofek}, {Petitpas}, {Riddle}, {Sharma}, {Smith}, {Stewart}, {Taggart}, {Tartaglia}, {Tzanidakis}, \& {Winters}}]{2023ApJ...949..120H}
{Ho}, A. Y.~Q., {Perley}, D.~A., {Gal-Yam}, A., {et~al.} 2023, \apj, 949, 120, \dodoi{10.3847/1538-4357/acc533}

\bibitem[{{Honda} {et~al.}(2015){Honda}, {Fujita}, {Negoro}, {Serino}, {Nakahira}, {Ueno}, {Tomida}, {Kimura}, {Ishikawa}, {Nakagawa}, {Mihara}, {Sugizaki}, {Shidatsu}, {Sugimoto}, {Takagi}, {Matsuoka}, {Kawai}, {Yoshii}, {Tachibana}, {Ono}, {Yoshida}, {Sakamoto}, {Kawakubo}, {Ohtsuki}, {Tsunemi}, {Imatani}, {Nakajima}, {Namba}, {Masumitsu}, {Tanaka}, {Ueda}, {Kawamuro}, {Hori}, {Tsuboi}, {Kanetou}, {Yamauchi}, {Itoh}, {Yamaoka}, \& {Morii}}]{2015GCN.17772....1H}
{Honda}, F., {Fujita}, M., {Negoro}, H., {et~al.} 2015, GRB Coordinates Network, 17772, 1

\bibitem[{{Hotokezaka} {et~al.}(2013){Hotokezaka}, {Kiuchi}, {Kyutoku}, {Okawa}, {Sekiguchi}, {Shibata}, \& {Taniguchi}}]{2013PhRvD..87b4001H}
{Hotokezaka}, K., {Kiuchi}, K., {Kyutoku}, K., {et~al.} 2013, \prd, 87, 024001, \dodoi{10.1103/PhysRevD.87.024001}

\bibitem[{{Hotokezaka} {et~al.}(2016){Hotokezaka}, {Wanajo}, {Tanaka}, {Bamba}, {Terada}, \& {Piran}}]{2016MNRAS.459...35H}
{Hotokezaka}, K., {Wanajo}, S., {Tanaka}, M., {et~al.} 2016, \mnras, 459, 35, \dodoi{10.1093/mnras/stw404}

\bibitem[{{Ioka} {et~al.}(2005){Ioka}, {Kobayashi}, \& {Zhang}}]{2005ApJ...631..429I}
{Ioka}, K., {Kobayashi}, S., \& {Zhang}, B. 2005, \apj, 631, 429, \dodoi{10.1086/432567}

\bibitem[{{Ioka} \& {Nakamura}(2018)}]{2018PTEP.2018d3E02I}
{Ioka}, K., \& {Nakamura}, T. 2018, Progress of Theoretical and Experimental Physics, 2018, 043E02, \dodoi{10.1093/ptep/pty036}

\bibitem[{{Ioka} {et~al.}(2006){Ioka}, {Toma}, {Yamazaki}, \& {Nakamura}}]{2006A&A...458....7I}
{Ioka}, K., {Toma}, K., {Yamazaki}, R., \& {Nakamura}, T. 2006, \aap, 458, 7, \dodoi{10.1051/0004-6361:20064939}

\bibitem[{{Ishizaki} {et~al.}(2021){Ishizaki}, {Kiuchi}, {Ioka}, \& {Wanajo}}]{2021ApJ...922..185I}
{Ishizaki}, W., {Kiuchi}, K., {Ioka}, K., \& {Wanajo}, S. 2021, \apj, 922, 185, \dodoi{10.3847/1538-4357/ac23d9}

\bibitem[{{Ivezi{\'c}} {et~al.}(2019){Ivezi{\'c}}, {Kahn}, {Tyson}, {Abel}, {Acosta}, {Allsman}, {Alonso}, {AlSayyad}, {Anderson}, {Andrew}, {Angel}, {Angeli}, {Ansari}, {Antilogus}, {Araujo}, {Armstrong}, {Arndt}, {Astier}, {Aubourg}, {Auza}, {Axelrod}, {Bard}, {Barr}, {Barrau}, {Bartlett}, {Bauer}, {Bauman}, {Baumont}, {Bechtol}, {Bechtol}, {Becker}, {Becla}, {Beldica}, {Bellavia}, {Bianco}, {Biswas}, {Blanc}, {Blazek}, {Blandford}, {Bloom}, {Bogart}, {Bond}, {Booth}, {Borgland}, {Borne}, {Bosch}, {Boutigny}, {Brackett}, {Bradshaw}, {Brandt}, {Brown}, {Bullock}, {Burchat}, {Burke}, {Cagnoli}, {Calabrese}, {Callahan}, {Callen}, {Carlin}, {Carlson}, {Chandrasekharan}, {Charles-Emerson}, {Chesley}, {Cheu}, {Chiang}, {Chiang}, {Chirino}, {Chow}, {Ciardi}, {Claver}, {Cohen-Tanugi}, {Cockrum}, {Coles}, {Connolly}, {Cook}, {Cooray}, {Covey}, {Cribbs}, {Cui}, {Cutri}, {Daly}, {Daniel}, {Daruich}, {Daubard}, {Daues}, {Dawson}, {Delgado}, {Dellapenna}, {de Peyster}, {de Val-Borro}, {Digel}, {Doherty}, {Dubois},
  {Dubois-Felsmann}, {Durech}, {Economou}, {Eifler}, {Eracleous}, {Emmons}, {Fausti Neto}, {Ferguson}, {Figueroa}, {Fisher-Levine}, {Focke}, {Foss}, {Frank}, {Freemon}, {Gangler}, {Gawiser}, {Geary}, {Gee}, {Geha}, {Gessner}, {Gibson}, {Gilmore}, {Glanzman}, {Glick}, {Goldina}, {Goldstein}, {Goodenow}, {Graham}, {Gressler}, {Gris}, {Guy}, {Guyonnet}, {Haller}, {Harris}, {Hascall}, {Haupt}, {Hernandez}, {Herrmann}, {Hileman}, {Hoblitt}, {Hodgson}, {Hogan}, {Howard}, {Huang}, {Huffer}, {Ingraham}, {Innes}, {Jacoby}, {Jain}, {Jammes}, {Jee}, {Jenness}, {Jernigan}, {Jevremovi{\'c}}, {Johns}, {Johnson}, {Johnson}, {Jones}, {Juramy-Gilles}, {Juri{\'c}}, {Kalirai}, {Kallivayalil}, {Kalmbach}, {Kantor}, {Karst}, {Kasliwal}, {Kelly}, {Kessler}, {Kinnison}, {Kirkby}, {Knox}, {Kotov}, {Krabbendam}, {Krughoff}, {Kub{\'a}nek}, {Kuczewski}, {Kulkarni}, {Ku}, {Kurita}, {Lage}, {Lambert}, {Lange}, {Langton}, {Le Guillou}, {Levine}, {Liang}, {Lim}, {Lintott}, {Long}, {Lopez}, {Lotz}, {Lupton}, {Lust}, {MacArthur}, {Mahabal},
  {Mandelbaum}, {Markiewicz}, {Marsh}, {Marshall}, {Marshall}, {May}, {McKercher}, {McQueen}, {Meyers}, {Migliore}, {Miller}, {Mills}, {Miraval}, {Moeyens}, {Moolekamp}, {Monet}, {Moniez}, {Monkewitz}, {Montgomery}, {Morrison}, {Mueller}, {Muller}, {Mu{\~n}oz Arancibia}, {Neill}, {Newbry}, {Nief}, {Nomerotski}, {Nordby}, {O'Connor}, {Oliver}, {Olivier}, {Olsen}, {O'Mullane}, {Ortiz}, {Osier}, {Owen}, {Pain}, {Palecek}, {Parejko}, {Parsons}, {Pease}, {Peterson}, {Peterson}, {Petravick}, {Libby Petrick}, {Petry}, {Pierfederici}, {Pietrowicz}, {Pike}, {Pinto}, {Plante}, {Plate}, {Plutchak}, {Price}, {Prouza}, {Radeka}, {Rajagopal}, {Rasmussen}, {Regnault}, {Reil}, {Reiss}, {Reuter}, {Ridgway}, {Riot}, {Ritz}, {Robinson}, {Roby}, {Roodman}, {Rosing}, {Roucelle}, {Rumore}, {Russo}, {Saha}, {Sassolas}, {Schalk}, {Schellart}, {Schindler}, {Schmidt}, {Schneider}, {Schneider}, {Schoening}, {Schumacher}, {Schwamb}, {Sebag}, {Selvy}, {Sembroski}, {Seppala}, {Serio}, {Serrano}, {Shaw}, {Shipsey}, {Sick}, {Silvestri},
  {Slater}, {Smith}, {Smith}, {Sobhani}, {Soldahl}, {Storrie-Lombardi}, {Stover}, {Strauss}, {Street}, {Stubbs}, {Sullivan}, {Sweeney}, {Swinbank}, {Szalay}, {Takacs}, {Tether}, {Thaler}, {Thayer}, {Thomas}, {Thornton}, {Thukral}, {Tice}, {Trilling}, {Turri}, {Van Berg}, {Vanden Berk}, {Vetter}, {Virieux}, {Vucina}, {Wahl}, {Walkowicz}, {Walsh}, {Walter}, {Wang}, {Wang}, {Warner}, {Wiecha}, {Willman}, {Winters}, {Wittman}, {Wolff}, {Wood-Vasey}, {Wu}, {Xin}, {Yoachim}, \& {Zhan}}]{2019ApJ...873..111I}
{Ivezi{\'c}}, {\v{Z}}., {Kahn}, S.~M., {Tyson}, J.~A., {et~al.} 2019, \apj, 873, 111, \dodoi{10.3847/1538-4357/ab042c}

\bibitem[{{Jin} {et~al.}(2020){Jin}, {Covino}, {Liao}, {Li}, {D'Avanzo}, {Fan}, \& {Wei}}]{2020NatAs...4...77J}
{Jin}, Z.-P., {Covino}, S., {Liao}, N.-H., {et~al.} 2020, Nature Astronomy, 4, 77, \dodoi{10.1038/s41550-019-0892-y}

\bibitem[{{Jin} {et~al.}(2015){Jin}, {Li}, {Cano}, {Covino}, {Fan}, \& {Wei}}]{2015ApJ...811L..22J}
{Jin}, Z.-P., {Li}, X., {Cano}, Z., {et~al.} 2015, \apjl, 811, L22, \dodoi{10.1088/2041-8205/811/2/L22}

\bibitem[{{Kagawa} {et~al.}(2019){Kagawa}, {Yonetoku}, {Sawano}, {Arimoto}, {Kisaka}, \& {Yamazaki}}]{2019ApJ...877..147K}
{Kagawa}, Y., {Yonetoku}, D., {Sawano}, T., {et~al.} 2019, \apj, 877, 147, \dodoi{10.3847/1538-4357/ab1bd6}

\bibitem[{{Kaneko} {et~al.}(2015){Kaneko}, {Bostanc{\i}}, {G{\"o}{\u{g}}{\"u}{\c{s}}}, \& {Lin}}]{2015MNRAS.452..824K}
{Kaneko}, Y., {Bostanc{\i}}, Z.~F., {G{\"o}{\u{g}}{\"u}{\c{s}}}, E., \& {Lin}, L. 2015, \mnras, 452, 824, \dodoi{10.1093/mnras/stv1286}

\bibitem[{{Kasen} {et~al.}(2013){Kasen}, {Badnell}, \& {Barnes}}]{2013ApJ...774...25K}
{Kasen}, D., {Badnell}, N.~R., \& {Barnes}, J. 2013, \apj, 774, 25, \dodoi{10.1088/0004-637X/774/1/25}

\bibitem[{{Kasliwal} {et~al.}(2017){Kasliwal}, {Nakar}, {Singer}, {Kaplan}, {Cook}, {Van Sistine}, {Lau}, {Fremling}, {Gottlieb}, {Jencson}, {Adams}, {Feindt}, {Hotokezaka}, {Ghosh}, {Perley}, {Yu}, {Piran}, {Allison}, {Anupama}, {Balasubramanian}, {Bannister}, {Bally}, {Barnes}, {Barway}, {Bellm}, {Bhalerao}, {Bhattacharya}, {Blagorodnova}, {Bloom}, {Brady}, {Cannella}, {Chatterjee}, {Cenko}, {Cobb}, {Copperwheat}, {Corsi}, {De}, {Dobie}, {Emery}, {Evans}, {Fox}, {Frail}, {Frohmaier}, {Goobar}, {Hallinan}, {Harrison}, {Helou}, {Hinderer}, {Ho}, {Horesh}, {Ip}, {Itoh}, {Kasen}, {Kim}, {Kuin}, {Kupfer}, {Lynch}, {Madsen}, {Mazzali}, {Miller}, {Mooley}, {Murphy}, {Ngeow}, {Nichols}, {Nissanke}, {Nugent}, {Ofek}, {Qi}, {Quimby}, {Rosswog}, {Rusu}, {Sadler}, {Schmidt}, {Sollerman}, {Steele}, {Williamson}, {Xu}, {Yan}, {Yatsu}, {Zhang}, \& {Zhao}}]{2017Sci...358.1559K}
{Kasliwal}, M.~M., {Nakar}, E., {Singer}, L.~P., {et~al.} 2017, Science, 358, 1559, \dodoi{10.1126/science.aap9455}

\bibitem[{{Kawaguchi} {et~al.}(2018){Kawaguchi}, {Shibata}, \& {Tanaka}}]{2018ApJ...865L..21K}
{Kawaguchi}, K., {Shibata}, M., \& {Tanaka}, M. 2018, \apjl, 865, L21, \dodoi{10.3847/2041-8213/aade02}

\bibitem[{Kilpatrick {et~al.}(2017)}]{Kilpatrick:2017mhz}
Kilpatrick, C.~D., {et~al.} 2017, Science, 358, 1583, \dodoi{10.1126/science.aaq0073}

\bibitem[{{Kisaka} \& {Ioka}(2015)}]{2015ApJ...804L..16K}
{Kisaka}, S., \& {Ioka}, K. 2015, \apjl, 804, L16, \dodoi{10.1088/2041-8205/804/1/L16}

\bibitem[{{Kisaka} {et~al.}(2016){Kisaka}, {Ioka}, \& {Nakar}}]{2016ApJ...818..104K}
{Kisaka}, S., {Ioka}, K., \& {Nakar}, E. 2016, \apj, 818, 104, \dodoi{10.3847/0004-637X/818/2/104}

\bibitem[{{Kisaka} {et~al.}(2017){Kisaka}, {Ioka}, \& {Sakamoto}}]{2017ApJ...846..142K}
{Kisaka}, S., {Ioka}, K., \& {Sakamoto}, T. 2017, \apj, 846, 142, \dodoi{10.3847/1538-4357/aa8775}

\bibitem[{{Kisaka} {et~al.}(2015){Kisaka}, {Ioka}, \& {Takami}}]{2015ApJ...802..119K}
{Kisaka}, S., {Ioka}, K., \& {Takami}, H. 2015, \apj, 802, 119, \dodoi{10.1088/0004-637X/802/2/119}

\bibitem[{{Kiuchi} {et~al.}(2019){Kiuchi}, {Kyutoku}, {Shibata}, \& {Taniguchi}}]{2019ApJ...876L..31K}
{Kiuchi}, K., {Kyutoku}, K., {Shibata}, M., \& {Taniguchi}, K. 2019, \apjl, 876, L31, \dodoi{10.3847/2041-8213/ab1e45}

\bibitem[{{Kouveliotou} {et~al.}(1993){Kouveliotou}, {Meegan}, {Fishman}, {Bhat}, {Briggs}, {Koshut}, {Paciesas}, \& {Pendleton}}]{1993ApJ...413L.101K}
{Kouveliotou}, C., {Meegan}, C.~A., {Fishman}, G.~J., {et~al.} 1993, \apjl, 413, L101, \dodoi{10.1086/186969}

\bibitem[{{Kulkarni}(2005)}]{2005astro.ph.10256K}
{Kulkarni}, S.~R. 2005, arXiv e-prints, astro.
\newblock \doarXiv{astro-ph/0510256}

\bibitem[{{Kulkarni} {et~al.}(2021){Kulkarni}, {Harrison}, {Grefenstette}, {Earnshaw}, {Andreoni}, {Berg}, {Bloom}, {Cenko}, {Chornock}, {Christiansen}, {Coughlin}, {Wuollet Criswell}, {Darvish}, {Das}, {De}, {Dessart}, {Dixon}, {Dorsman}, {El-Badry}, {Evans}, {Ford}, {Fremling}, {Gansicke}, {Gezari}, {Goetberg}, {Green}, {Graham}, {Heida}, {Ho}, {Jaodand}, {Johns-Krull}, {Kasliwal}, {Lazzarini}, {Lu}, {Margutti}, {Martin}, {Masters}, {McKernan}, {Naze}, {Nissanke}, {Parazin}, {Perley}, {Phinney}, {Piro}, {Raaijmakers}, {Rauw}, {Rodriguez}, {Sana}, {Senchyna}, {Singer}, {Spake}, {Stassun}, {Stern}, {Teplitz}, {Weisz}, \& {Yao}}]{2021arXiv211115608K}
{Kulkarni}, S.~R., {Harrison}, F.~A., {Grefenstette}, B.~W., {et~al.} 2021, arXiv e-prints, arXiv:2111.15608, \dodoi{10.48550/arXiv.2111.15608}

\bibitem[{{Lamb} \& {Kobayashi}(2016)}]{2016ApJ...829..112L}
{Lamb}, G.~P., \& {Kobayashi}, S. 2016, \apj, 829, 112, \dodoi{10.3847/0004-637X/829/2/112}

\bibitem[{{Lamb} \& {Kobayashi}(2017)}]{2017MNRAS.472.4953L}
---. 2017, \mnras, 472, 4953, \dodoi{10.1093/mnras/stx2345}

\bibitem[{{Lamb} \& {Kobayashi}(2018)}]{2018MNRAS.478..733L}
---. 2018, \mnras, 478, 733, \dodoi{10.1093/mnras/sty1108}

\bibitem[{{Lamb} {et~al.}(2022){Lamb}, {Nativi}, {Rosswog}, {Kann}, {Levan}, {Lundman}, \& {Tanvir}}]{2022Univ....8..612L}
{Lamb}, G.~P., {Nativi}, L., {Rosswog}, S., {et~al.} 2022, Universe, 8, 612, \dodoi{10.3390/universe8120612}

\bibitem[{{Lamb} {et~al.}(2019){Lamb}, {Tanvir}, {Levan}, {de Ugarte Postigo}, {Kawaguchi}, {Corsi}, {Evans}, {Gompertz}, {Malesani}, {Page}, {Wiersema}, {Rosswog}, {Shibata}, {Tanaka}, {van der Horst}, {Cano}, {Fynbo}, {Fruchter}, {Greiner}, {Heintz}, {Higgins}, {Hjorth}, {Izzo}, {Jakobsson}, {Kann}, {O'Brien}, {Perley}, {Pian}, {Pugliese}, {Starling}, {Th{\"o}ne}, {Watson}, {Wijers}, \& {Xu}}]{2019ApJ...883...48L}
{Lamb}, G.~P., {Tanvir}, N.~R., {Levan}, A.~J., {et~al.} 2019, \apj, 883, 48, \dodoi{10.3847/1538-4357/ab38bb}

\bibitem[{{Laskar} {et~al.}(2022){Laskar}, {Escorial}, {Schroeder}, {Fong}, {Berger}, {Veres}, {Bhandari}, {Rastinejad}, {Kilpatrick}, {Tohuvavohu}, {Margutti}, {Alexander}, {DeLaunay}, {Kennea}, {Nugent}, {Paterson}, \& {Williams}}]{2022ApJ...935L..11L}
{Laskar}, T., {Escorial}, A.~R., {Schroeder}, G., {et~al.} 2022, \apjl, 935, L11, \dodoi{10.3847/2041-8213/ac8421}

\bibitem[{{Lazzati} {et~al.}(2017){Lazzati}, {L{\'o}pez-C{\'a}mara}, {Cantiello}, {Morsony}, {Perna}, \& {Workman}}]{2017ApJ...848L...6L}
{Lazzati}, D., {L{\'o}pez-C{\'a}mara}, D., {Cantiello}, M., {et~al.} 2017, \apjl, 848, L6, \dodoi{10.3847/2041-8213/aa8f3d}

\bibitem[{{Lazzati} {et~al.}(2018){Lazzati}, {Perna}, {Morsony}, {Lopez-Camara}, {Cantiello}, {Ciolfi}, {Giacomazzo}, \& {Workman}}]{2018PhRvL.120x1103L}
{Lazzati}, D., {Perna}, R., {Morsony}, B.~J., {et~al.} 2018, \prl, 120, 241103, \dodoi{10.1103/PhysRevLett.120.241103}

\bibitem[{{Levan} {et~al.}(2023){Levan}, {Gompertz}, {Salafia}, {Bulla}, {Burns}, {Hotokezaka}, {Izzo}, {Lamb}, {Malesani}, {Oates}, {Ravasio}, {Rouco Escorial}, {Schneider}, {Sarin}, {Schulze}, {Tanvir}, {Ackley}, {Anderson}, {Brammer}, {Christensen}, {Dhillon}, {Evans}, {Fausnaugh}, {Fong}, {Fruchter}, {Fryer}, {Fynbo}, {Gaspari}, {Heintz}, {Hjorth}, {Kennea}, {Kennedy}, {Laskar}, {Leloudas}, {Mandel}, {Martin-Carrillo}, {Metzger}, {Nicholl}, {Nugent}, {Palmerio}, {Pugliese}, {Rastinejad}, {Rhodes}, {Rossi}, {Smartt}, {Stevance}, {Tohuvavohu}, {van der Horst}, {Vergani}, {Watson}, {Barclay}, {Bhirombhakdi}, {Breedt}, {Breeveld}, {Brown}, {Campana}, {Chrimes}, {D'Avanzo}, {D'Elia}, {De Pasquale}, {Dyer}, {Galloway}, {Garbutt}, {Green}, {Hartmann}, {Jakobsson}, {Kerry}, {Langeroodi}, {Leung}, {Littlefair}, {Munday}, {O'Brien}, {Parsons}, {Pelisoli}, {Saccardi}, {Sahman}, {Salvaterra}, {Sbarufatti}, {Steeghs}, {Tagliaferri}, {Th{\"o}ne}, {de Ugarte Postigo}, \& {Kann}}]{2023arXiv230702098L}
{Levan}, A., {Gompertz}, B.~P., {Salafia}, O.~S., {et~al.} 2023, arXiv e-prints, arXiv:2307.02098, \dodoi{10.48550/arXiv.2307.02098}

\bibitem[{{Li} \& {Paczy{\'n}ski}(1998)}]{1998ApJ...507L..59L}
{Li}, L.-X., \& {Paczy{\'n}ski}, B. 1998, \apjl, 507, L59, \dodoi{10.1086/311680}

\bibitem[{{Liang} {et~al.}(2007){Liang}, {Zhang}, \& {Zhang}}]{2007ApJ...670..565L}
{Liang}, E.-W., {Zhang}, B.-B., \& {Zhang}, B. 2007, \apj, 670, 565, \dodoi{10.1086/521870}

\bibitem[{{Lien} {et~al.}(2016){Lien}, {Sakamoto}, {Barthelmy}, {Baumgartner}, {Cannizzo}, {Chen}, {Collins}, {Cummings}, {Gehrels}, {Krimm}, {Markwardt}, {Palmer}, {Stamatikos}, {Troja}, \& {Ukwatta}}]{2016ApJ...829....7L}
{Lien}, A., {Sakamoto}, T., {Barthelmy}, S.~D., {et~al.} 2016, \apj, 829, 7, \dodoi{10.3847/0004-637X/829/1/7}

\bibitem[{{Lyman} {et~al.}(2018){Lyman}, {Lamb}, {Levan}, {Mandel}, {Tanvir}, {Kobayashi}, {Gompertz}, {Hjorth}, {Fruchter}, {Kangas}, {Steeghs}, {Steele}, {Cano}, {Copperwheat}, {Evans}, {Fynbo}, {Gall}, {Im}, {Izzo}, {Jakobsson}, {Milvang-Jensen}, {O'Brien}, {Osborne}, {Palazzi}, {Perley}, {Pian}, {Rosswog}, {Rowlinson}, {Schulze}, {Stanway}, {Sutton}, {Th{\"o}ne}, {de Ugarte Postigo}, {Watson}, {Wiersema}, \& {Wijers}}]{2018NatAs...2..751L}
{Lyman}, J.~D., {Lamb}, G.~P., {Levan}, A.~J., {et~al.} 2018, Nature Astronomy, 2, 751, \dodoi{10.1038/s41550-018-0511-3}

\bibitem[{{MacFadyen} \& {Woosley}(1999)}]{1999ApJ...524..262M}
{MacFadyen}, A.~I., \& {Woosley}, S.~E. 1999, \apj, 524, 262, \dodoi{10.1086/307790}

\bibitem[{{Maggiore} {et~al.}(2020){Maggiore}, {Van Den Broeck}, {Bartolo}, {Belgacem}, {Bertacca}, {Bizouard}, {Branchesi}, {Clesse}, {Foffa}, {Garc{\'\i}a-Bellido}, {Grimm}, {Harms}, {Hinderer}, {Matarrese}, {Palomba}, {Peloso}, {Ricciardone}, \& {Sakellariadou}}]{2020JCAP...03..050M}
{Maggiore}, M., {Van Den Broeck}, C., {Bartolo}, N., {et~al.} 2020, \jcap, 2020, 050, \dodoi{10.1088/1475-7516/2020/03/050}

\bibitem[{{Mandel} \& {Broekgaarden}(2022)}]{2022LRR....25....1M}
{Mandel}, I., \& {Broekgaarden}, F.~S. 2022, Living Reviews in Relativity, 25, 1, \dodoi{10.1007/s41114-021-00034-3}

\bibitem[{{Margutti} {et~al.}(2017){Margutti}, {Berger}, {Fong}, {Guidorzi}, {Alexander}, {Metzger}, {Blanchard}, {Cowperthwaite}, {Chornock}, {Eftekhari}, {Nicholl}, {Villar}, {Williams}, {Annis}, {Brown}, {Chen}, {Doctor}, {Frieman}, {Holz}, {Sako}, \& {Soares-Santos}}]{2017ApJ...848L..20M}
{Margutti}, R., {Berger}, E., {Fong}, W., {et~al.} 2017, \apjl, 848, L20, \dodoi{10.3847/2041-8213/aa9057}

\bibitem[{{Mart{\'{\i}}} {et~al.}(1997){Mart{\'{\i}}}, {M{\"u}ller}, {Font}, {Ib{\'a}{\~n}ez}, \& {Marquina}}]{1997ApJ...479..151M}
{Mart{\'{\i}}}, J.~M., {M{\"u}ller}, E., {Font}, J.~A., {Ib{\'a}{\~n}ez}, J.~M.~Z., \& {Marquina}, A. 1997, \apj, 479, 151, \dodoi{10.1086/303842}

\bibitem[{{Matsumoto} {et~al.}(2018){Matsumoto}, {Ioka}, {Kisaka}, \& {Nakar}}]{2018ApJ...861...55M}
{Matsumoto}, T., {Ioka}, K., {Kisaka}, S., \& {Nakar}, E. 2018, \apj, 861, 55, \dodoi{10.3847/1538-4357/aac4a8}

\bibitem[{{Matsumoto} \& {Kimura}(2018)}]{2018ApJ...866L..16M}
{Matsumoto}, T., \& {Kimura}, S.~S. 2018, \apjl, 866, L16, \dodoi{10.3847/2041-8213/aae51b}

\bibitem[{{Matsumoto} {et~al.}(2020){Matsumoto}, {Kimura}, {Murase}, \& {M{\'e}sz{\'a}ros}}]{2020MNRAS.493..783M}
{Matsumoto}, T., {Kimura}, S.~S., {Murase}, K., \& {M{\'e}sz{\'a}ros}, P. 2020, \mnras, 493, 783, \dodoi{10.1093/mnras/staa305}

\bibitem[{{Matsuoka} {et~al.}(2009){Matsuoka}, {Kawasaki}, {Ueno}, {Tomida}, {Kohama}, {Suzuki}, {Adachi}, {Ishikawa}, {Mihara}, {Sugizaki}, {Isobe}, {Nakagawa}, {Tsunemi}, {Miyata}, {Kawai}, {Kataoka}, {Morii}, {Yoshida}, {Negoro}, {Nakajima}, {Ueda}, {Chujo}, {Yamaoka}, {Yamazaki}, {Nakahira}, {You}, {Ishiwata}, {Miyoshi}, {Eguchi}, {Hiroi}, {Katayama}, \& {Ebisawa}}]{2009PASJ...61..999M}
{Matsuoka}, M., {Kawasaki}, K., {Ueno}, S., {et~al.} 2009, \pasj, 61, 999, \dodoi{10.1093/pasj/61.5.999}

\bibitem[{{Matzner}(2003)}]{2003MNRAS.345..575M}
{Matzner}, C.~D. 2003, \mnras, 345, 575, \dodoi{10.1046/j.1365-8711.2003.06969.x}

\bibitem[{{Mei} {et~al.}(2022){Mei}, {Banerjee}, {Oganesyan}, {Salafia}, {Giarratana}, {Branchesi}, {D'Avanzo}, {Campana}, {Ghirlanda}, {Ronchini}, {Shukla}, \& {Tiwari}}]{2022Natur.612..236M}
{Mei}, A., {Banerjee}, B., {Oganesyan}, G., {et~al.} 2022, \nat, 612, 236, \dodoi{10.1038/s41586-022-05404-7}

\bibitem[{{Metzger}(2019)}]{2019LRR....23....1M}
{Metzger}, B.~D. 2019, Living Reviews in Relativity, 23, 1, \dodoi{10.1007/s41114-019-0024-0}

\bibitem[{{Metzger} \& {Fern{\'a}ndez}(2014)}]{2014MNRAS.441.3444M}
{Metzger}, B.~D., \& {Fern{\'a}ndez}, R. 2014, \mnras, 441, 3444, \dodoi{10.1093/mnras/stu802}

\bibitem[{{Metzger} {et~al.}(2018){Metzger}, {Thompson}, \& {Quataert}}]{2018ApJ...856..101M}
{Metzger}, B.~D., {Thompson}, T.~A., \& {Quataert}, E. 2018, \apj, 856, 101, \dodoi{10.3847/1538-4357/aab095}

\bibitem[{{Metzger} {et~al.}(2010){Metzger}, {Mart{\'\i}nez-Pinedo}, {Darbha}, {Quataert}, {Arcones}, {Kasen}, {Thomas}, {Nugent}, {Panov}, \& {Zinner}}]{2010MNRAS.406.2650M}
{Metzger}, B.~D., {Mart{\'\i}nez-Pinedo}, G., {Darbha}, S., {et~al.} 2010, \mnras, 406, 2650, \dodoi{10.1111/j.1365-2966.2010.16864.x}

\bibitem[{{Mizuta} \& {Ioka}(2013)}]{2013ApJ...777..162M}
{Mizuta}, A., \& {Ioka}, K. 2013, \apj, 777, 162, \dodoi{10.1088/0004-637X/777/2/162}

\bibitem[{{Mooley} {et~al.}(2018){Mooley}, {Deller}, {Gottlieb}, {Nakar}, {Hallinan}, {Bourke}, {Frail}, {Horesh}, {Corsi}, \& {Hotokezaka}}]{2018Natur.561..355M}
{Mooley}, K.~P., {Deller}, A.~T., {Gottlieb}, O., {et~al.} 2018, \nat, 561, 355, \dodoi{10.1038/s41586-018-0486-3}

\bibitem[{{Morokuma} {et~al.}(2015){Morokuma}, {Tominaga}, {Tanaka}, {Sarugaku}, \& {Kawai}}]{2015ATel.7960....1M}
{Morokuma}, T., {Tominaga}, N., {Tanaka}, M., {Sarugaku}, Y., \& {Kawai}, N. 2015, The Astronomer's Telegram, 7960, 1

\bibitem[{{Murguia-Berthier} {et~al.}(2014){Murguia-Berthier}, {Montes}, {Ramirez-Ruiz}, {De Colle}, \& {Lee}}]{2014ApJ...788L...8M}
{Murguia-Berthier}, A., {Montes}, G., {Ramirez-Ruiz}, E., {De Colle}, F., \& {Lee}, W.~H. 2014, \apjl, 788, L8, \dodoi{10.1088/2041-8205/788/1/L8}

\bibitem[{{Nagakura} {et~al.}(2014){Nagakura}, {Hotokezaka}, {Sekiguchi}, {Shibata}, \& {Ioka}}]{2014ApJ...784L..28N}
{Nagakura}, H., {Hotokezaka}, K., {Sekiguchi}, Y., {Shibata}, M., \& {Ioka}, K. 2014, \apjl, 784, L28, \dodoi{10.1088/2041-8205/784/2/L28}

\bibitem[{{Nakar} \& {Piran}(2017)}]{2017ApJ...834...28N}
{Nakar}, E., \& {Piran}, T. 2017, \apj, 834, 28, \dodoi{10.3847/1538-4357/834/1/28}

\bibitem[{{Nakar} \& {Sari}(2010)}]{2010ApJ...725..904N}
{Nakar}, E., \& {Sari}, R. 2010, \apj, 725, 904, \dodoi{10.1088/0004-637X/725/1/904}

\bibitem[{{Nakar} \& {Sari}(2012)}]{2012ApJ...747...88N}
---. 2012, \apj, 747, 88, \dodoi{10.1088/0004-637X/747/2/88}

\bibitem[{{Nathanail} {et~al.}(2020){Nathanail}, {Gill}, {Porth}, {Fromm}, \& {Rezzolla}}]{2020MNRAS.495.3780N}
{Nathanail}, A., {Gill}, R., {Porth}, O., {Fromm}, C.~M., \& {Rezzolla}, L. 2020, \mnras, 495, 3780, \dodoi{10.1093/mnras/staa1454}

\bibitem[{{Nathanail} {et~al.}(2021){Nathanail}, {Gill}, {Porth}, {Fromm}, \& {Rezzolla}}]{2021MNRAS.502.1843N}
---. 2021, \mnras, 502, 1843, \dodoi{10.1093/mnras/stab115}

\bibitem[{{Nativi} {et~al.}(2021){Nativi}, {Bulla}, {Rosswog}, {Lundman}, {Kowal}, {Gizzi}, {Lamb}, \& {Perego}}]{2021MNRAS.500.1772N}
{Nativi}, L., {Bulla}, M., {Rosswog}, S., {et~al.} 2021, \mnras, 500, 1772, \dodoi{10.1093/mnras/staa3337}

\bibitem[{{Nicholl} {et~al.}(2017){Nicholl}, {Berger}, {Kasen}, {Metzger}, {Elias}, {Brice{\~n}o}, {Alexander}, {Blanchard}, {Chornock}, {Cowperthwaite}, {Eftekhari}, {Fong}, {Margutti}, {Villar}, {Williams}, {Brown}, {Annis}, {Bahramian}, {Brout}, {Brown}, {Chen}, {Clemens}, {Dennihy}, {Dunlap}, {Holz}, {Marchesini}, {Massaro}, {Moskowitz}, {Pelisoli}, {Rest}, {Ricci}, {Sako}, {Soares-Santos}, \& {Strader}}]{2017ApJ...848L..18N}
{Nicholl}, M., {Berger}, E., {Kasen}, D., {et~al.} 2017, \apjl, 848, L18, \dodoi{10.3847/2041-8213/aa9029}

\bibitem[{{Norris} \& {Bonnell}(2006)}]{2006ApJ...643..266N}
{Norris}, J.~P., \& {Bonnell}, J.~T. 2006, \apj, 643, 266, \dodoi{10.1086/502796}

\bibitem[{{Nousek} {et~al.}(2006){Nousek}, {Kouveliotou}, {Grupe}, {Page}, {Granot}, {Ramirez-Ruiz}, {Patel}, {Burrows}, {Mangano}, {Barthelmy}, {Beardmore}, {Campana}, {Capalbi}, {Chincarini}, {Cusumano}, {Falcone}, {Gehrels}, {Giommi}, {Goad}, {Godet}, {Hurkett}, {Kennea}, {Moretti}, {O'Brien}, {Osborne}, {Romano}, {Tagliaferri}, \& {Wells}}]{2006ApJ...642..389N}
{Nousek}, J.~A., {Kouveliotou}, C., {Grupe}, D., {et~al.} 2006, \apj, 642, 389, \dodoi{10.1086/500724}

\bibitem[{{Oganesyan} {et~al.}(2020){Oganesyan}, {Ascenzi}, {Branchesi}, {Salafia}, {Dall'Osso}, \& {Ghirlanda}}]{2020ApJ...893...88O}
{Oganesyan}, G., {Ascenzi}, S., {Branchesi}, M., {et~al.} 2020, \apj, 893, 88, \dodoi{10.3847/1538-4357/ab8221}

\bibitem[{{Oshikiri} {et~al.}(2023){Oshikiri}, {Tanaka}, {Tominaga}, {Morokuma}, {Takahashi}, {Tampo}, {Hamidani}, {Arima}, {Arimatsu}, {Kasuga}, {Kobayashi}, {Kondo}, {Mori}, {Niino}, {Ohsawa}, {Okumura}, {Sako}, \& {Takahashi}}]{2023arXiv231010066O}
{Oshikiri}, K., {Tanaka}, M., {Tominaga}, N., {et~al.} 2023, arXiv e-prints, arXiv:2310.10066, \dodoi{10.48550/arXiv.2310.10066}

\bibitem[{{Paczynski}(1986)}]{1986ApJ...308L..43P}
{Paczynski}, B. 1986, \apjl, 308, L43, \dodoi{10.1086/184740}

\bibitem[{{Panaitescu} {et~al.}(2006){Panaitescu}, {M{\'e}sz{\'a}ros}, {Gehrels}, {Burrows}, \& {Nousek}}]{2006MNRAS.366.1357P}
{Panaitescu}, A., {M{\'e}sz{\'a}ros}, P., {Gehrels}, N., {Burrows}, D., \& {Nousek}, J. 2006, \mnras, 366, 1357, \dodoi{10.1111/j.1365-2966.2005.09900.x}

\bibitem[{Pian {et~al.}(2017)}]{Pian:2017gtc}
Pian, E., {et~al.} 2017, Nature, 551, 67, \dodoi{10.1038/nature24298}

\bibitem[{{Piro} \& {Kollmeier}(2018)}]{2018ApJ...855..103P}
{Piro}, A.~L., \& {Kollmeier}, J.~A. 2018, \apj, 855, 103, \dodoi{10.3847/1538-4357/aaaab3}

\bibitem[{{Punturo} {et~al.}(2010{\natexlab{a}}){Punturo}, {Abernathy}, {Acernese}, {Allen}, {Andersson}, {Arun}, {Barone}, {Barr}, {Barsuglia}, {Beker}, {Beveridge}, {Birindelli}, {Bose}, {Bosi}, {Braccini}, {Bradaschia}, {Bulik}, {Calloni}, {Cella}, {Chassande Mottin}, {Chelkowski}, {Chincarini}, {Clark}, {Coccia}, {Colacino}, {Colas}, {Cumming}, {Cunningham}, {Cuoco}, {Danilishin}, {Danzmann}, {De Luca}, {De Salvo}, {Dent}, {Derosa}, {Di Fiore}, {Di Virgilio}, {Doets}, {Fafone}, {Falferi}, {Flaminio}, {Franc}, {Frasconi}, {Freise}, {Fulda}, {Gair}, {Gemme}, {Gennai}, {Giazotto}, {Glampedakis}, {Granata}, {Grote}, {Guidi}, {Hammond}, {Hannam}, {Harms}, {Heinert}, {Hendry}, {Heng}, {Hennes}, {Hild}, {Hough}, {Husa}, {Huttner}, {Jones}, {Khalili}, {Kokeyama}, {Kokkotas}, {Krishnan}, {Lorenzini}, {L{\"u}ck}, {Majorana}, {Mandel}, {Mandic}, {Martin}, {Michel}, {Minenkov}, {Morgado}, {Mosca}, {Mours}, {M{\"u}ller-Ebhardt}, {Murray}, {Nawrodt}, {Nelson}, {Oshaughnessy}, {Ott}, {Palomba}, {Paoli}, {Parguez},
  {Pasqualetti}, {Passaquieti}, {Passuello}, {Pinard}, {Poggiani}, {Popolizio}, {Prato}, {Puppo}, {Rabeling}, {Rapagnani}, {Read}, {Regimbau}, {Rehbein}, {Reid}, {Rezzolla}, {Ricci}, {Richard}, {Rocchi}, {Rowan}, {R{\"u}diger}, {Sassolas}, {Sathyaprakash}, {Schnabel}, {Schwarz}, {Seidel}, {Sintes}, {Somiya}, {Speirits}, {Strain}, {Strigin}, {Sutton}, {Tarabrin}, {van den Brand}, {van Leewen}, {van Veggel}, {van den Broeck}, {Vecchio}, {Veitch}, {Vetrano}, {Vicere}, {Vyatchanin}, {Willke}, {Woan}, {Wolfango}, \& {Yamamoto}}]{2010CQGra..27h4007P}
{Punturo}, M., {Abernathy}, M., {Acernese}, F., {et~al.} 2010{\natexlab{a}}, Classical and Quantum Gravity, 27, 084007, \dodoi{10.1088/0264-9381/27/8/084007}

\bibitem[{{Punturo} {et~al.}(2010{\natexlab{b}}){Punturo}, {Abernathy}, {Acernese}, {Allen}, {Andersson}, {Arun}, {Barone}, {Barr}, {Barsuglia}, {Beker}, {Beveridge}, {Birindelli}, {Bose}, {Bosi}, {Braccini}, {Bradaschia}, {Bulik}, {Calloni}, {Cella}, {Chassande Mottin}, {Chelkowski}, {Chincarini}, {Clark}, {Coccia}, {Colacino}, {Colas}, {Cumming}, {Cunningham}, {Cuoco}, {Danilishin}, {Danzmann}, {De Luca}, {De Salvo}, {Dent}, {De Rosa}, {Di Fiore}, {Di Virgilio}, {Doets}, {Fafone}, {Falferi}, {Flaminio}, {Franc}, {Frasconi}, {Freise}, {Fulda}, {Gair}, {Gemme}, {Gennai}, {Giazotto}, {Glampedakis}, {Granata}, {Grote}, {Guidi}, {Hammond}, {Hannam}, {Harms}, {Heinert}, {Hendry}, {Heng}, {Hennes}, {Hild}, {Hough}, {Husa}, {Huttner}, {Jones}, {Khalili}, {Kokeyama}, {Kokkotas}, {Krishnan}, {Lorenzini}, {L{\"u}ck}, {Majorana}, {Mandel}, {Mandic}, {Martin}, {Michel}, {Minenkov}, {Morgado}, {Mosca}, {Mours}, {M{\"u}ller{\textendash}Ebhardt}, {Murray}, {Nawrodt}, {Nelson}, {Oshaughnessy}, {Ott}, {Palomba}, {Paoli},
  {Parguez}, {Pasqualetti}, {Passaquieti}, {Passuello}, {Pinard}, {Poggiani}, {Popolizio}, {Prato}, {Puppo}, {Rabeling}, {Rapagnani}, {Read}, {Regimbau}, {Rehbein}, {Reid}, {Rezzolla}, {Ricci}, {Richard}, {Rocchi}, {Rowan}, {R{\"u}diger}, {Sassolas}, {Sathyaprakash}, {Schnabel}, {Schwarz}, {Seidel}, {Sintes}, {Somiya}, {Speirits}, {Strain}, {Strigin}, {Sutton}, {Tarabrin}, {Th{\"u}ring}, {van den Brand}, {van Leewen}, {van Veggel}, {van den Broeck}, {Vecchio}, {Veitch}, {Vetrano}, {Vicere}, {Vyatchanin}, {Willke}, {Woan}, {Wolfango}, \& {Yamamoto}}]{2010CQGra..27s4002P}
---. 2010{\natexlab{b}}, Classical and Quantum Gravity, 27, 194002, \dodoi{10.1088/0264-9381/27/19/194002}

\bibitem[{{Quirola-V{\'a}squez} {et~al.}(2022){Quirola-V{\'a}squez}, {Bauer}, {Jonker}, {Brandt}, {Yang}, {Levan}, {Xue}, {Eappachen}, {Zheng}, \& {Luo}}]{2022A&A...663A.168Q}
{Quirola-V{\'a}squez}, J., {Bauer}, F.~E., {Jonker}, P.~G., {et~al.} 2022, \aap, 663, A168, \dodoi{10.1051/0004-6361/202243047}

\bibitem[{{Quirola-V{\'a}squez} {et~al.}(2023){Quirola-V{\'a}squez}, {Bauer}, {Jonker}, {Brandt}, {Yang}, {Levan}, {Xue}, {Eappachen}, {Camacho}, {Ravasio}, {Zheng}, \& {Luo}}]{2023A&A...675A..44Q}
---. 2023, \aap, 675, A44, \dodoi{10.1051/0004-6361/202345912}

\bibitem[{{Radice} {et~al.}(2016){Radice}, {Galeazzi}, {Lippuner}, {Roberts}, {Ott}, \& {Rezzolla}}]{2016MNRAS.460.3255R}
{Radice}, D., {Galeazzi}, F., {Lippuner}, J., {et~al.} 2016, \mnras, 460, 3255, \dodoi{10.1093/mnras/stw1227}

\bibitem[{{Rastinejad} {et~al.}(2022){Rastinejad}, {Gompertz}, {Levan}, {Fong}, {Nicholl}, {Lamb}, {Malesani}, {Nugent}, {Oates}, {Tanvir}, {de Ugarte Postigo}, {Kilpatrick}, {Moore}, {Metzger}, {Ravasio}, {Rossi}, {Schroeder}, {Jencson}, {Sand}, {Smith}, {Ag{\"u}{\'\i} Fern{\'a}ndez}, {Berger}, {Blanchard}, {Chornock}, {Cobb}, {De Pasquale}, {Fynbo}, {Izzo}, {Kann}, {Laskar}, {Marini}, {Paterson}, {Escorial}, {Sears}, \& {Th{\"o}ne}}]{2022Natur.612..223R}
{Rastinejad}, J.~C., {Gompertz}, B.~P., {Levan}, A.~J., {et~al.} 2022, \nat, 612, 223, \dodoi{10.1038/s41586-022-05390-w}

\bibitem[{{Reitze} {et~al.}(2019){Reitze}, {Adhikari}, {Ballmer}, {Barish}, {Barsotti}, {Billingsley}, {Brown}, {Chen}, {Coyne}, {Eisenstein}, {Evans}, {Fritschel}, {Hall}, {Lazzarini}, {Lovelace}, {Read}, {Sathyaprakash}, {Shoemaker}, {Smith}, {Torrie}, {Vitale}, {Weiss}, {Wipf}, \& {Zucker}}]{2019BAAS...51g..35R}
{Reitze}, D., {Adhikari}, R.~X., {Ballmer}, S., {et~al.} 2019, in Bulletin of the American Astronomical Society, Vol.~51, 35, \dodoi{10.48550/arXiv.1907.04833}

\bibitem[{{Roming} {et~al.}(2005){Roming}, {Kennedy}, {Mason}, {Nousek}, {Ahr}, {Bingham}, {Broos}, {Carter}, {Hancock}, {Huckle}, {Hunsberger}, {Kawakami}, {Killough}, {Koch}, {McLelland}, {Smith}, {Smith}, {Soto}, {Boyd}, {Breeveld}, {Holland}, {Ivanushkina}, {Pryzby}, {Still}, \& {Stock}}]{2005SSRv..120...95R}
{Roming}, P. W.~A., {Kennedy}, T.~E., {Mason}, K.~O., {et~al.} 2005, \ssr, 120, 95, \dodoi{10.1007/s11214-005-5095-4}

\bibitem[{{Rossi} {et~al.}(2020){Rossi}, {Stratta}, {Maiorano}, {Spighi}, {Masetti}, {Palazzi}, {Gardini}, {Melandri}, {Nicastro}, {Pian}, {Branchesi}, {Dadina}, {Testa}, {Brocato}, {Benetti}, {Ciolfi}, {Covino}, {D'Elia}, {Grado}, {Izzo}, {Perego}, {Piranomonte}, {Salvaterra}, {Selsing}, {Tomasella}, {Yang}, {Vergani}, {Amati}, \& {Stephen}}]{2020MNRAS.493.3379R}
{Rossi}, A., {Stratta}, G., {Maiorano}, E., {et~al.} 2020, \mnras, 493, 3379, \dodoi{10.1093/mnras/staa479}

\bibitem[{{Rouco Escorial} {et~al.}(2023){Rouco Escorial}, {Fong}, {Berger}, {Laskar}, {Margutti}, {Schroeder}, {Rastinejad}, {Cornish}, {Popp}, {Lally}, {Nugent}, {Paterson}, {Metzger}, {Chornock}, {Alexander}, {Cendes}, \& {Eftekhari}}]{2023ApJ...959...13R}
{Rouco Escorial}, A., {Fong}, W., {Berger}, E., {et~al.} 2023, \apj, 959, 13, \dodoi{10.3847/1538-4357/acf830}

\bibitem[{{Ruderman}(1975)}]{1975NYASA.262..164R}
{Ruderman}, M. 1975, in Seventh Texas Symposium on Relativistic Astrophysics, ed. P.~G. {Bergman}, E.~J. {Fenyves}, \& L.~{Motz}, Vol. 262, 164--180, \dodoi{10.1111/j.1749-6632.1975.tb31430.x}

\bibitem[{{Rybicki} \& {Lightman}(1979)}]{1979rpa..book.....R}
{Rybicki}, G.~B., \& {Lightman}, A.~P. 1979, {Radiative processes in astrophysics}

\bibitem[{{Sagiv} {et~al.}(2014){Sagiv}, {Gal-Yam}, {Ofek}, {Waxman}, {Aharonson}, {Kulkarni}, {Nakar}, {Maoz}, {Trakhtenbrot}, {Phinney}, {Topaz}, {Beichman}, {Murthy}, \& {Worden}}]{2014AJ....147...79S}
{Sagiv}, I., {Gal-Yam}, A., {Ofek}, E.~O., {et~al.} 2014, \aj, 147, 79, \dodoi{10.1088/0004-6256/147/4/79}

\bibitem[{{Sakamoto} {et~al.}(2005){Sakamoto}, {Lamb}, {Kawai}, {Yoshida}, {Graziani}, {Fenimore}, {Donaghy}, {Matsuoka}, {Suzuki}, {Ricker}, {Atteia}, {Shirasaki}, {Tamagawa}, {Torii}, {Galassi}, {Doty}, {Vanderspek}, {Crew}, {Villasenor}, {Butler}, {Prigozhin}, {Jernigan}, {Barraud}, {Boer}, {Dezalay}, {Olive}, {Hurley}, {Levine}, {Monnelly}, {Martel}, {Morgan}, {Woosley}, {Cline}, {Braga}, {Manchanda}, {Pizzichini}, {Takagishi}, \& {Yamauchi}}]{2005ApJ...629..311S}
{Sakamoto}, T., {Lamb}, D.~Q., {Kawai}, N., {et~al.} 2005, \apj, 629, 311, \dodoi{10.1086/431235}

\bibitem[{{Sakamoto} {et~al.}(2008){Sakamoto}, {Hullinger}, {Sato}, {Yamazaki}, {Barbier}, {Barthelmy}, {Cummings}, {Fenimore}, {Gehrels}, {Krimm}, {Lamb}, {Markwardt}, {Osborne}, {Palmer}, {Parsons}, {Stamatikos}, \& {Tueller}}]{2008ApJ...679..570S}
{Sakamoto}, T., {Hullinger}, D., {Sato}, G., {et~al.} 2008, \apj, 679, 570, \dodoi{10.1086/586884}

\bibitem[{{Sari} {et~al.}(1998){Sari}, {Piran}, \& {Narayan}}]{1998ApJ...497L..17S}
{Sari}, R., {Piran}, T., \& {Narayan}, R. 1998, \apjl, 497, L17, \dodoi{10.1086/311269}

\bibitem[{{Schlegel} {et~al.}(1998){Schlegel}, {Finkbeiner}, \& {Davis}}]{1998ApJ...500..525S}
{Schlegel}, D.~J., {Finkbeiner}, D.~P., \& {Davis}, M. 1998, \apj, 500, 525, \dodoi{10.1086/305772}

\bibitem[{{Serino} {et~al.}(2014){Serino}, {Sakamoto}, {Kawai}, {Yoshida}, {Ohno}, {Ogawa}, {Nishimura}, {Fukushima}, {Higa}, {Ishikawa}, {Ishikawa}, {Kawamuro}, {Kimura}, {Matsuoka}, {Mihara}, {Morii}, {Nakagawa}, {Nakahira}, {Nakajima}, {Nakano}, {Negoro}, {Onodera}, {Sasaki}, {Shidatsu}, {Sugimoto}, {Sugizaki}, {Suwa}, {Suzuki}, {Tachibana}, {Takagi}, {Toizumi}, {Tomida}, {Tsuboi}, {Tsunemi}, {Ueda}, {Ueno}, {Usui}, {Yamada}, {Yamamoto}, {Yamaoka}, {Yamauchi}, {Yoshidome}, \& {Yoshii}}]{2014PASJ...66...87S}
{Serino}, M., {Sakamoto}, T., {Kawai}, N., {et~al.} 2014, \pasj, 66, 87, \dodoi{10.1093/pasj/psu063}

\bibitem[{Shappee {et~al.}(2017)}]{Shappee:2017zly}
Shappee, B.~J., {et~al.} 2017, Science, 358, 1574, \dodoi{10.1126/science.aaq0186}

\bibitem[{{Shibata}(1999)}]{1999PhRvD..60j4052S}
{Shibata}, M. 1999, \prd, 60, 104052, \dodoi{10.1103/PhysRevD.60.104052}

\bibitem[{{Shibata} {et~al.}(2017){Shibata}, {Fujibayashi}, {Hotokezaka}, {Kiuchi}, {Kyutoku}, {Sekiguchi}, \& {Tanaka}}]{2017PhRvD..96l3012S}
{Shibata}, M., {Fujibayashi}, S., {Hotokezaka}, K., {et~al.} 2017, \prd, 96, 123012, \dodoi{10.1103/PhysRevD.96.123012}

\bibitem[{{Shibata} \& {Hotokezaka}(2019)}]{2019ARNPS..69...41S}
{Shibata}, M., \& {Hotokezaka}, K. 2019, Annual Review of Nuclear and Particle Science, 69, 41, \dodoi{10.1146/annurev-nucl-101918-023625}

\bibitem[{{Shibata} \& {Ury{\= u}}(2000)}]{2000PhRvD..61f4001S}
{Shibata}, M., \& {Ury{\= u}}, K.~{\= o}. 2000, \prd, 61, 064001, \dodoi{10.1103/PhysRevD.61.064001}

\bibitem[{{Shrestha} {et~al.}(2023){Shrestha}, {Bulla}, {Nativi}, {Markin}, {Rosswog}, \& {Dietrich}}]{2023MNRAS.523.2990S}
{Shrestha}, M., {Bulla}, M., {Nativi}, L., {et~al.} 2023, \mnras, 523, 2990, \dodoi{10.1093/mnras/stad1583}

\bibitem[{{Smartt} {et~al.}(2017){Smartt}, {Chen}, {Jerkstrand}, {Coughlin}, {Kankare}, {Sim}, {Fraser}, {Inserra}, {Maguire}, {Chambers}, {Huber}, {Kr{\"u}hler}, {Leloudas}, {Magee}, {Shingles}, {Smith}, {Young}, {Tonry}, {Kotak}, {Gal-Yam}, {Lyman}, {Homan}, {Agliozzo}, {Anderson}, {Angus}, {Ashall}, {Barbarino}, {Bauer}, {Berton}, {Botticella}, {Bulla}, {Bulger}, {Cannizzaro}, {Cano}, {Cartier}, {Cikota}, {Clark}, {De Cia}, {Della Valle}, {Denneau}, {Dennefeld}, {Dessart}, {Dimitriadis}, {Elias-Rosa}, {Firth}, {Flewelling}, {Fl{\"o}rs}, {Franckowiak}, {Frohmaier}, {Galbany}, {Gonz{\'a}lez-Gait{\'a}n}, {Greiner}, {Gromadzki}, {Guelbenzu}, {Guti{\'e}rrez}, {Hamanowicz}, {Hanlon}, {Harmanen}, {Heintz}, {Heinze}, {Hernandez}, {Hodgkin}, {Hook}, {Izzo}, {James}, {Jonker}, {Kerzendorf}, {Klose}, {Kostrzewa-Rutkowska}, {Kowalski}, {Kromer}, {Kuncarayakti}, {Lawrence}, {Lowe}, {Magnier}, {Manulis}, {Martin-Carrillo}, {Mattila}, {McBrien}, {M{\"u}ller}, {Nordin}, {O'Neill}, {Onori}, {Palmerio}, {Pastorello},
  {Patat}, {Pignata}, {Podsiadlowski}, {Pumo}, {Prentice}, {Rau}, {Razza}, {Rest}, {Reynolds}, {Roy}, {Ruiter}, {Rybicki}, {Salmon}, {Schady}, {Schultz}, {Schweyer}, {Seitenzahl}, {Smith}, {Sollerman}, {Stalder}, {Stubbs}, {Sullivan}, {Szegedi}, {Taddia}, {Taubenberger}, {Terreran}, {van Soelen}, {Vos}, {Wainscoat}, {Walton}, {Waters}, {Weiland}, {Willman}, {Wiseman}, {Wright}, {Wyrzykowski}, \& {Yaron}}]{2017Natur.551...75S}
{Smartt}, S.~J., {Chen}, T.~W., {Jerkstrand}, A., {et~al.} 2017, \nat, 551, 75, \dodoi{10.1038/nature24303}

\bibitem[{{Soares-Santos} {et~al.}(2017){Soares-Santos}, {Holz}, {Annis}, {Chornock}, {Herner}, {Berger}, {Brout}, {Chen}, {Kessler}, {Sako}, {Allam}, {Tucker}, {Butler}, {Palmese}, {Doctor}, {Diehl}, {Frieman}, {Yanny}, {Lin}, {Scolnic}, {Cowperthwaite}, {Neilsen}, {Marriner}, {Kuropatkin}, {Hartley}, {Paz-Chinch{\'o}n}, {Alexander}, {Balbinot}, {Blanchard}, {Brown}, {Carlin}, {Conselice}, {Cook}, {Drlica-Wagner}, {Drout}, {Durret}, {Eftekhari}, {Farr}, {Finley}, {Foley}, {Fong}, {Fryer}, {Garc{\'\i}a-Bellido}, {Gill}, {Gruendl}, {Hanna}, {Kasen}, {Li}, {Lopes}, {Louren{\c{c}}o}, {Margutti}, {Marshall}, {Matheson}, {Medina}, {Metzger}, {Mu{\~n}oz}, {Muir}, {Nicholl}, {Quataert}, {Rest}, {Sauseda}, {Schlegel}, {Secco}, {Sobreira}, {Stebbins}, {Villar}, {Vivas}, {Walker}, {Wester}, {Williams}, {Zenteno}, {Zhang}, {Abbott}, {Abdalla}, {Banerji}, {Bechtol}, {Benoit-L{\'e}vy}, {Bertin}, {Brooks}, {Buckley-Geer}, {Burke}, {Carnero Rosell}, {Carrasco Kind}, {Carretero}, {Castander}, {Crocce}, {Cunha}, {D'Andrea},
  {da Costa}, {Davis}, {Desai}, {Dietrich}, {Doel}, {Eifler}, {Fernandez}, {Flaugher}, {Fosalba}, {Gaztanaga}, {Gerdes}, {Giannantonio}, {Goldstein}, {Gruen}, {Gschwend}, {Gutierrez}, {Honscheid}, {Jain}, {James}, {Jeltema}, {Johnson}, {Johnson}, {Kent}, {Krause}, {Kron}, {Kuehn}, {Kuhlmann}, {Lahav}, {Lima}, {Maia}, {March}, {McMahon}, {Menanteau}, {Miquel}, {Mohr}, {Nichol}, {Nord}, {Ogando}, {Petravick}, {Plazas}, {Romer}, {Roodman}, {Rykoff}, {Sanchez}, {Scarpine}, {Schubnell}, {Sevilla-Noarbe}, {Smith}, {Smith}, {Suchyta}, {Swanson}, {Tarle}, {Thomas}, {Thomas}, {Troxel}, {Vikram}, {Wechsler}, {Weller}, {Dark Energy Survey}, \& {Dark Energy Camera GW-EM Collaboration}}]{2017ApJ...848L..16S}
{Soares-Santos}, M., {Holz}, D.~E., {Annis}, J., {et~al.} 2017, \apjl, 848, L16, \dodoi{10.3847/2041-8213/aa9059}

\bibitem[{{Sun} {et~al.}(2023){Sun}, {Wang}, {Yang}, {Zhang}, {Xiong}, {Yin}, {Liu}, {Li}, {Xue}, {Yan}, {Zhang}, {Tan}, {Pan}, {Liu}, {Cheng}, {Zhang}, {Hu}, {Zheng}, {An}, {Cai}, {Hu}, {Jin}, {Li}, {Li}, {Liu}, {Liu}, {Peng}, {Song}, {Sun}, {Sun}, {Wang}, {Wen}, {Xiao}, {Yi}, {Zhang}, {Zhang}, {Zhang}, {Zhang}, {Zhao}, {Zheng}, {Ling}, {Zhang}, {Yuan}, \& {Zhang}}]{2023arXiv230705689S}
{Sun}, H., {Wang}, C.~W., {Yang}, J., {et~al.} 2023, arXiv e-prints, arXiv:2307.05689, \dodoi{10.48550/arXiv.2307.05689}

\bibitem[{{Tanaka} \& {Hotokezaka}(2013)}]{2013ApJ...775..113T}
{Tanaka}, M., \& {Hotokezaka}, K. 2013, \apj, 775, 113, \dodoi{10.1088/0004-637X/775/2/113}

\bibitem[{{Tanaka} {et~al.}(2020){Tanaka}, {Kato}, {Gaigalas}, \& {Kawaguchi}}]{2020MNRAS.496.1369T}
{Tanaka}, M., {Kato}, D., {Gaigalas}, G., \& {Kawaguchi}, K. 2020, \mnras, 496, 1369, \dodoi{10.1093/mnras/staa1576}

\bibitem[{{Tanaka} {et~al.}(2017){Tanaka}, {Utsumi}, {Mazzali}, {Tominaga}, {Yoshida}, {Sekiguchi}, {Morokuma}, {Motohara}, {Ohta}, {Kawabata}, {Abe}, {Aoki}, {Asakura}, {Baar}, {Barway}, {Bond}, {Doi}, {Fujiyoshi}, {Furusawa}, {Honda}, {Itoh}, {Kawabata}, {Kawai}, {Kim}, {Lee}, {Miyazaki}, {Morihana}, {Nagashima}, {Nagayama}, {Nakaoka}, {Nakata}, {Ohsawa}, {Ohshima}, {Okita}, {Saito}, {Sumi}, {Tajitsu}, {Takahashi}, {Takayama}, {Tamura}, {Tanaka}, {Terai}, {Tristram}, {Yasuda}, \& {Zenko}}]{2017PASJ...69..102T}
{Tanaka}, M., {Utsumi}, Y., {Mazzali}, P.~A., {et~al.} 2017, \pasj, 69, 102, \dodoi{10.1093/pasj/psx121}

\bibitem[{{Tanvir} {et~al.}(2013){Tanvir}, {Levan}, {Fruchter}, {Hjorth}, {Hounsell}, {Wiersema}, \& {Tunnicliffe}}]{2013Natur.500..547T}
{Tanvir}, N.~R., {Levan}, A.~J., {Fruchter}, A.~S., {et~al.} 2013, \nat, 500, 547, \dodoi{10.1038/nature12505}

\bibitem[{{Troja} {et~al.}(2007){Troja}, {Cusumano}, {O'Brien}, {Zhang}, {Sbarufatti}, {Mangano}, {Willingale}, {Chincarini}, {Osborne}, {Marshall}, {Burrows}, {Campana}, {Gehrels}, {Guidorzi}, {Krimm}, {La Parola}, {Liang}, {Mineo}, {Moretti}, {Page}, {Romano}, {Tagliaferri}, {Zhang}, {Page}, \& {Schady}}]{2007ApJ...665..599T}
{Troja}, E., {Cusumano}, G., {O'Brien}, P.~T., {et~al.} 2007, \apj, 665, 599, \dodoi{10.1086/519450}

\bibitem[{{Troja} {et~al.}(2018){Troja}, {Ryan}, {Piro}, {van Eerten}, {Cenko}, {Yoon}, {Lee}, {Im}, {Sakamoto}, {Gatkine}, {Kutyrev}, \& {Veilleux}}]{2018NatCo...9.4089T}
{Troja}, E., {Ryan}, G., {Piro}, L., {et~al.} 2018, Nature Communications, 9, 4089, \dodoi{10.1038/s41467-018-06558-7}

\bibitem[{{Troja} {et~al.}(2019){Troja}, {van Eerten}, {Ryan}, {Ricci}, {Burgess}, {Wieringa}, {Piro}, {Cenko}, \& {Sakamoto}}]{2019MNRAS.489.1919T}
{Troja}, E., {van Eerten}, H., {Ryan}, G., {et~al.} 2019, \mnras, 489, 1919, \dodoi{10.1093/mnras/stz2248}

\bibitem[{{Troja} {et~al.}(2022){Troja}, {Fryer}, {O'Connor}, {Ryan}, {Dichiara}, {Kumar}, {Ito}, {Gupta}, {Wollaeger}, {Norris}, {Kawai}, {Butler}, {Aryan}, {Misra}, {Hosokawa}, {Murata}, {Niwano}, {Pandey}, {Kutyrev}, {van Eerten}, {Chase}, {Hu}, {Caballero-Garcia}, \& {Castro-Tirado}}]{2022Natur.612..228T}
{Troja}, E., {Fryer}, C.~L., {O'Connor}, B., {et~al.} 2022, \nat, 612, 228, \dodoi{10.1038/s41586-022-05327-3}

\bibitem[{{Uchida} {et~al.}(2014){Uchida}, {Nakahira}, {Ueno}, {Tomida}, {Kimura}, {Ishikawa}, {Nakagawa}, {Mihara}, {Sugizaki}, {Morii}, {Sugimoto}, {Takagi}, {Yoshikawa}, {Matsuoka}, {Kawai}, {Yoshii}, {Tachibana}, {Yoshida}, {Sakamoto}, {Kawakubo}, {Ohtsuki}, {Tsunemi}, {Negoro}, {Nakajima}, {Fukushima}, {Onodera}, {Suzuki}, {Namba}, {Fujita}, {Honda}, {Ueda}, {Shidatsu}, {Kawamuro}, {Hori}, {Tsuboi}, {Kawagoe}, {Yamauchi}, {Morooka}, \& {Yamaoka}}]{2014GCN.16686....1U}
{Uchida}, D., {Nakahira}, S., {Ueno}, S., {et~al.} 2014, GRB Coordinates Network, 16686, 1

\bibitem[{{Urata} {et~al.}(2019){Urata}, {Toma}, {Huang}, {Asada}, {Nagai}, {Takahashi}, {Petitpas}, {Tashiro}, \& {Yamaoka}}]{2019ApJ...884L..58U}
{Urata}, Y., {Toma}, K., {Huang}, K., {et~al.} 2019, \apjl, 884, L58, \dodoi{10.3847/2041-8213/ab48f3}

\bibitem[{{Urrutia} {et~al.}(2023){Urrutia}, {De Colle}, \& {L{\'o}pez-C{\'a}mara}}]{2023MNRAS.518.5145U}
{Urrutia}, G., {De Colle}, F., \& {L{\'o}pez-C{\'a}mara}, D. 2023, \mnras, 518, 5145, \dodoi{10.1093/mnras/stac3401}

\bibitem[{{Utsumi} {et~al.}(2017){Utsumi}, {Tanaka}, {Tominaga}, {Yoshida}, {Barway}, {Nagayama}, {Zenko}, {Aoki}, {Fujiyoshi}, {Furusawa}, {Kawabata}, {Koshida}, {Lee}, {Morokuma}, {Motohara}, {Nakata}, {Ohsawa}, {Ohta}, {Okita}, {Tajitsu}, {Tanaka}, {Terai}, {Yasuda}, {Abe}, {Asakura}, {Bond}, {Miyazaki}, {Sumi}, {Tristram}, {Honda}, {Itoh}, {Itoh}, {Kawabata}, {Morihana}, {Nagashima}, {Nakaoka}, {Ohshima}, {Takahashi}, {Takayama}, {Aoki}, {Baar}, {Doi}, {Finet}, {Kanda}, {Kawai}, {Kim}, {Kuroda}, {Liu}, {Matsubayashi}, {Murata}, {Nagai}, {Saito}, {Saito}, {Sako}, {Sekiguchi}, {Tamura}, {Tanaka}, {Uemura}, \& {Yamaguchi}}]{2017PASJ...69..101U}
{Utsumi}, Y., {Tanaka}, M., {Tominaga}, N., {et~al.} 2017, \pasj, 69, 101, \dodoi{10.1093/pasj/psx118}

\bibitem[{{Valenti} {et~al.}(2017){Valenti}, {Sand}, {Yang}, {Cappellaro}, {Tartaglia}, {Corsi}, {Jha}, {Reichart}, {Haislip}, \& {Kouprianov}}]{2017ApJ...848L..24V}
{Valenti}, S., {Sand}, D.~J., {Yang}, S., {et~al.} 2017, \apjl, 848, L24, \dodoi{10.3847/2041-8213/aa8edf}

\bibitem[{{Villar} {et~al.}(2017){Villar}, {Guillochon}, {Berger}, {Metzger}, {Cowperthwaite}, {Nicholl}, {Alexander}, {Blanchard}, {Chornock}, {Eftekhari}, {Fong}, {Margutti}, \& {Williams}}]{2017ApJ...851L..21V}
{Villar}, V.~A., {Guillochon}, J., {Berger}, E., {et~al.} 2017, \apjl, 851, L21, \dodoi{10.3847/2041-8213/aa9c84}

\bibitem[{{Wanajo} {et~al.}(2014){Wanajo}, {Sekiguchi}, {Nishimura}, {Kiuchi}, {Kyutoku}, \& {Shibata}}]{2014ApJ...789L..39W}
{Wanajo}, S., {Sekiguchi}, Y., {Nishimura}, N., {et~al.} 2014, \apjl, 789, L39, \dodoi{10.1088/2041-8205/789/2/L39}

\bibitem[{{Waxman} {et~al.}(2018){Waxman}, {Ofek}, {Kushnir}, \& {Gal-Yam}}]{2018MNRAS.481.3423W}
{Waxman}, E., {Ofek}, E.~O., {Kushnir}, D., \& {Gal-Yam}, A. 2018, \mnras, 481, 3423, \dodoi{10.1093/mnras/sty2441}

\bibitem[{{Yamazaki} {et~al.}(2002){Yamazaki}, {Ioka}, \& {Nakamura}}]{2002ApJ...571L..31Y}
{Yamazaki}, R., {Ioka}, K., \& {Nakamura}, T. 2002, \apjl, 571, L31, \dodoi{10.1086/341225}

\bibitem[{{Yang} {et~al.}(2015){Yang}, {Jin}, {Li}, {Covino}, {Zheng}, {Hotokezaka}, {Fan}, {Piran}, \& {Wei}}]{2015NatCo...6.7323Y}
{Yang}, B., {Jin}, Z.-P., {Li}, X., {et~al.} 2015, Nature Communications, 6, 7323, \dodoi{10.1038/ncomms8323}

\bibitem[{{Yang} {et~al.}(2022){Yang}, {Ai}, {Zhang}, {Zhang}, {Liu}, {Wang}, {Yang}, {Yin}, {Li}, \& {L{\"u}}}]{2022Natur.612..232Y}
{Yang}, J., {Ai}, S., {Zhang}, B.-B., {et~al.} 2022, \nat, 612, 232, \dodoi{10.1038/s41586-022-05403-8}

\bibitem[{{Yang} {et~al.}(2023){Yang}, {Troja}, {O'Connor}, {Fryer}, {Im}, {Durbak}, {Paek}, {Ricci}, {De Bom}, {Gillanders}, {Castro-Tirado}, {Peng}, {Dichiara}, {Ryan}, {van Eerten}, {Dai}, {Chang}, {Choi}, {De}, {Hu}, {Kilpatrick}, {Kutyrev}, {Jeong}, {Lee}, {Makler}, {Navarete}, \& {P{\'e}rez-Garc{\'\i}a}}]{2023arXiv230800638Y}
{Yang}, Y.-H., {Troja}, E., {O'Connor}, B., {et~al.} 2023, arXiv e-prints, arXiv:2308.00638, \dodoi{10.48550/arXiv.2308.00638}

\bibitem[{{Yonetoku} {et~al.}(2020){Yonetoku}, {Mihara}, {Doi}, {Sakamoto}, {Tsumura}, {Ioka}, {Amaya}, {Arimoto}, {Enoto}, {Fujii}, {Goto}, {Gunji}, {Hiraga}, {Ikeda}, {Kawai}, {Kurosawa}, {Li}, {Maeda}, {Mitsuishi}, {Murakami}, {Nakagawa}, {Ogino}, {Ohno}, {Sawano}, {Sei}, {Serino}, {Sugita}, {Tamagawa}, {Tamura}, {Tanaka}, {Tanimori}, {Tashiro}, {Tomida}, {Wang}, {Yamaguchi}, {Yamamoto}, {Yamaoka}, {Yamauchi}, {Yatsu}, {Yoshida}, {Yuhi}, {Akitaya}, {Fukui}, {Ita}, {Kaneda}, {Kawabata}, {Kawata}, {Kurimata}, {Matsumoto}, {Matsuura}, {Miyasaka}, {Motohara}, {Narita}, {Noda}, {Ohashi}, {Okita}, {Sano}, {Tanaka}, {Urata}, {Wada}, {Yamaguchi}, {Yanagisawa}, {Yoshida}, {Asano}, {Inayoshi}, {Inoue}, {Ito}, {Izumiura}, {Kawanaka}, {Kinugawa}, {Kisaka}, {Kiuchi}, {Matsumoto}, {Mizuta}, {Murase}, {Nagakura}, {Nagataki}, {Nakada}, {Nakamura}, {Niino}, {Suwa}, {Takahashi}, {Tanaka}, {Toma}, {Totani}, {Yamazaki}, \& {Yokoyama}}]{2020SPIE11444E..2ZY}
{Yonetoku}, D., {Mihara}, T., {Doi}, A., {et~al.} 2020, in Society of Photo-Optical Instrumentation Engineers (SPIE) Conference Series, Vol. 11444, Society of Photo-Optical Instrumentation Engineers (SPIE) Conference Series, 114442Z, \dodoi{10.1117/12.2560603}

\bibitem[{{Yuan} {et~al.}(2022{\natexlab{a}}){Yuan}, {Zhang}, {Chen}, \& {Ling}}]{2022arXiv220909763Y}
{Yuan}, W., {Zhang}, C., {Chen}, Y., \& {Ling}, Z. 2022{\natexlab{a}}, arXiv e-prints, arXiv:2209.09763, \dodoi{10.48550/arXiv.2209.09763}

\bibitem[{{Yuan} {et~al.}(2022{\natexlab{b}}){Yuan}, {Zhang}, {Chen}, \& {Ling}}]{2022hxga.book...86Y}
---. 2022{\natexlab{b}}, in Handbook of X-ray and Gamma-ray Astrophysics, 86, \dodoi{10.1007/978-981-16-4544-0_151-1}

\bibitem[{{Zhang} {et~al.}(2006){Zhang}, {Fan}, {Dyks}, {Kobayashi}, {M{\'e}sz{\'a}ros}, {Burrows}, {Nousek}, \& {Gehrels}}]{2006ApJ...642..354Z}
{Zhang}, B., {Fan}, Y.~Z., {Dyks}, J., {et~al.} 2006, \apj, 642, 354, \dodoi{10.1086/500723}

\bibitem[{{Zhu} {et~al.}(2022){Zhu}, {Wang}, {Sun}, {Yang}, {Li}, {Hu}, {Qin}, \& {Wu}}]{2022ApJ...936L..10Z}
{Zhu}, J.-P., {Wang}, X.~I., {Sun}, H., {et~al.} 2022, \apjl, 936, L10, \dodoi{10.3847/2041-8213/ac85ad}

\bibitem[{{Zhu} {et~al.}(2023){Zhu}, {Wu}, {Yang}, {Liu}, {Zhang}, {Song}, {Gao}, {Cao}, {Yu}, {Kang}, \& {Shao}}]{2023ApJ...942...88Z}
{Zhu}, J.-P., {Wu}, S., {Yang}, Y.-P., {et~al.} 2023, \apj, 942, 88, \dodoi{10.3847/1538-4357/aca527}

\bibitem[{{Zou} {et~al.}(2018){Zou}, {Wang}, {Moharana}, {Liao}, {Chen}, {Wu}, {Lei}, \& {Wang}}]{2018ApJ...852L...1Z}
{Zou}, Y.-C., {Wang}, F.-F., {Moharana}, R., {et~al.} 2018, \apjl, 852, L1, \dodoi{10.3847/2041-8213/aaa123}

\end{thebibliography}
\bibliographystyle{aasjournal}


\appendix
\section{Analytic model for jet propagation}
\label{ap:jet}
\subsection{Basic model}
\label{ap:basic model}
We follow the same physical model for the jet-cocoon in \cite{2021MNRAS.500..627H} with the same approximations. 
In this model, the jet-cocoon is determined using the following system of equations (initially derived by \citealt{2011ApJ...740..100B} for a static medium) that is solved analytically:
\begin{align}
\label{eq:S1}
  \frac{dr_c(t)}{dt} =& c\beta_\perp(t) , \\
\label{eq:S2}
    \beta_{\perp}(t) =&
    \sqrt{\frac{P_{c}(t)}{{\rho}_{e}(r_h/2,t) c^{2}}} +\left[\frac{r_c(t)}{r_m(t)}\right]\beta_m, \\
    \label{eq:S3}
    P_{{c}}(t) =& \:\:\:\:\:\: \frac{E_{i}(t)}{3\:V_{{c}}(t)} \:\:=  \eta\frac{L_j\left(1-\langle{\beta_h}\rangle \right) \:(t-t_0)}{2 \pi r_c^{2}(t) r_{{h}}(t)} , \\
    \label{eq:S4}
    \Sigma_j(t) =& \pi r_h^2(t) \theta_j^2(t) = \frac{L_j \theta_0^2}{4 c P_c(t)} ,
\end{align}
where $r_c(t)$ and $r_h(t)$ are the semi-minor and semi-major axis of the ellipsoidally shaped cocoon [$r_h(t)$ is also the jet head radius] so that the cocoon volume (including both hemispheres) is written as:
\begin{equation}
    V_c(t) =\frac{4\pi}{3}{r_c(t)^2r_h(t)} ,
    \label{eq:Vc}
\end{equation}
Also, $\beta_\perp(t)$ is the lateral expansion velocity of cocoon's semi-minor axis, 
$P_c(t)$ is the pressure inside the cocoon (determined by jet-shock heating), 
$\rho_e(r,t)$ is the mass-density of the ejecta, 
$r_m(t)$ is the outer radius of the ejecta and $\beta_m$ is its velocity, 
$E_i(t)$ is the internal energy in the cocoon, 
$\eta$ is a parameter to quantify the fraction of internal energy in the cocoon (taken as $\eta\sim 0.5$; i.e., energy equipartition of energy is assumed after the passage of the jet-shock),
$\langle{\beta_h}\rangle=\frac{r_h-r_0}{c(t-t_0)}$ is the average jet head velocity (from $t_0$ to $t$),
$L_j$ is the jet power (one side),
$\Sigma_j(t)$ is the cross section of the jet with its opening angle as $\theta_j(t)$.

Solving this system of equations gives the spatial (e.g., morphology and volume) and the macroscopic physical quantities (i.e., total mass, internal energy) that defines the jet and the cocoon.
For more details see \cite{2020MNRAS.491.3192H} and \cite{2021MNRAS.500..627H}.

\subsection{Late jet propagation}
\label{sec:late jet approx}
After the prompt phase, engine activity (and jet propagation) is considered to continue to much later times.
Ideally, one should solve jet propagation with hydrodynamical simulations. 
However, solving jet propagation at such later (and long) timescales and the required large computational domain make it extremely challenging.
Therefore, the best available approach is to solve this problem analytically with ideas from simulations of prompt jets.

We assume that as the prompt jet is turned off at the end of the prompt phase (temporarily before the extended phase), it takes a brief moment, for the extended jet to be launched (through the injection nozzle with $\beta=\beta_0$) [see Table \ref{tab:models}]; this would allow the prompt-jet cocoon material to redistribute (closing the hole made by the prompt-jet). 
One may assume that the late jet would later propagate through the path (i.e., hole) created by the prompt jet.
However, this seems unlikely considering the results of hydrodynamical simulations:
\begin{itemize}
    \item First, it has been shown in \cite{2023MNRAS.520.1111H} that most ($\sim 95-99\%$) of the prompt-jet cocoon is trapped inside the ejecta. Furthermore, numerical simulations show that the trapped cocoon (in particular its inner part) is pressurized after the jet breakout. 
    Hence, this trapped cocoon is expected to close the prompt-jet-made path, as soon as the prompt-jet-engine is turned off (see bottom panels in Figure 2 of \citealt{2023MNRAS.520.1111H}); 
    therefore, late jets are expected to have to propagate (and penetrate) through this trapped cocoon, with the mass-density comparable to that of the ejecta, once again.
    \item Mass ejection is expected to dominate at later times $t\sim 1-10$ s (see \citealt{2018ApJ...860...64F}; and Figure 5 in \citealt{2020ApJ...901..122F} in particular). 
    This implies that even if the prompt jet has created a path through the ejecta in the polar direction, the post-merger ejecta tries to fill this path.  
    \item Observations of the extended/plateau emission indicate that the intrinsic luminosity of late time jets is orders of magnitude less than that of prompt jets (see Figure \ref{fig:intro} and Table \ref{tab:models}). 
    Considering that the ejecta radius is much larger at these times, and the post-merger mass ejection,
    it is difficult to consider that such significantly weaker jets will penetrate (i.e., moving with $\sim c$) and breakout without interaction.
\end{itemize}
As a result, it is reasonable to assume that late jets propagation forms a jet-head structure, in a similar manner to prompt jets, 
in particular, when there is even a slight interval between the prompt jet and the late jet.
Hence, this jet-head structure is expected to shock (and heat) the ejecta, forming another cocoon component.
Therefore, we apply the same analytical model in §\ref{ap:basic model} for late jets.

\section{Analytic model for cocoon breakout and escape}
\label{ap:es}
We follow the analytic model presented in \cite{2023MNRAS.520.1111H} with the same approximations.
The model is based on the parameter $\alpha$, which is the ratio of the energy density of the cocoon (jet energy + kinetic energy of the expanding ejecta) over the energy density of the expanding ejecta (kinetic energy of the ejecta). 
$\alpha$ can be written as a function of the initial condition, breakout time $t_b$, and cocoon volume (over the ejecta volume: $V_c/V_e$), as:
\begin{eqnarray}
\alpha \approx 1+\frac{3\theta_0^2 L_{iso,0}(t_b-t_0)\left\{1-\beta_m\frac{t_b}{(t_b-t_0)}\right\}}{(V_c/V_e)\beta_m^2M_{e}c^2} .
\label{eq:alpha tb}
\end{eqnarray}
This parameter is crucial to determine the part of the cocoon that can escape the ejecta.
Then, 
the cocoon part that is able to escape has the fractions of the mass, energy (kinetic+internal), and internal energy, which can be written
as a function of $\alpha$, as bellow [similarly to equations (44), (52), and (53) in \cite{2023MNRAS.520.1111H}]:
\begin{eqnarray}
\frac{M_{c}^{es}}{M_{c}}\approx& \frac{ \frac{1}{\sqrt{\alpha}}-1+\ln\left(\sqrt{\alpha}\right)}{\ln \left(\frac{\beta_{m}}{\beta_{0}}\right)-1} ,\label{eq:es mass}\\
\frac{E_{c}^{es}}{E_c} \approx& \left(\frac{1+\alpha}{2\alpha-(\alpha-1)/f_{mix}}\right)\left(1+\frac{2}{\alpha^{3/2}}-\frac{3}{\alpha}\right) ,\\
\frac{E_{c,i}^{es}}{E_{c,i}} \approx& \frac{1}{f_{mix}}  \left(\frac{1+\alpha}{2\alpha-(\alpha-1)/f_{mix}}\right)\left(1+\frac{2}{\alpha^{3/2}}-\frac{3}{\alpha}\right) ,
\label{eq:es all}
\end{eqnarray}
where $f_{mix}\sim 2/3$ (or $\sim 1$ for failed jet case) is a mixing parameter for the shocked jet and ejecta.
The cocoon mass, energy, and internal energy, can be found at the breakout time as $\frac{M_c}{M_{e}}\approx 2 \left[\ln\left(\frac{r_b}{r_0}\right)-1\right] \frac{V_c}{V_{e}}
$, $E_c \approx \alpha E_e \left(\frac{V_c}{V_e}\right)$, and $E_{c,i}\approx 3P_c V_c$, respectively.
It should be noted that the fraction of the escaped cocoon, in particularly in terms of mass, is very small ($\sim0.5 -5\%$).
After the breakout, and until $t_1 \sim 2-3 t_b$, these quantities (all the cocoon's and escaped cocoon's) are expected to grow by a factor of $f_g\sim 2$ (due to the post-breakout cocoon growth).

Additionally as shown in \cite{2023MNRAS.520.1111H}, the escaped cocoon takes the shape of a cone with $\theta_c^{es}$ as its opening angle.
A reasonable approximation of this angle is found (empirically) as determined by the solid angle of the cocoon ellipsoidal at its midpoint (at $r_b/2$, at the breakout time; while accounting for the lateral expansion at $t>t_b$): 
\begin{eqnarray}
\theta_c^{es} \sim \arctan\left[\frac{r_c \sqrt{f_g}}{r_b/2}\right] .
    \label{eq:theta c es}
\end{eqnarray}
The term $\sqrt{f_g}$ accounts for the growth of the cocoon after the breakout (in terms of mass and energy; see \citealt{2023MNRAS.520.1111H}).
This is explained by the cocoon's volume growth (by a factor $f_g$), and more specifically by the lateral expansion of the cocoon (in $r_c$, in the comoving frame of the ejecta) by a factor of $\sqrt{f_g}$ (as $V_c\propto  r_c^2$).

\section{Analytic model for the escaped cocoon's cooling emission}
\label{ap:cooling}
We follow the analytic model in \cite{2023MNRAS.524.4841H}, originally for the prompt jet case,
with the exception that key approximations, $t\gg t_b$ and $r\approx c\beta t$ (practical for the prompt jet case), are not ideal for late jets (as $t_b$ is much larger), and have to be bypassed.

\begin{figure}
    \centering
    \includegraphics[width=0.95\linewidth]{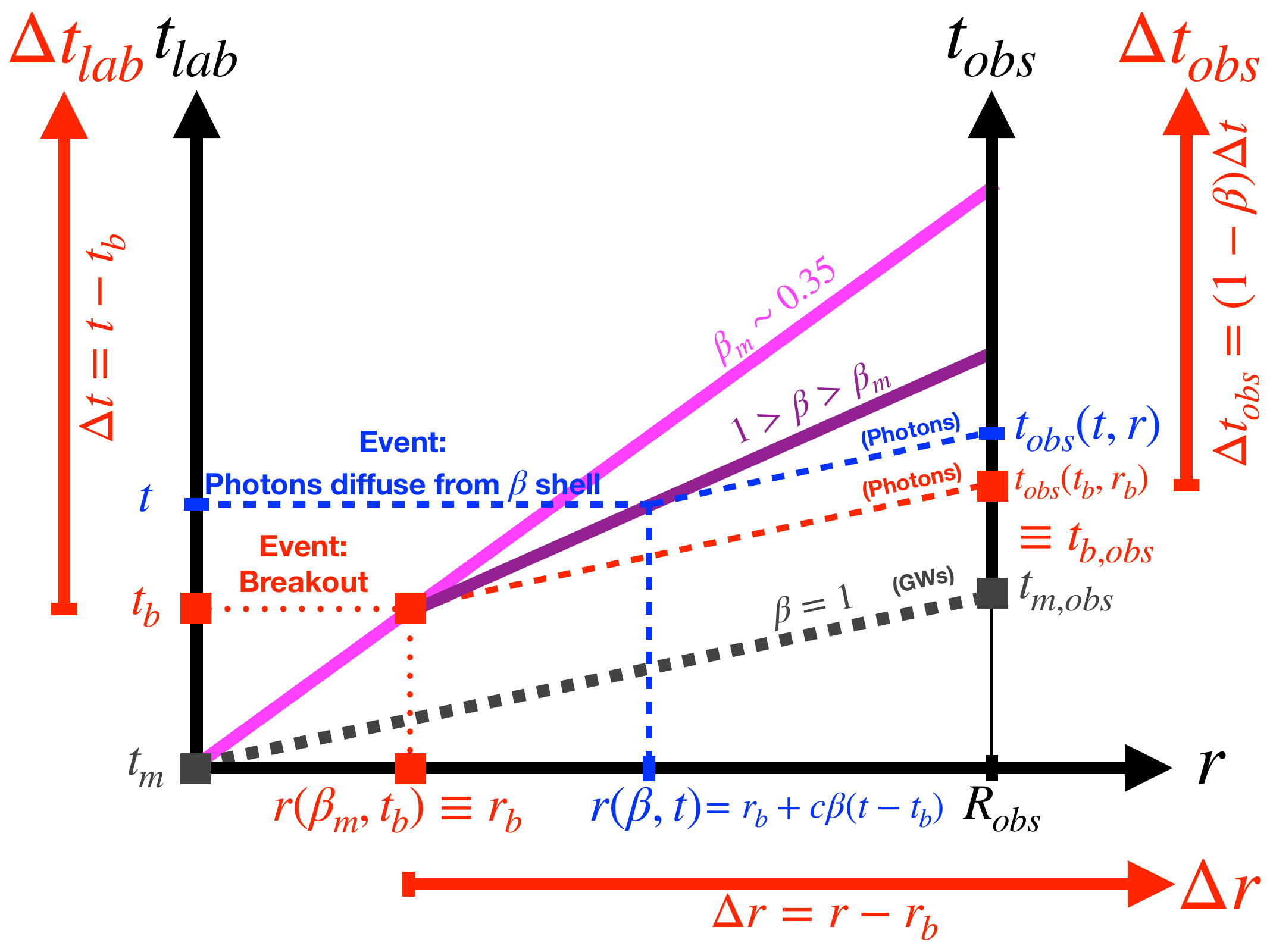} 
  \caption{Illustration of the relation between the laboratory time (since the merger time; with $t_m=0$) $t_{lab}\equiv t$,
  the radius $r$ (from the central engine's center),
  and the observed time $t_{obs}$ (for an observer at $R_{obs}$, and since the observed merger time; $t_{m,obs}=0$) [black solid lines].
  Colored solid lines show the homologous expansion of the outer edge of the ejecta (magenta) and a given shell of the escaped cocoon with a velocity $\beta$ (purple).
  Dashed lines show radiation (GWs in grey, breakout emission in red, and cooling emission from the shell moving with $\beta$ in blue).
  The event of jet breakout is shown in red, in each of the three coordinates.
  The alternative coordinates used here with the event of jet breakout as their starting point ($\Delta t$, $\Delta r$, and $\Delta t_{obs}$) are shown with red solid lines. 
  The choice of colors here is arbitrary with no connection to observed colors.
  }
  \label{fig:time} 
\end{figure}

For that, the formulations is generalized.
In Figure \ref{fig:time} we show the laboratory time, radius, and observed time, with the merger as the reference ($t$, $r$, and $t_{obs}$; as used in \citealt{2023MNRAS.524.4841H}).
The alternative coordinates are also shown (in red), with the jet breakout time (or radius) as the new reference, as follows:

\begin{eqnarray}
\Delta t =& t -t_b    ,\\
\Delta r =& r -r_b \approx c\beta\Delta t    ,\\
\Delta t_{obs} =& t_{obs} -t_{b,obs}    ,
\label{eq:delta}
\end{eqnarray}
with $r_b$ as the breakout radius (in the central engine frame), $t_{b,obs}=t_b(1-\beta_m)$ as the observed breakout time since the merger time ($t_{m,obs}=0$).

For the cocoon breakout, the following picture is considered.
As the escaped cocoon is originally located in the ellipsoidal cap of the pre-breakout cocoon, and as this cap is usually a small fraction of the cocoon \citep{2023MNRAS.520.1111H}, we approximately consider the escaped cocoon's breakout take places in the polar axis of the ejecta (at the point $r=r_b$ and $\theta\sim 0$).
Considering that the spread of the escaped cocoon can be approximated to a cone structure with the opening angle $\theta_c^{es}$ [see equation (\ref{eq:theta c es})], 
the base of this cone has been set at the same point ($r=r_b$ and $\theta=0$ in the central engine frame).
Note that, until the free-expansion phase is reached ($t\gtrsim t_1$, introduced next), our analytic model cannot be applied (due to the singularity in around $t\sim t_b$).

For the rest, the same procedure in \cite{2023MNRAS.524.4841H} is followed.
The escaped cocoon is assumed to accelerate outward after the breakout time, and reach the free expansion at $t=t_1\sim 2 t_b$ ($t=t_1\sim 1.5 t_b$ for the plateau jets where $t_b$ is long enough).
The volume of the escaped cocoon up to a certain velocity $\beta$ can be found as:
\begin{equation}
    V(\beta>\beta_m,\Delta t)=\frac{\Omega}{3}\left[\left(c\beta\Delta t\right)^3-\left(c\beta_m\Delta t\right)^3\right] ,
    \label{eq:vol}
\end{equation}
where $\Omega =4\pi(1-\cos\theta_c^{es})$.

The escaped cocoon is divided into two parts: relativistic ($\beta_{out}>\beta>\beta_t$) where $\Gamma\beta\sim \Gamma$, and non-relativistic  ($\beta_{t}>\beta>\beta_m$) where $\Gamma\beta\sim \beta$ (with $\Gamma_{out}\approx 10$, $\beta_t\approx 0.8$, and $\beta_m\approx 0.35$), each with a different power-law distribution for the mass-density (indices are $l\approx 0$ and $m\approx 8$, respectively)
and internal energy density (indices are $-3$ and $2$, respectively; see \citealt{2023MNRAS.524.4841H}).

Hence, the mass-densities can be found as:
\begin{equation}
    \rho_{c}(\Gamma\beta, \Delta t) \equiv \begin{cases}
\rho_{c, r}(\Gamma, \Delta t)=\rho_t(\Delta t)\left(\frac{\Gamma}{\Gamma_t}\right)^{-l} \\
\rho_{c, n r}(\beta, \Delta t)=\rho_t(\Delta t)\left(\frac{\beta}{\beta_t}\right)^{-m}
\end{cases}
\end{equation}
with $\rho_t(\Delta t) =\left[\frac{M_{c, r}^{e s}(2+l)}{\Omega(c \Delta t)^3 \Gamma_t^l}\right]\left\{\Gamma_t^{-(2+l)}-\Gamma_{\text {out }}^{-(2+l)}\right\}^{-1}
=\left[\frac{M_{c, n r}^{e s}(m-3)}{\Omega(c \Delta t)^3 \beta_t^m}\right]\left\{\beta_m^{-(m-3)}-\beta_t^{-(m-3)}\right\}^{-1} $.
These expression are equivalent to those in \cite{2023MNRAS.524.4841H} with $t$ substituted by $\Delta t$.

Similarly, assuming grey opacity ($\kappa =$ Const.) gives the observed time of a given shell $\Gamma_d\beta_d$ as a function of its optical depth, as:
\begin{equation}
\begin{split}
\Delta t_{o b s}(\Gamma_d\beta_d)\approx
\begin{cases}
 \left\{\frac{1}{\tau_r}\left[\frac{\kappa M_{c, r}^{e s}}{c^2 \Omega} \frac{2+l}{4(6+l)}\right] \frac{\Gamma_d^{-(6+l)}-\Gamma_{o u t}^{-(6+l)}}{\Gamma_t^{-(2+l)}-\Gamma_{{out }}^{-(2+l)}}\right\}^{\frac{1}{2}} \\
\text{(Rela. cocoon)},\\
\\
 \left\{\frac{1}{\tau_{n r}} \frac{\kappa M_{c, n r}^{e s} \beta_m^{m-3}}{\Omega c^2}\left[\frac{(1-\beta_d)^2}{\beta_d^{m-1}}-\frac{\left(1-\beta_t\right)^2}{\beta_t^{m-1}}\right]\right\}^{\frac{1}{2}}\\
\text{(N. Rela.  cocoon)} .\\
\end{cases} 
\end{split}
\label{eq:tobs R NR} 
\end{equation}
$\tau_r$ and $\tau_{nr}$ are the optical depth.
We use the same criteria as in \cite{2023MNRAS.524.4841H} for photon diffusion from a thin shell: $\tau_{r}=1$; and $\tau_{nr} \sim \frac{1}{\Delta\beta'}\sim \frac{20}{\beta_d}$.

With the assumption that the free-expansion is reached at $\sim t_1\sim 2t_b$, adiabatic cooling starts after $t_1$\footnote{\label{foot:t1} This is consistent with numerical simulations in \cite{2023MNRAS.520.1111H}. 
Also, this is a similar effect to the geometrical effect discussed in \cite{2010ApJ...725..904N}.
One likely explanation is that, 
as the escaped cocoon fluid accelerates during its escape, 
and before it converges to a homologous structure at $t_1$ (free-expansion), 
internal shocks happen, between the unorganized escaping cocoon shells, enhancing the internal energy content. 
Therefore, the full effect of the adiabatic cooling is not seen until the free expansion is reached, as these interactions can no longer happen. 
As shown in \cite{2023MNRAS.520.1111H}, simulations show that internal energy is boosted by $f_g\sim 2$, and since $t_1\sim 2 t_b$, one can approximate the adiabatic expansion to start after $t_1$.
This approximation is not expected to affect the cocoon cooling emission, as photons are mostly released at $t\gg t_1$.
}.
One can find [using equation (\ref{eq:vol})] that
$V(>\beta_m,t>t_1) \propto \Delta t^3$ and the coefficient for internal energy loss due to adiabatic cooling is $\left[\frac{V(\beta>\beta_m,\Delta t_1)}{V(\beta>\beta_m,\Delta t)}\right]^\frac{1}{3}=\frac{\Delta t_1}{\Delta t}=\frac{t_1-t_b}{t-t_b}\propto \frac{1}{\Delta t}$.
Hence, the evolution of the internal energy can be found as:
\begin{equation}
    E_i(\Gamma\beta,t>t_1)=E_i(\Gamma\beta,t_1)\frac{\Delta t_1}{\Delta t} .
    \label{eq:adia}
\end{equation}

Then, the internal energy beyond a certain cocoon shell (moving with $\Gamma\beta$) and at a given time ($\Delta t$) can be found (for the relativistic and non-relativistic cocoon parts, respectively) as:
\begin{equation}
    \begin{cases}
E_{c,i,r}^j(>\Gamma,\Delta t)=E_{c,i,r}^j(\Delta t)\left[\frac{\ln(\Gamma_{out}/\Gamma_{})}{\ln(\Gamma_{out}/\Gamma_t)}\right],\\
E_{c,i,nr}^j(>\beta,\Delta t)=E_{c,i,nr}^j(\Delta t)\left[\frac{\ln(\beta_{t}/\beta_{})}{\ln(\beta_{t}/\beta_m)}\right],\\
\end{cases}
\label{eq:Ej R NR}
\end{equation}
where $E_{c,i,r}^j(\Delta t)$ and $E_{c,i,nr}^j(\Delta t)$ are the total internal energies (from jet heating) at a given time ($\Delta t$), in the relativistic, and non-relativistic parts of the cocoon, respectively.
Here too, these expressions are equivalent to those in \cite{2023MNRAS.524.4841H} (with $t$ substituted by $\Delta t$).

Next, using the time evolution of the diffusion velocity, the internal energy density [found using equations (\ref{eq:vol}) and (\ref{eq:Ej R NR})] is used to estimate the radiated energy (in the form of blackbody photons); i.e., cooling luminosity.
This is determined differentially (avoiding any one-zone approximations due to their inaccuracy), while taking into account the effect of the observer time and the relativistic beaming effect.
The cooling bolometric luminosity can be found using:
\begin{equation}
L_{bol}^j(\Delta t_{obs})=
\frac{\partial E_{c,i}^j(>\Gamma_d\beta_d,\Delta t)}{\partial \Gamma_d\beta_d }\frac{\partial \Gamma_d\beta_d }{\partial \Delta t}\times\frac{d\Delta t_{}}{d\Delta t_{obs}}\times\frac{4\pi}{\Omega_{EM}} ,
\label{eq:Lbl j}
\end{equation}
where $\Omega_{EM} \approx 4\pi(1-\cos[\max\{\arcsin[{\Gamma_d}^{-1}],\theta_{c}^{es}\}])$ is the solid angle of the emission coming from the shell moving with $\Gamma_d$.
The observed time can also be calculated as a function of the laboratory time as:
\begin{eqnarray}
\Delta t_{obs} =(1-\beta)\Delta t .
\label{eq:t obs lab}
\end{eqnarray}
With equations (\ref{eq:tobs R NR}) and (\ref{eq:Ej R NR}), the full analytic expression of the jet-shock powered bolometric luminosity can be found as a function of $\Delta t_{obs}$ (by varying the diffusion velocity $\Gamma_d\beta_d$) as:
\begin{equation}
\begin{split}
L_{b l}^j\left(\Delta t_{o b s}\right) \approx 
    \begin{cases}
 &\frac{E_{c, i, r}^j\left(>\Gamma_d, t\right)\left[\frac{1 / 3}{\ln \left(\Gamma_{o u t} / \Gamma_d\right)}\right]}{\Delta t_{o b s}} \times \frac{4 \pi}{\Omega_{E M}} \\
 & [\Gamma_{out}>\Gamma_d\gg 1] \\
 &\text{(Rela. cocoon)} ,\\
 \\
&\frac{E_{c, i, n r}^j\left(>\beta_d, t\right)}{\Delta t_{o b s}} \frac{4 \pi}{\Omega_{E M}} \\
& [\beta_d\lesssim \beta_t] \\
&\text{(N. Rela. cocoon)},\\
 &\frac{E_{c, i, n r}^j\left(>\beta_d, t\right)}{\Delta t_{o b s}}\left[\frac{\frac{4 \pi}{\Omega_{E M}}}{\ln \left(\beta_t / \beta_d\right)\left(\frac{m-2}{2}+\frac{\beta_d}{1-\beta_d}\right)}\right] \\
 & [\beta_d\gtrsim \beta_m] \\
 &\text{(N. Rela. cocoon)}.\\
\end{cases}
\end{split}
\label{eq:Lbl full}
\end{equation}

The photospheric radius $r_{ph}$ (and velocity $\Gamma_{ph}\beta_{ph}$) and the observed temperature can be found using equations (\ref{eq:rph}) and (\ref{eq:Teff}), after applying the criteria $\tau_{ph}=1$ for the optical depth.
All these expressions are similar to those found in \cite{2023MNRAS.524.4841H}, with $t$, $r_{ph}$, and $t_{obs}$, being substituted with $\Delta t$, $\Delta r_{ph}$, and $\Delta t_{obs}$, respectively.
The observed blackbody flux density can be found (for a given frequency $\nu$) as
\begin{equation}
F_\nu\left(T_{o b s}, \nu\right)=\frac{\pi B_\nu}{\sigma T_{o b s}^4} \frac{L_{b l}}{4 \pi D^2},     
\end{equation}
where $B_\nu\left(T_{o b s}\right)=\frac{2 h \nu^3}{c^2} \frac{1}{e^{\frac{h \nu}{k_b T_{o b s}}}-1}$ and $D$ is the luminosity distance ($D=40$ Mpc here for reference).

\section{Analytic model for the trapped cocoon's cooling emission}
\label{ap:trapped}
In addition to cooling emission from the escaped cocoon, here we present an analytic estimation of the cooling emission from the jet-shock-heated trapped cocoon (i.e., $\beta<\beta_m$).
The same method is used to calculate the cooling emission from the escaped cocoon as in the following.
Notable differences are the following: 
\begin{itemize}
    \item The morphology of the trapped cocoon's shape, although better represented by that of an ellipsoidal at the breakout, is approximated to take the form of a cone with a solid angle $\Omega_c$ at much later times $t\gg t_b$, so that:
    \begin{equation}
    \Omega_{c}=\frac{M_c}{M_e}. 
    \label{eq:Omegac}
    \end{equation}
    The opening angle of the trapped cocoon is larger than the opening angle of the escaped cocoon (by about a factor $\sim 2$).
    \item As the escaped cocoon mass is negligible compared to the overall mass of the cocoon, we approximate that mass-density of the trapped cocoon is comparable to the mass-density of the ejecta (see Figure 2 in \citealt{2023MNRAS.524.4841H}).
    Hence, the trapped cocoon's mass-density profile is fitted with the same power-law index $n$, as the ejecta.
    \item When estimating the optical depth of the trapped cocoon, optical thickness of the outer escaped cocoon is neglected.
    \item Since the velocity of the trapped cocoon (and the ejecta) is significantly smaller, the non-relativistic limit $\beta\ll 1$ is applied, allowing us to express the observed time and the laboratory time interchangeably; $t_{obs}\sim t$.
    \item Cooling emission from the trapped cocoon is approximated to start once the bolometric luminosity of the trapped cocoon dominates (over the escaped cocoon's).
\end{itemize}

The mass-density of the trapped cocoon can be found as
\begin{equation}
\rho_c(\beta<\beta_m,t) = \frac{M_c(3-n)}{\Omega_c(ct)^3[\beta_m^{3-n}-\beta_0^{3-n}]}\beta^{-n}.    
\end{equation}
Calculating the optical depth as $\tau\approx \int_\beta^{\beta_{o u t}} \kappa\rho (ct)d \beta$ gives the following expression of time, at which the cooling emission from the $\beta_d$ shell is observed:
\begin{equation}
    t_{o b s}(\beta_d) \approx\left\{\frac{1}{\tau_{tr}} \frac{\kappa M_{c}\left(\frac{3-n}{n-1}\right) }{\Omega_c c^2(\beta_m^{3-n}-\beta_0^{3-n})}\left[\frac{1}{\beta_d^{n-1}}-\frac{1}{\beta_m^{n-1}}\right]\right\}^{\frac{1}{2}} .
    \label{eq:time tr}
\end{equation}
The criteria for photon diffusion from the $\beta_d$ shell is (see \citealt{2015ApJ...802..119K,2023MNRAS.524.4841H}):
\begin{equation}
\tau_{tr} \sim \frac{1}{\beta_m-\beta_d} .    
\end{equation}

For the internal energy density of the trapped cocoon, taking into account that the pressure inside the trapped cocoon should roughly be constant, and after accounting for adiabatic internal energy loss and volume increase (in a homologously expanding system), one can find that $P_c(t)\propto V_c^{-\frac{4}{3}}(t)\propto t^{-4}$ while $P_c\propto \beta^0$.  
Hence
\begin{equation}
E_{c,i}^{tr}(>\beta,t)=E_{c,i}^{tr}(>\beta_0,t_1)\frac{t_1}{t}\frac{\beta_m^3-\beta^3}{\beta_m^3-\beta_0^3}, 
\label{eq:tr Ei}
\end{equation}
with $t_1\sim 2 t_b$ is the time from which the cocoon is considered to be expanding freely, and $E_{c,i}^{tr}(t_1,>\beta_0)= E_{c,i} - E_{c,i}^{es}$ is the total internal energy in the trapped cocoon (accounting for the escaped internal energy).

In the non-relativistic limit, the cooling emission can be found as:
\begin{equation}
    L_{bol}^{tr}\approx \frac{\partial E_{c,i}^{tr}(>\beta,t)}{\partial\beta} \frac{\partial \beta}{\partial t} .
\end{equation}
With $\frac{\partial E_{c,i}^{tr}(>\beta,t)}{\partial\beta} \approx \frac{-3\beta^2}{\beta_m^3-\beta^3}E_{c,i}^{tr}(>\beta,t)$ [using equation (\ref{eq:tr Ei})], 
and $\frac{\partial t}{\partial\beta} = -\frac{t}{2}\left[\frac{(n-1)\beta^{-n}}{\beta^{1-n}-\beta_m^{1-n}}+\frac{1}{\beta_m-\beta}\right]$ [using equation (\ref{eq:time tr})],
the bolometric cooling luminosity can be found as:
\begin{equation}
    L_{bol}^{tr}=\frac{E_{c,i}^{tr}(>\beta,t_{lab})}{t_{obs}}\frac{\left(\frac{6\beta^2}{\beta_m^3-\beta^3}\right)}{\left[\frac{(n-1)\beta^{-n}}{\beta^{1-n}-\beta_m^{1-n}}+\frac{1}{\beta_m-\beta}\right]} .
    \label{eq:Lbl tr}
\end{equation}

The effective temperature can then be found using Stefan-Boltzmann law:
\begin{equation}
    T_{KN}(t_{})  = \left\{\frac{L_{KN}^{}(t_{})}{\Omega_c \sigma r_{ph}^2 }\right\}^{\frac{1}{4}} ,
    \label{eq:T tr}
\end{equation}
with $r_{ph}=c\beta_{ph}t$, and $\beta_{ph}$ found for $\tau_{tr}=\tau_{ph}= 1$ in equation (\ref{eq:time tr}).

\section{Analytic model for the r-process powered kilonova}
\label{ap:KN}
We present a simple analytic model for the r-process powered KN emission from the merger ejecta.
The model shares several similarities to the one in \cite{2015ApJ...802..119K}.
The aim of this model is to estimate the early KN emission and give a relative comparison to the brightness of the cocoon emission.

Here too, the approximation of a diffusion shell ($\beta_d$), is used (see §\ref{ap:cooling}).
The same mass-density and velocity profile is considered for the KN ejecta, although here, focusing on the early part of the KN (i.e., blue KN), peaking  at $\sim 1$ day (in terms of magnitudes), only the blue component $M_{KN}\sim 0.02 \msun$ is considered.

First, similarly to §\ref{ap:trapped}, the optical depth is calculated to find the diffusion velocity $\beta_d$ as a function of time (assuming the same grey opacity $\kappa=1$ cm$^{2}$ {g}$^{-1}$).
The mass of the KN ejecta beyond the velocity $\beta_d$ can be found as:
\begin{equation}
    M_{KN}(>\beta)=M_{KN}\left(\frac{\beta_m^{3-n}-\beta^{3-n}}{\beta_m^{3-n}-\beta_0^{3-n}}\right) .
\end{equation}
At later times ($t\gg t_h$), the energy deposition (from r-process) in the optically thin KN ejecta is
\begin{equation}
    \dot{E}_{rp}(>\beta_d,t)=\dot{\varepsilon}_0M_e(>\beta_d)\left(\frac{t}{t_h}\right)^{-k} ,
    \label{eq:KN Edep}
\end{equation}
where $k\sim 1.3$, and $\dot{\varepsilon}_0\sim 1\times 10^{18}$ erg g$^{-1}$ s$^{-1}$ is the heating rate at $t_h\sim 0.1$ s  (or $\sim2\times 10^{10}$ erg g$^{-1}$ s$^{-1}$; see \citealt{2014ApJ...789L..39W} and \citealt{2021ApJ...922..185I}).

The bolometric luminosity of the KN is found as a function of time as the sum of two terms:
\begin{equation}
    L_{KN}(t)=L_{KN}(\geqslant\beta_d,t)+L_{KN}(<\beta_d,t) ,
        \label{eq:Lbl KN}
\end{equation}
where $L_{KN}(\geqslant\beta_d,t)$ is emission from the optically thin part of the KN ejecta, and $L_{KN}(<\beta_d,t)$ is the cooling emission from the optically thick region due to the inward motion (in Lagrangian coordinate) of the diffusion shell.

Taking into account that only a fraction of this deposited energy is thermelized, the first term can be found as:
\begin{equation}
    L_{KN}(\geqslant\beta_d,t)\approx f_{tot} \dot{E}_{rp}(\geqslant\beta_d,t),
\end{equation}
where we consider the thermalisation efficiency around 1 day (around the peak of the blue KN) $f_{tot}\sim 0.4$ (\citealt{2016MNRAS.459...35H}; \citealt{2016ApJ...829..110B}).

For the second term, the released internal energy can be found using 
$L_{KN}(<\beta,t)= f_{tot}\frac{\partial E_{rp}(>\beta,t)}{\partial\beta} \frac{\partial \beta}{\partial t}$
and the remaining r-process energy in the optically-thick region is found with the differential equation $\frac{\partial E_{rp}(>\beta,t)}{\partial t}=-\frac{E_{rp}(>\beta,t)}{t}+\dot{E}_{rp}(>\beta,t)$, with $\dot{E}_{rp}$ being the r-process energy deposition [see equation (\ref{eq:KN Edep})].
Following the same procedure as in §\ref{ap:cooling} and §\ref{ap:trapped}, analytic derivation gives:
\begin{equation}
    L_{KN}(<\beta_d,t)=f_{tot}\frac{\dot{E}(>\beta_d,t)}{2-k}
\left[\frac{\frac{2(3-n)\beta^{2-n}}{\beta_m^{3-n}-\beta_d^{3-n}}}{\frac{(n-1)\beta_d^{-n}}{\beta_d^{1-n}-\beta_m^{1-n}}+\frac{1}{\beta_m-\beta_d}}\right] .
\end{equation}

The effective temperature can then be found from Stefan-Boltzmann law [same as in equation (\ref{eq:T tr})]:
\begin{equation}
    T_{KN}(t_{})  = \left\{\frac{L_{KN}^{}(t_{})}{4\pi \sigma r_{ph}^2 }\right\}^{\frac{1}{4}} .
\end{equation}
Note that here too, as $\beta \ll 1$, the observed time and the laboratory time can be used interchangeably $t_{obs}\sim t$.

\label{lastpage}

\listofchanges

\end{document}